\documentclass[11pt,a4paper]{article}
\pdfoutput=1
\usepackage{jcappub}
\usepackage{afterpage}

\usepackage{color}
\usepackage{amsfonts, amsmath}
\usepackage{booktabs}
\usepackage[english]{babel}
\usepackage{colortbl}
\usepackage{epsfig}
\usepackage{epstopdf}
\usepackage{graphicx}
\usepackage{subfig}
\usepackage{lipsum}
\usepackage{hyperref}
\usepackage{listings}
\usepackage{setspace}
\usepackage{tabularx}

\DeclareFixedFont{\ttb}{T1}{txtt}{bx}{n}{8} 
\DeclareFixedFont{\ttm}{T1}{txtt}{m}{n}{8}  
\DeclareFixedFont{\ttc}{T1}{txtt}{m}{it}{8}  

\usepackage{color}
\definecolor{deepblue}{rgb}{0,0,0.5}
\definecolor{deepred}{rgb}{0.6,0,0}
\definecolor{deepgreen}{rgb}{0,0.5,0}
\definecolor{aurometalsaurus}{rgb}{0.43, 0.5, 0.5}

\newcommand\pythonstyle{\lstset{
language=Python,
commentstyle=\ttc\color{aurometalsaurus},
basicstyle=\ttm,
otherkeywords={self},             
keywordstyle=\ttb\color{deepblue},
emphstyle=\ttb\color{deepred},    
stringstyle=\color{deepgreen},
frame=shadowbox,                         
rulesepcolor=\color{black},
showstringspaces=false,            %
numbers=left,                  
numberstyle=\footnotesize\color{aurometalsaurus},     
numbersep=8pt
}}

\lstnewenvironment{python}[1][]
{
\pythonstyle
\setstretch{0.8}
\lstset{#1}
}
{}

\usepackage[T1]{fontenc} 
\usepackage[load-configurations=astronomy, range-units=brackets, range-phrase=-, per-mode=reciprocal, mode=math]{siunitx}

\newcommand{\vb}[1]{\mathbf{#1}}
\newcommand{\kth}[1]{k_{#1}}
\newcommand{\vx}{\vb{x}}
\newcommand{\vk}{\vb{k}}
\newcommand{\vM}{\vb{M}}
\newcommand{\vW}{\vb{W}}
\newcommand{\vC}{\vb{C}}
\def\kMpc{\, h \, {\rm Mpc}^{-1}}
\newcommand{\eq}[1]{eq.~(\ref{#1})}
\newcommand{\vbunit}[1]{\hat{\vb{#1}}}

\newcommand{\fig}[1]{figure~\ref{#1}}

\newcommand{\code}{\texttt}

\newcommand{\pvmhid}[1]{}

\newcommand{\Ptrue}{\bold{P}^{\rm true}}
\newcommand{\Pconv}{\bold{P}^{\rm conv}}
\newcommand{\Pconvo}{\bold{P}^{\rm conv}_{\rm o}}
\newcommand{\PconvoT}{\bold{P}^{\rm conv, T}_{\rm o}}
\newcommand{\Pdeconv}{\bold{P}^{\rm true}_{\rm o}}

\newcommand{\invCconv}{\bold{C}^{-1}_{\rm conv}}
\newcommand{\invCdeconv}{\bold{C}^{-1}_{\rm deconv}}
\newcommand{\Ptruefs}{\bold{P}^{\rm true, flat\text{-}sky}}
\newcommand{\PtruefsT}{\bold{P}^{\rm true, flat\text{-}sky, T}}
\newcommand{\tilW}{\tilde{W}}
\newcommand{\vtilW}{\vb{\tilW}}

\newcommand{\nbw}{\bar{n}_w}

\title{Unified galaxy power spectrum
measurements from 6dFGS, BOSS, and eBOSS}

\author[a]{Florian Beutler,}
\author[b]{Patrick McDonald}

\affiliation[a]{Institute for Astronomy, University of Edinburgh, Royal Observatory, Blackford Hill, Edinburgh EH9 3HJ, UK}
\affiliation[b]{Lawrence Berkeley National Lab, 1 Cyclotron Rd, Berkeley CA 94720, USA}

\emailAdd{florian.beutler@ed.ac.uk}
\emailAdd{pvmcdonald@lbl.gov}

\abstract{We make use of recent developments in the analysis of galaxy redshift
surveys to present an easy to use matrix-based analysis framework for the 
galaxy power spectrum multipoles, including wide-angle effects and the survey 
window function. We employ this framework to derive the deconvolved power 
spectrum multipoles of 6dFGS DR3, BOSS DR12 and the eBOSS DR16 quasar sample. 
As an alternative to the standard analysis, the deconvolved power spectrum 
multipoles can be used to perform a data analysis agnostic of survey specific 
aspects, like the window function. 
We show that in the case of the BOSS dataset, the Baryon Acoustic Oscillation 
(BAO) analysis using the deconvolved power spectra results in the same 
likelihood as the standard analysis. 
To facilitate the analysis based on both the convolved and deconvolved power 
spectrum measurements, we provide the window function matrices, wide-angle 
matrices, covariance matrices and the power spectrum multipole measurements 
for the datasets mentioned above. Together with this paper we publish a 
\code{Python}-based toolbox to calculate the different analysis components. 
The appendix contains a detailed user guide with examples for how a 
cosmological analysis of these datasets could be implemented. We hope that our 
work makes the analysis of galaxy survey datasets more accessible to the wider cosmology community.}

\begin{document}
\maketitle
\flushbottom

\section{Introduction}
\label{sec:intro}

In the last two decades the analysis of galaxy redshift survey datasets has become one of the most powerful tools to constrain cosmological models~\citep{Alam2016:1607.03155v1,Collaboration]2020:2007.08991v1}. The next generation of galaxy redshift surveys, such as the Dark Energy Spectroscopic Instrument (DESI~\citep{2016:1611.00036v2}) and the Euclid space mission~\citep{Laureijs2011:1110.3193v1}, aim to observe $20$ to $40$ million galaxy redshifts, increasing the largest current dataset (BOSS~\citep{Dawson2012:1208.0022v3}, 1 million galaxies) by more than an order of magnitude. While observations of the Cosmic Microwave Background (CMB) have reached the sample variance limit for some observables, galaxy surveys are far away from this limit. 
Extracting cosmological information from galaxy surveys does pose significant challenges. Most modes which can be accessed by galaxy surveys are contaminated by non-linear contributions along with redshift-space distortions and a complicated relation between galaxy density and matter density. Many different avenues are currently investigated to tackle these challenges (e.g.~\citep{Taruya2010:1006.0699v1,Senatore2014:1409.1225v1,McDonald2009:0902.0991v1,Seljak2011:1109.1888v1}). Here we will ignore these issues, but focus on the analysis formalism itself.

Extracting cosmological information from the galaxy power spectrum comes 
with technical complications, such as wide-angle effects 
(e.g., \citep{Reimberg2015:1506.06596v2,Castorina2017:1709.09730v2,
Beutler2018:1810.05051v3})
and the survey 
window function, which convolves the measured power spectrum. While the 
survey window is usually known, this convolution can add significant 
modeling challenges. The current standard analysis derives the survey 
window function from the data and convolves any power spectrum model with 
the survey window before comparing it with the power spectrum  
measurement~\citep{Beutler2013:1312.4611v2,Wilson2015:1511.07799v2,Beutler2016:1607.03150v1}. Instead of convolving the model we can also deconvolve the measured power spectrum.
Past attempts of deconvolution were done on a mode by mode 
basis~\citep{Tegmark2001:astro-ph/0111575v3} 
or assumed the global plane parallel approximation (often called distant 
observer approximation~\citep{Sato2010:1010.0289v2,Sato2013:1308.3551v2,
Outram2001:astro-ph/0106012v2}). Here we present a deconvolution formalism, 
which can be applied to a wide-angle 
multipole representation of the power spectrum. 
Recently, \cite{2021PhRvD.103j3504P} advocated deconvolution in the 
context of going back to the quadratic
estimator formalism originally derived in
\cite{1998PhRvD..57.2117B,1998ApJ...503..492S}.
For these
estimators one can naturally quote 
results convolved with a window, or deconvolved, or something in between
\cite{2002MNRAS.335..887T,2006PhRvD..74l3507T}.
Note that there is a technical difference between that 
formalism, which assumes the $k$ bands in which the quadratic averaging is 
performed are the same as the $k$ bands used to model the theory power, so that
the window matrix is automatically square and invertible, and the formalism
in this and other recent papers, 
in which the theory bands can have arbitrarily fine $k$ resolution, 
independent of the observational averaging band width. 

First we express the power spectrum analysis as two matrix multiplications, one matrix accounting for wide-angle effects and a second matrix accounting for the window function. We derive these matrices for some of the largest galaxy redshift surveys currently available (6dFGS DR3, BOSS DR12 and the eBOSS DR16 QSOs sample) and make these matrices available together with the power spectrum measurements~\footnote{\url{https://fbeutler.github.io/hub/deconv_paper.html}}. The appendix of this paper provides a step-by-step guide for a galaxy survey analysis, including \code{Python}-based examples. The aim is to make galaxy redshift survey datasets more easily accessible for the wider cosmology community.

As an example application we perform a BAO analysis on the deconvolved power spectrum of BOSS DR12. We show that with the products derived in this paper, the likelihood derived from the convolved and deconvolved power spectrum multipoles is identical.

This paper is organized as follows. We start with a review of the current formulation of the survey window function, which we turn into a matrix multiplication in section~\ref{sec:window}. In section~\ref{sec:wamatrix} we derive a similar matrix accounting for wide-angle effects. In section~\ref{sec:deconvolution} we deriving our deconvolution procedure. In section~\ref{sec:datasets} we introduce the 6dFGS, BOSS and eBOSS datasets followed by an application of our matrix based deconvolution procedure in section~\ref{sec:analysis}. We conclude in section~\ref{sec:conclusion}. Appendix~\ref{app:userguide} provides a detailed user guide with \code{Python}-based examples. In appendix~\ref{app:Cls} we discuss the correlation matrices needed for the window function calculations. We also show consistency between the equations used in this paper with the equations in~\citep{Beutler2013:1312.4611v2} and~\citep{Beutler2018:1810.05051v3} in appendix~\ref{app:deriveW}. In appendix~\ref{app:dipole_der} we derive analytic equations for wide-angle effects. Finally, appendix~\ref{app:deconvolution} contains a summary of the convolved and deconvolved power spectrum measurements for the different datasets.

Throughout the paper we use a (fiducial) flat $\Lambda$CDM cosmology with $\Omega_m=0.31$ when transferring observables (RA, DEC, z) into cartesian coordinates (x, y, z). When analysing mock datasets we use the cosmological parameters of the underlying simulations, which are different for each of the galaxy samples studied here (see section~\ref{sec:datasets}). The \code{Python} code used to calculate the wide-angle and window function matrices discussed in this paper is available at \url{https://github.com/fbeutler/pk_tools}.

\section{The survey window function}
\label{sec:window}

Most studies involving the galaxy power spectrum include the survey window 
function by convolving the model for the power spectrum before comparing it to 
the measurement. 
Any asymmetry in the survey window distributes power between the multipoles. This can bias measurements of anisotropic observables like redshift-space distortions, if not taken into account correctly. The convolution of the power 
spectrum multipoles with the window function multipoles has been laid out in \citep{Beutler2013:1312.4611v2} and is given by 
\begin{equation}
    P^{\rm conv}_{\ell}(k) = \int dk' k'^2\sum_{\ell'}W_{\ell \ell'}(k,k')P_{\ell'}^{\rm true}(k') - \frac{Q_{\ell}(k)}{Q_{0}(0)}\int dk'k'^2\sum_{\ell'}W_{0\ell'}(0,k')P^{\rm true}_{\ell'}(k')\, ,
    \label{eq:conv}
\end{equation}
where $W_{\ell \ell'}(k,k')$  describes the contribution of multipole $\ell'$ 
to multipole $\ell$ due to the survey 
window~\footnote{Note that~\citep{Beutler2013:1312.4611v2} 
uses a slightly different nomenclature with 
$|W(k,k')|^2_{\ell\ell'} \equiv W_{\ell \ell'}(k,k')$}. 
The second term on the right hand side describes the integral constraint 
correction~\citep{Peacock1991} and includes the 1D survey window function multipoles in Fourier space $Q_{\ell}(k)$. While the treatment of the integral constraint in \eq{eq:conv} is a good approximation, \citep{Mattia2019:1904.08851v1} showed that there are additional contributions depending on how the reference (random) catalog has been generated.

While~\cite{Beutler2013:1312.4611v2} calculated $W_{\ell \ell'}(k,k')$ through pair counting (see eq. 33 of that reference), recently it was 
pointed out that this quantity can be obtained using a double Bessel 
integral~\citep{D'Amico2019:1909.05271v1} 
\begin{equation}
    W_{\ell \ell'}(k,k') = \frac{2}{\pi}(-i)^{\ell}i^{\ell'}\int ds \,s^{2} j_{\ell}(ks)j_{\ell'}(k's)\sum_LC_{\ell\ell'L}Q_{L}(s)\, ,
    \label{eq:2Dwindow_nowa}
\end{equation}
where the matrices $C_{\ell\ell'L}$ are given in section 3.1 
of~\citep{D'Amico2019:1909.05271v1}. 
The 1D window function multipoles in 
configuration-space are defined as
\begin{equation}
    Q_{L}(s) = \frac{(2L + 1)}{A}\int \frac{d\Omega_{\vb{s}}}{4\pi}\int 
d\vb{s}_1 \nbw(\vb{s}_1)\nbw(\vb{s}+\vb{s}_1)
\mathcal{L}_{L}(\vbunit{s}\cdot\vbunit{d})\, ,
\label{eq:QLs}
\end{equation}
with $\nbw(\vb{x}) = \bar{n}(\vb{x})w(\vb{x})$, where 
$\bar{n}(\vb{x})$ is the local mean density of galaxies (i.e., expected 
number after selection effects), and $w(\vb{x})$ represents a weight 
applied to the density measured at $\vb{x}$ (completeness and signal to noise 
weight). 
The normalization factor $A$ here is the same number that appears in the power
spectrum estimate, e.g., eq.~(3) of \cite{Beutler2016:1607.03150v1}. However,
we do not use the standard value of $A$ given in eq.~(13) of 
\cite{Beutler2016:1607.03150v1}. Instead we 
set $A$ to the value necessary to enforce 
$Q_{0}(s\rightarrow 0) \equiv 1$. The same $A$ value is 
then also
used to normalize the power spectrum estimate, as required for
consistency (as recently emphasized by \citep{Mattia2019:1904.08851v1}). 
We use this definition because we find it gives much closer to 
the desired unit normalized window, i.e., window where integration over 
theory $k$ gives 1. This is desirable because it makes the convolved power
close to the true power, i.e., just smeared slightly by the window instead of 
adding an offset in the overall normalization.
Note that, at the continuum limit of~\eq{eq:QLs}, 
our normalization convention can be written as
\begin{equation}
    A = \int d\vb{x}\,\nbw^2(\vb{x})\, .
\end{equation}
However, the traditional definition as represented by
eq.~(13) of \cite{Beutler2016:1607.03150v1} is {\it also} intended to 
represent this equation! In the traditional calculation the integral over
position and the $\bar{n}(\vb{x})$ factors are represented by a sum
over randoms times $\bar{n}(z)$, where $\bar{n}(z)$ is supposed to be a
redshift dependent average of $\bar{n}(\vb{x})$. This is an approximation
that in practice only 
achieves the goal of $Q_0(s\rightarrow 0)=1$ to $\sim 10$\%. 
We guarantee $Q_0(s\rightarrow 0)=1$ by computing
$Q_L(s)$ first and using it directly to fix $A$. 
For comparison with past
results we give the relative normalization between the two methods in
Table~\ref{tab:survey_properties}. To be clear, if the normalisation between the power spectrum and window function are consistent, it will not impact any likelihood analysis, so it is essentially cosmetic.

In the following section we will demonstrate how our window function formalism can be extended to include wide-angle effects before moving on to develop a matrix 
based convolution and deconvolution procedure. 

\subsection{Including wide-angle effects}
\label{sec:wa}

\begin{figure}[t]
    \centering
    \includegraphics[width=1.0\textwidth]{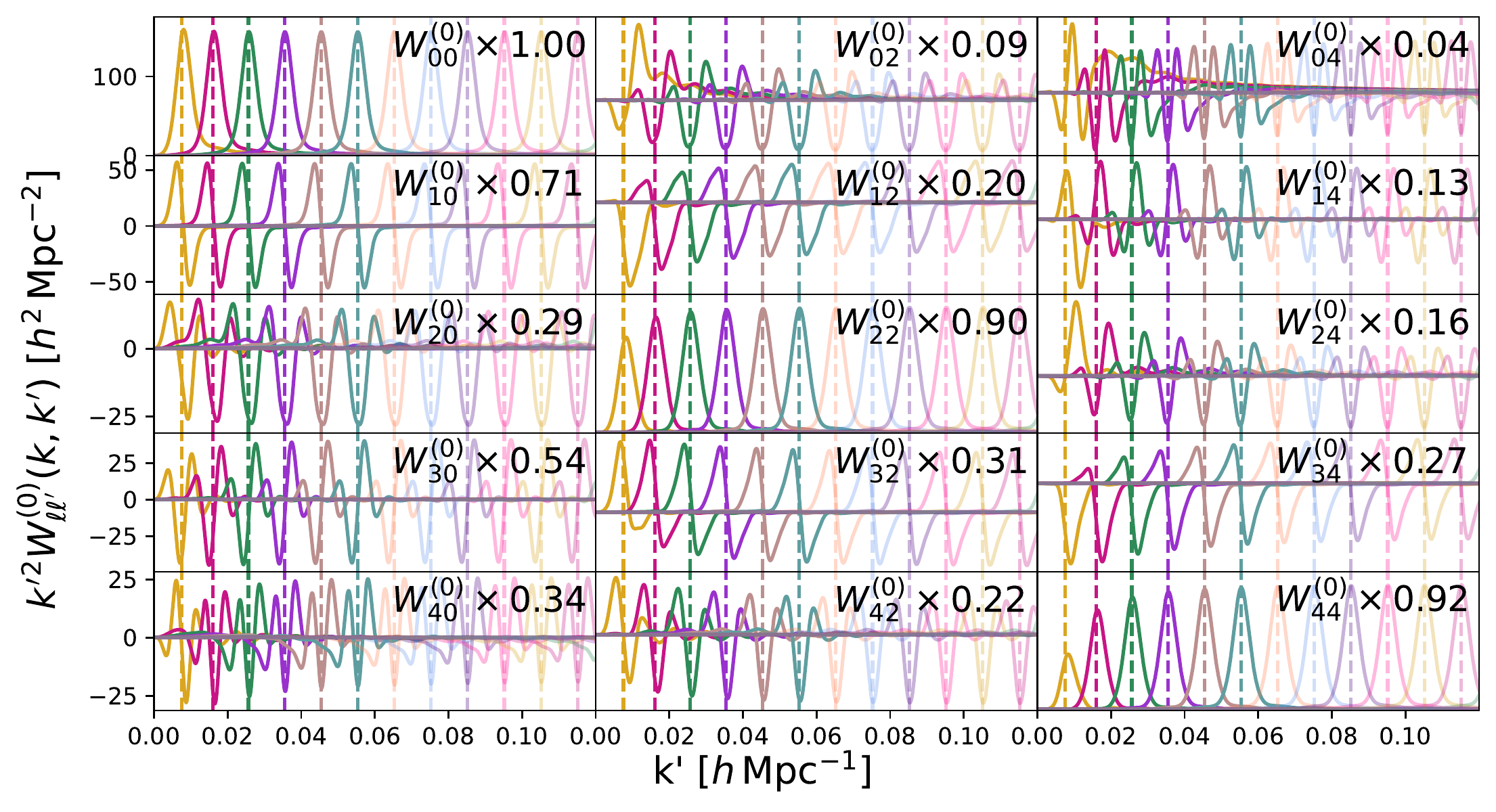}
    \caption{The window function multipoles of the low redshift bin 
$0.2 < z < 0.5$ of BOSS DR12, NGC, calculated following \eq{eq:2Dwindow} at 
zero order in the wide-angle expansion ($n=0$). The corresponding window 
function multipoles at first order in the wide-angle expansion ($n=1$) are 
shown in \fig{fig:wll1_NGC_z1}. The three columns correspond to the 
contributions of the three even multipoles to the five non-zero multipoles, 
including the dipole and octopole. Each window function is plotted multiple times 
as a function of $k'$ with fixed values of 
$k$ (vertical dashed lines). 
This window function does not account for any bin averaging, but is evaluated at specific values $k$ with $16\,384$ values of $k'$ (here we focus on $0 < k' < 0.12\kMpc$, while in practice we calculate the full range of $0 < k' < 0.4\kMpc$). For visual purposes, each of the sub-panels is re-normalized by the factor written in the right hand corner. The final window function matrix used for the actual analysis requires bin averaging in both $k_{\rm o}$ and $k_{\rm th}$ (see eq.~\ref{eq:intW} for more details).
}
    \label{fig:wll0_NGC_z1}
\end{figure}

\begin{figure}[t]
    \centering
    \includegraphics[width=0.75\textwidth]{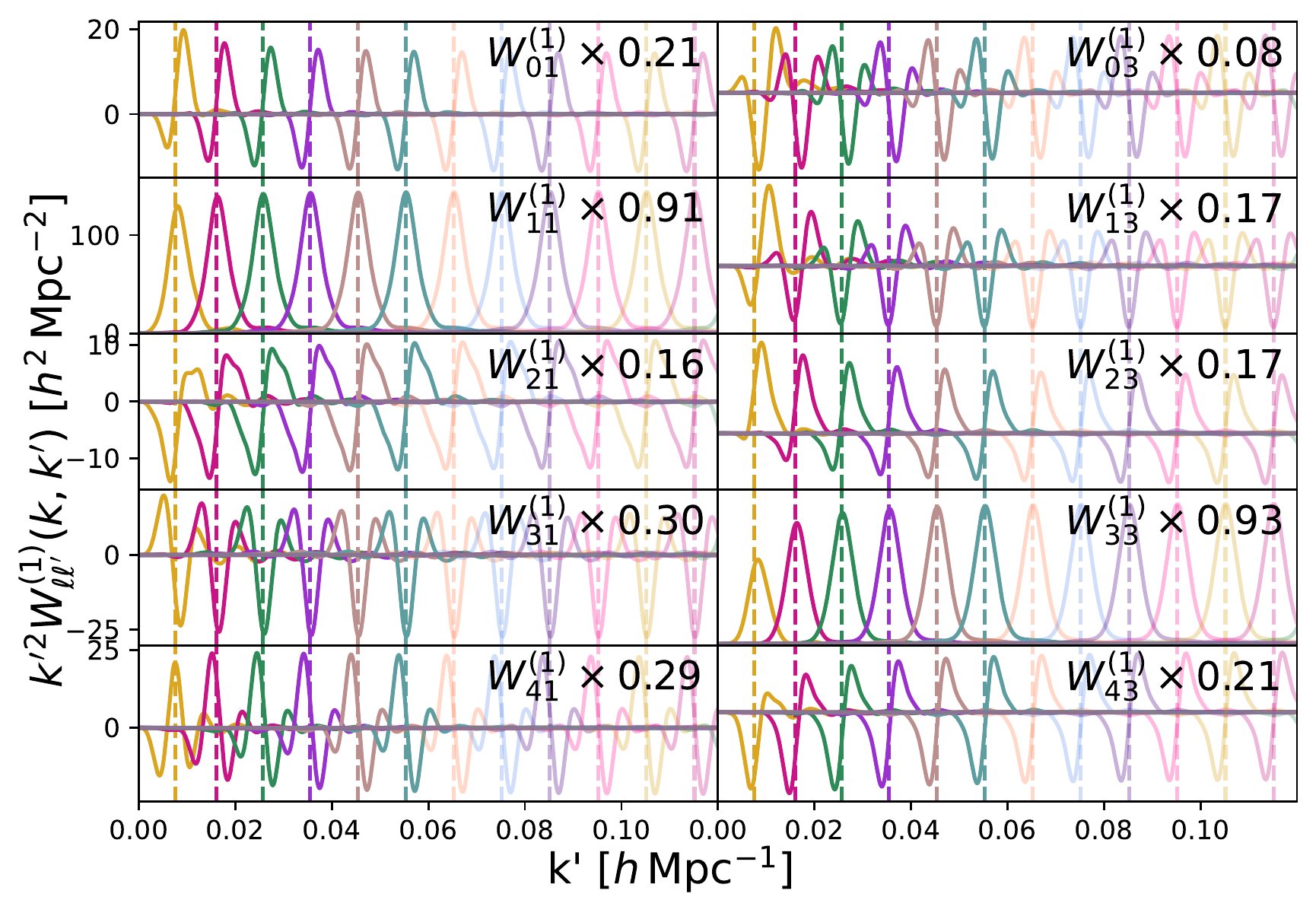}
    \caption{The window function multipoles of the low redshift bin $0.2 < z < 0.5$ of BOSS DR12, NGC, calculated following \eq{eq:2Dwindow} at first order in the wide-angle expansion ($n=1$). The corresponding window function multipoles at zero order in the wide-angle expansion ($n=0$) are shown in \fig{fig:wll0_NGC_z1}. The two columns correspond to the contributions of the two odd multipoles to the five non-zero multipoles. Each window function is plotted multiple times as a function of $k'$ with fixed values of $k$ (vertical dashed lines). For visual purposes, each of the sub-panels is re-normalized by the factor written in the right hand corner.}
    \label{fig:wll1_NGC_z1}
\end{figure}

Any measurement of the anisotropic power spectrum has to make a choice 
regarding the line-of-sight (LOS) direction for a galaxy pair
(see, e.g., 
\citep{Reimberg2015:1506.06596v2,Castorina2017:1709.09730v2,Beutler2018:1810.05051v3}). 
Based on this choice the triangle configuration between the observer and the 
galaxy pair is reduced to a separation amplitude $s$ and a multipole dependent 
weighting given by the Legendre polynomials $\mathcal{L}_{\ell}$. This 
geometric simplification introduces wide-angle effects in the power spectrum 
multipoles $\ell>0$ as well as the associated window function multipoles. 
Crucially, wide-angle effects can break the symmetry between the galaxy pair 
and hence can introduce odd multipoles, like the dipole and octopole. These 
wide-angle effects can be absorbed into the window function formalism 
described above. Based on~\citep{Beutler2018:1810.05051v3} we can 
extend~\eq{eq:2Dwindow_nowa} to get
\begin{equation}
    W^{(n)}_{\ell \ell'}(k,k') = \frac{2}{\pi}(-i)^{\ell}i^{\ell'}\int ds \,s^{2} j_{\ell}(ks)j_{\ell'}(k's)\sum_LC^{(n)}_{\ell\ell'L}Q^{(n)}_{L}(s)\, ,
    \label{eq:2Dwindow}
\end{equation}
where the index $n$ describes the order in the wide-angle expansion and the $C^{(n)}_{\ell\ell'L}$ matrices are given in appendix~\ref{app:Cls}. Eq.~(\ref{eq:2Dwindow}) includes new window function multipoles $Q^{(n)}_{L}$ at each order of the wide-angle expansion $n$ given by 
\begin{equation}
    Q^{(n)}_{\ell}(s) = \frac{(2\ell + 1)}{A}\int \frac{d\Omega_{\vb{s}}}{4\pi}\int d\vb{s}_1(s_1)^{-n}W(\vb{s}_1)W(\vb{s}+\vb{s}_1)\mathcal{L}_{\ell}(\vbunit{s}\cdot\vbunit{d})\, .
    \label{eq:window}
\end{equation}
A derivation of \eq{eq:2Dwindow} is included in Appendix~\ref{app:W_with_wa}. The constraints imposed by FFTs usually enforce the end-point LOS definition $\vbunit{d} = \vbunit{s}_1$ where the line of sight is chosen along the distance vector of one of the galaxies (see \citep{Beutler2018:1810.05051v3} for details).

Eq.~(\ref{eq:2Dwindow}) can be calculated using one 1D Fourier transforms (FT) 
for each $k$ bin with the complexity 
$\mathcal{O}(N_k\times N_{k'}\log N_{k'})$, where $N_k$ and $N_{k'}$ are the 
number of bins in $k$ and $k'$, respectively. Additionally we need to calculate 
the $Q_{\ell}(k)$ using a 3D FT, which has the complexity 
$\mathcal{O}(N\log N)$, where $N$ is the number of grid cells in which the 
random galaxies are binned (see appendix E.1 
of ~\cite{Beutler2018:1810.05051v3}). We then obtain $Q_{\ell}(s)$ through 
Hankel transforms. In this paper we will use~\eq{eq:2Dwindow} to calculate 
the window function using $N_{k'} = 16\,384$ and $N_{k} = 400$ 
(from $0 < k < 0.4\kMpc$ in bins of $\Delta k = 0.001\kMpc$), which typically takes 
$< 1$ minute on a single state of the art laptop. 

Figure~\ref{fig:wll0_NGC_z1} shows the window function multipoles 
$W^{(n)}_{\ell \ell'}(k,k')$ for the low redshift bin of BOSS DR12 in the 
North Galactic Cap (NGC). This figure only shows the window function 
multipoles at zero order in the wide-angle expansion ($n=0$ in 
eq.~\ref{eq:2Dwindow}). At this order, only even multipoles $\ell'$ are 
generated, meaning we have a contribution from the monopole (left column), 
quadrupole (middle column) and hexadecapole (right column). Each window 
function is plotted several times with fixed $k$ (dashed lines). These plots agree with the plots shown 
in Figure 7 of \citep{Beutler2013:1312.4611v2}, where these multipoles have 
first been investigated, but they now include contributions to the odd 
multipoles. 

Figure~\ref{fig:wll1_NGC_z1} shows the corresponding window function multipoles at first order in the wide-angle expansion ($n=1$ in eq.~\ref{eq:2Dwindow}), in which case we only have odd contributions from the dipole (left column) and octopole (right column). Also note that any window function multipole $W^{(n)}_{\ell\ell'}$ in which the combination $\ell + \ell'$ is a odd number, results in a complex quantity, to account for the fact that the estimated odd power spectrum multipoles are complex~\citep{McDonald2009:0907.5220v1,Beutler2018:1810.05051v3}.

Based on the discussion above we can extend~\eq{eq:conv} in section~\ref{sec:window}, which describes the convolution of the power spectrum multipoles with the window function multipoles by including wide-angle effects:
\begin{equation}
    \begin{split}
        P^{\rm conv}_{\ell}(k) &= \int dk' \sum_{\ell',n}k'^{2-n}W_{\ell \ell'}^{(n)}(k,k')P^{(n), \rm true}_{\ell'}(k')\\
        &-\frac{Q^{(0)}_{\ell}(k)}{Q^{(0)}_{0}(0)}\int dk'\sum_{\ell',n}k'^{2-n}W_{0 \ell'}^{(n)}(0,k')P^{(n), \rm true}_{\ell'}(k')\, . 
    \end{split}
    \label{eq:conv_wa}
\end{equation}
We now define a new multipole expansion including wide-angle effects as \begin{equation}
    P(k,\mu) = \sum_{\ell,n}(kd)^{-n}P_{\ell}^{(n)}(k)\mathcal{L}_{\ell}(\vbunit{k}\cdot\vbunit{d})\, ,
    \label{eq:multipoles_n}
\end{equation}
where $\vb{d}$ is the LOS vector and $d=|\vbunit{d}|$. Based on this definition the power spectrum multipoles, $P^{(n)}_{\ell}(k)$, at different order in the wide-angle expansions $n$ are given 
by~\citep{Castorina2017:1709.09730v2,Beutler2018:1810.05051v3}
\begin{equation}
    P^{(n)}_{\ell}(k) = 4\pi (-i)^{\ell}\int ds\,s^2(ks)^{n}\xi^{(n)}_{\ell}(s)j_{\ell}(ks)\, ,
    \label{eq:hankel_n}
\end{equation}
where $\xi^{(n)}_{\ell}$ are the configuration space multipoles at wide-angle 
order $n$.

For the remainder of this paper we will use the end-point LOS definition, meaning the LOS for each galaxy pair is oriented along the distance vector of one of the galaxies ($\vb{d} = \vb{s}_1$). In this case the zero order term ($n=0$) is equivalent to the commonly used power spectrum multipoles and at this order only the even multipoles are present.
At $n=1$ only the odd multipoles are present and can be obtained as linear combinations of the even multipoles in configuration space
\begin{align}
    \xi^{(1)}_1 &= -\frac{3}{5}\xi_2^{(0)}(s)\, ,\\
    \xi^{(1)}_3 &= \frac{3}{5}\xi_2^{(0)}(s) - \frac{10}{9}\xi_4^{(0)}(s)\, .
    \label{eq:odd_cor}
\end{align}
The second order terms ($n=2$) are given in eq. (2.16) - (2.18) of~\citep{Beutler2018:1810.05051v3}. The analysis in this paper will be limited to $n\leq 1$, even though including higher order terms ($n>1$) is straightforward. Note that the dipole and octopole power spectra can be calculated directly in Fourier space without any need for a Hankel transform (see also eq.~3.56 of \citep{Beutler2020:2004.08014v1}):
\begin{align}
    P^{(1)}_1(k) &= -i\frac{3}{5}\left[3P^{(0)}_2(k) + k\partial_{k}P^{(0)}_2(k)\right]\, ,\\
    P^{(1)}_3(k) &= -i\left[\frac{3}{5}\left(2P^{(0)}_2(k) - k\partial_{k}P^{(0)}_2(k)\right) + \frac{10}{9}\left(5P^{(0)}_{4}(k) + k\partial_{k}P^{(0)}_4(k)\right)\right]\, .
\end{align}
We included a derivation of these equations in appendix~\ref{app:dipole_der}. Using linear theory we obtain
\begin{align}
    P^{(1)}_1(k) &= -if\left(\frac{4}{5}b_1+\frac{12}{35}f\right)\left[3P_m(k) + k\partial_{k}P_m(k)\right],\label{eq:dipole_analytic}\\
    P^{(1)}_3(k) &= -i4f\left[\frac{1}{5}\left(b_1 + \frac{3}{7}f\right)\left(2P_m(k) - k\partial_{k}P_m(k)\right) + \frac{4}{63}f\left(5P_m(k) + k\partial_{k}P_m(k)\right)\right]\, .\label{eq:octopole_analytic}
\end{align}

\subsection{Window function convolution as matrix multiplication}
\label{sec:window_matrix}

\begin{figure}[t]
    \centering
    \includegraphics[width=0.55\textwidth]{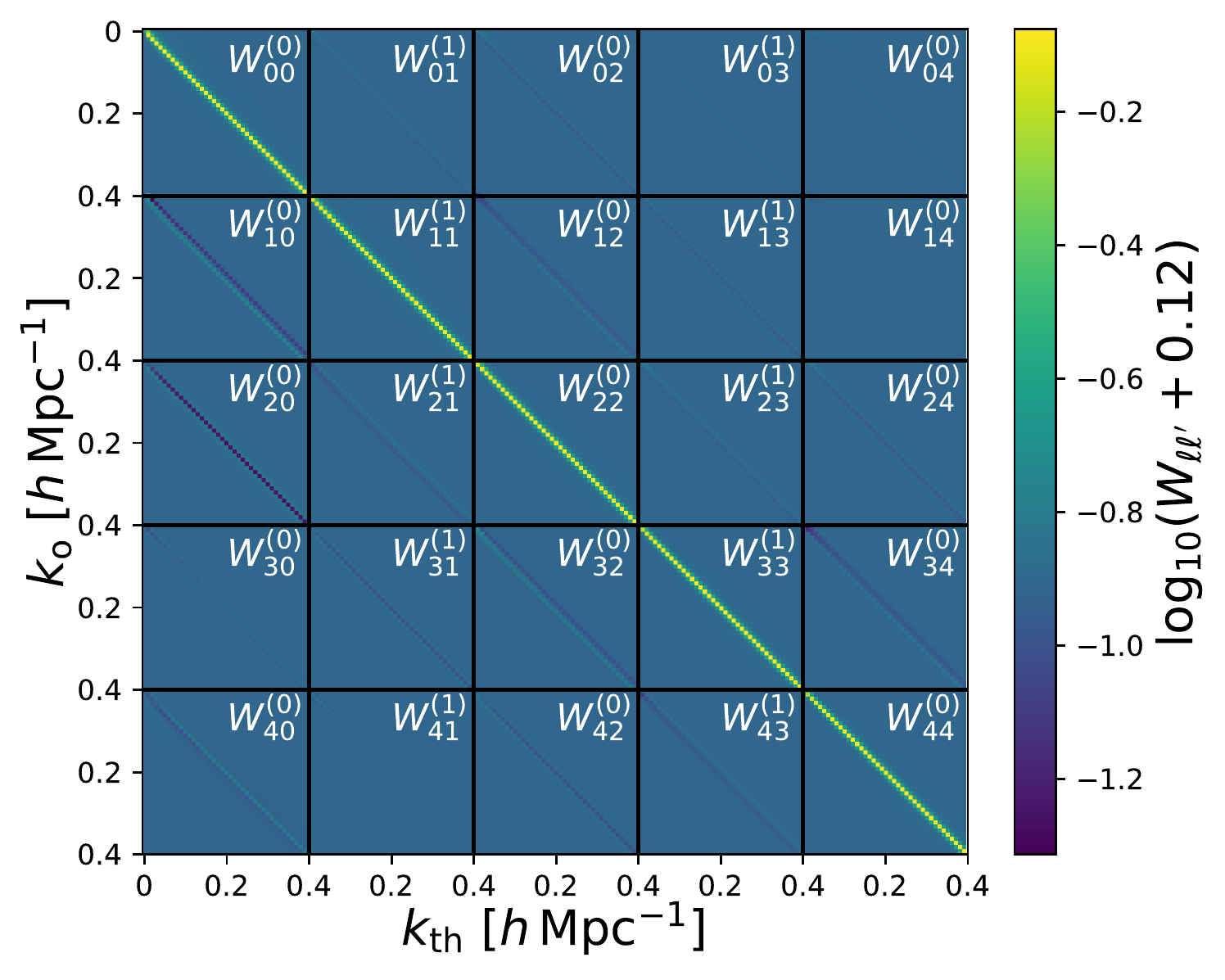}
    \includegraphics[width=0.42\textwidth]{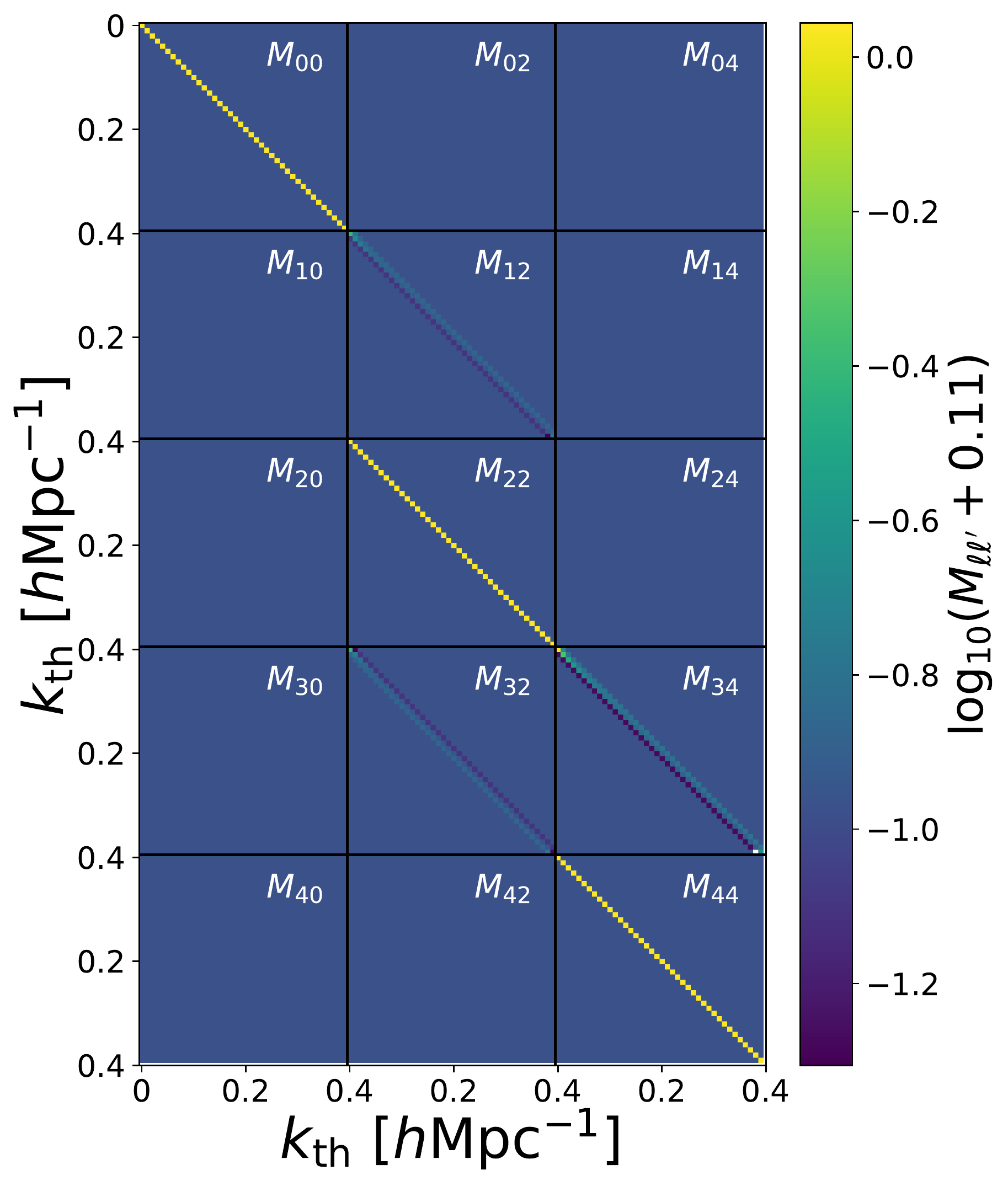}
    \caption{The window function matrix ($W_{\ell\ell'}$, see section~\ref{sec:window_matrix}) and wide-angle matrix ($M_{\ell\ell'}$, section~\ref{sec:wamatrix}) of BOSS DR12 NGC in the low redshift bin. The wide-angle transformation matrix maps the 
theoretical power spectrum multipoles with three even multipoles to a vector with five multipoles (only accounting for wide-angle effects at order $n=1$). The window function maps the five (theoretical) power spectrum multipoles (in bins of $\Delta k_{\rm th}$) to the observed/convolved 
power spectrum multipoles (in bins of $\Delta k_{\rm o}$), which can be compared to the data. 
For plotting purposes here we use $\Delta k_{\rm th} = \Delta k_{\rm o} = 0.01\kMpc$, 
while the analysis in this paper uses 
$\Delta k_{\rm o} = 0.01\kMpc$ and $\Delta k_{\rm th} = 0.001\kMpc$. The window function matrix is not symmetric, since the contribution of the monopole to the dipole 
is different to the contribution of the dipole to the monopole. The diagonal 
elements are not $1$, since they correspond to the bin integral of the window 
function within the bin $\Delta k_{\rm th}$ (see eq.~\ref{eq:intW}), which is 
only $1$ for $\Delta k \gg k_f$ as shown in \fig{fig:Ik}. The z-axis is 
logarithmic for better contrast.
\pvmhid{finewindowfigures.py}
}
    \label{fig:Wkokt}
\end{figure}

If we assume that we calculate our power spectrum model in bins of $\Delta k_{\rm th}$ and intend to compare this model with power spectrum measurements in bins of `observed' bandpowers $\Delta k_{\rm o}$, the window function needs to be integrated over $k$ and $k'$ as
\begin{equation}
    W^{(n)}_{\ell \ell'}(k_{\rm o},k_{\rm th}) = \int dk\,k^2\Theta(k_{\rm o}, k)\int dk'\,k'^{2-n} W^{(n)}_{\ell \ell'}(k,k')\Theta(k_{\rm th}, k')\, ,
    \label{eq:intW}
\end{equation}
where the step function $\Theta$ is defined as
\begin{equation}
    \Theta(k_{\rm x}, k) = \begin{cases}1\;\;\;\;\; \text{if } |k_x - k| < \frac{\Delta k_{\rm x}}{2} \\ 0\;\;\;\;\; \text{otherwise.}\end{cases}
    \label{eq:step}
\end{equation}
We will test the bin averaging employed in our analysis in the next subsection. Before we get to that, let us define our matrix nomenclature, starting with the vectors
\begin{equation}
    \Ptrue = \begin{pmatrix}
        P_0^{(0),\rm true}(k)\\
        P_1^{(1),\rm true}(k)\\
        P_2^{(0),\rm true}(k)\\
        P_3^{(1),\rm true}(k)\\
        P_4^{(0),\rm true}(k)
    \end{pmatrix}\, ,\;\;\;
    \Pconv = \begin{pmatrix}
        P_0^{(0),\rm conv}(k)\\
        P_1^{(1),\rm conv}(k)\\
        P_2^{(0),\rm conv}(k)\\
        P_3^{(1),\rm conv}(k)\\
        P_4^{(0),\rm conv}(k)
    \end{pmatrix}\, ,\;\;\;
    \bold{Q} = 
    \begin{pmatrix}
        Q^{(0)}_{0}(k)\\
        Q^{(1)}_{1}(k)\\
        Q^{(0)}_{2}(k)\\
        Q^{(1)}_{3}(k)\\
        Q^{(0)}_{4}(k)
    \end{pmatrix}\, ,
\end{equation}
where $\bold{P}^{\rm true}$ is a vector of model power spectrum multipoles and the convolution matrix is given by
\begin{equation}
    \bold{W} = 
    \begin{pmatrix}
        W_{0 0}^{(0)} & W_{0 1}^{(1)} & W_{0 2}^{(0)} & W_{0 3}^{(1)} & W_{0 4}^{(0)}\\
        W_{1 0}^{(0)} & W_{1 1}^{(1)} & W_{1 2}^{(0)} & W_{1 3}^{(1)} & W_{1 4}^{(0)}\\
        W_{2 0}^{(0)} & W_{2 1}^{(1)} & W_{2 2}^{(0)} & W_{2 3}^{(1)} & W_{2 4}^{(0)}\\
        W_{3 0}^{(0)} & W_{3 1}^{(1)} & W_{3 2}^{(0)} & W_{3 3}^{(1)} & W_{3 4}^{(0)}\\
        W_{4 0}^{(0)} & W_{4 1}^{(1)} & W_{4 2}^{(0)} & W_{4 3}^{(1)} & W_{4 4}^{(0)}
    \end{pmatrix}\, ,
\end{equation}
where each sub-matrix $W_{\ell\ell'}^{(n)}$ is of size $N_{\rm th}\times N_{\rm o}$. Based on the definitions above, we can write the convolution of the power 
spectrum multipoles with the window function multipoles as a matrix 
multiplication
\begin{align}
    \Pconv &= \bold{W}\Ptrue - 
\frac{\bold{Q}}{Q_0(0)}\bold{W}_{0\ell'}(0,k')\Ptrue_{\ell'}(k')\, ,
    \label{eq:mconvolution}
\end{align}
where $\frac{\bold{Q}}{Q_0(0)}\bold{W}_{0\ell'}(0,k')$ is a matrix representing the integral 
constraint correction. 
For the remainder of this paper we will absorb the integral constraint into $\bold{W}$ meaning we re-define
\begin{equation}
    \bold{W} \equiv \bold{W} - \frac{\bold{Q}}{Q_0(0)}\bold{W}_{0\ell'}(0,k')\, .
\end{equation}
This works exactly because the integral constraint effect is linear in the
power spectrum just like the standard window function. 
In this paper the window function matrix $\bold{W}$ is generally defined in the k-range $0.0 < k < 0.4\kMpc$, while the recommended $k$-range for any fit is $0.01 < k < 0.3\kMpc$. The increased $k$-range accounts for the redistribution of power due to the window function, which connects modes inside the fitting range with modes beyond that range.
The equations above include wide-angle effects at order $n=0$ and $n=1$, even though this formalism could easily be extended to $n>1$. Figure~\ref{fig:Wkokt} shows the matrix $\bold{W}$ after performing the integral in \eq{eq:intW} (and including the integral constraint correction).

\subsection{Bin averaging the window function}
\label{sec:binning_tests}

In \eq{eq:intW} we wrote down how to take the window function (as derived in eq.~\ref{eq:2Dwindow} and plotted in \fig{fig:wll0_NGC_z1} and \ref{fig:wll1_NGC_z1}) an05d account for the bin averaging to match the layout of the data vector. The default choices for our window functions are $\Delta k_{\rm th} = 0.001\kMpc$ and $\Delta k_{\rm o} = 0.01\kMpc$. Here we will demonstrate how we implemented the bin averaging of the window function and we will test certain assumptions made in our formalism. 

\subsubsection{The observational binning $\Delta k_{\rm o}$}

\begin{figure}[t]
    \centering
    \includegraphics[width=0.32\textwidth]{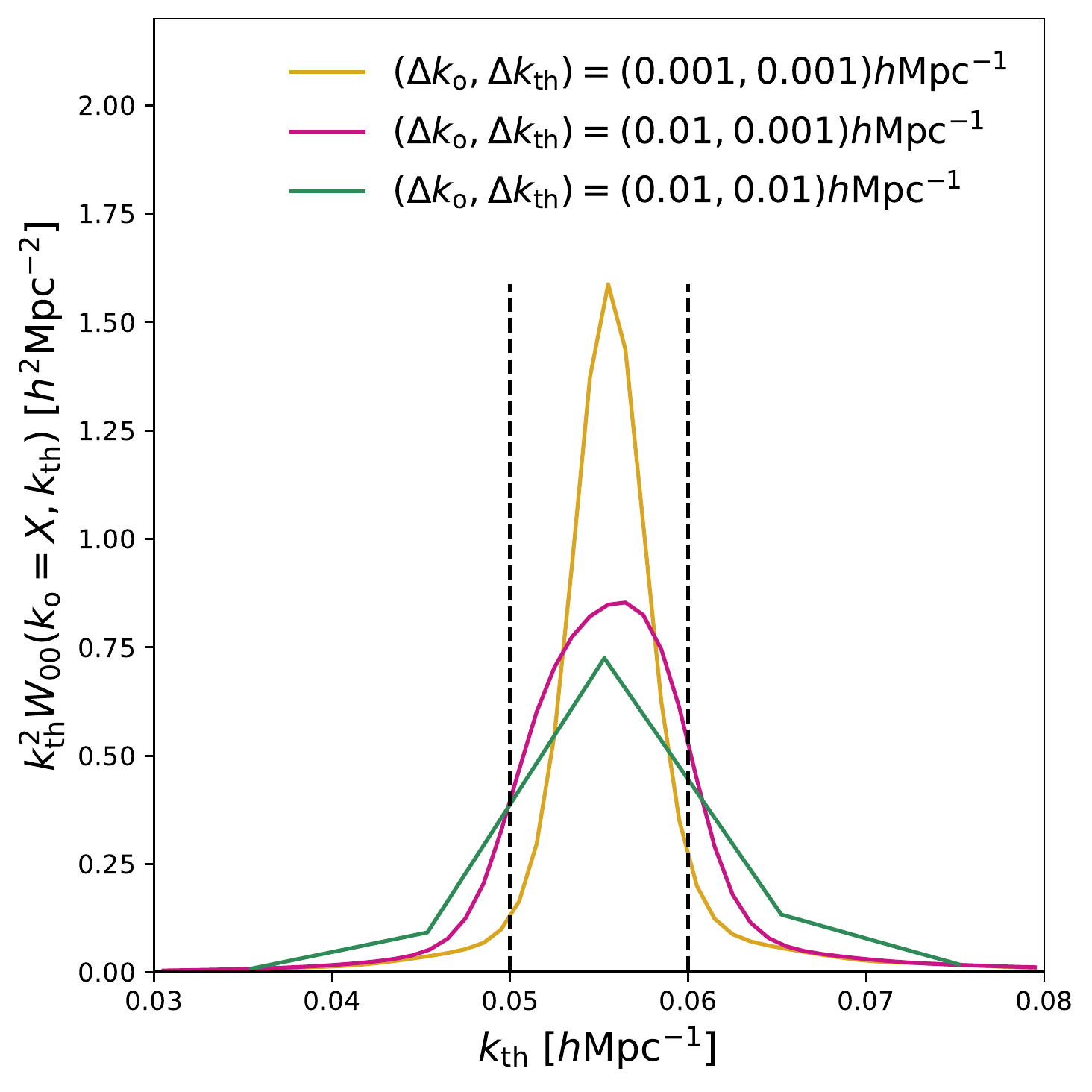}
    \includegraphics[width=0.32\textwidth]{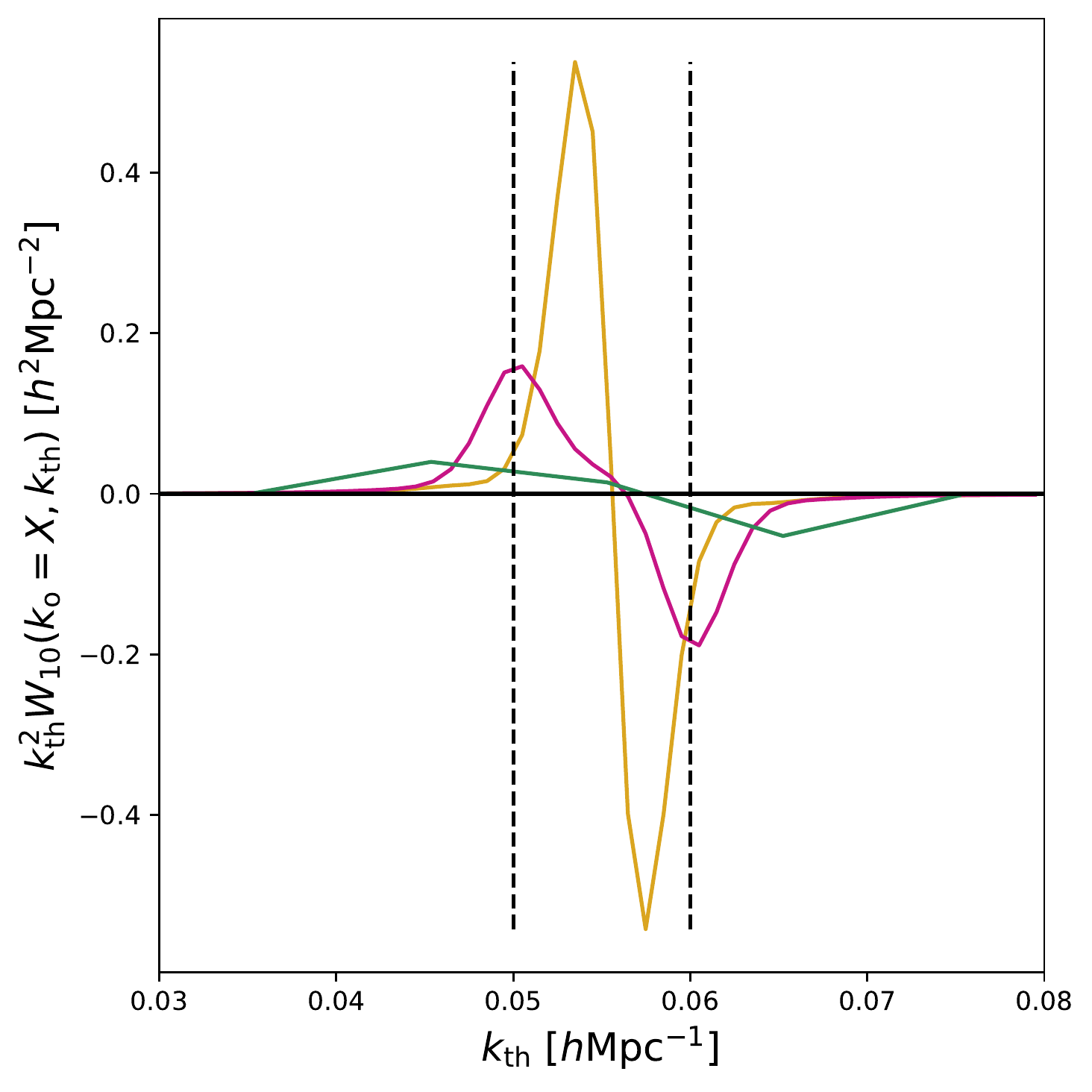}
    \includegraphics[width=0.32\textwidth]{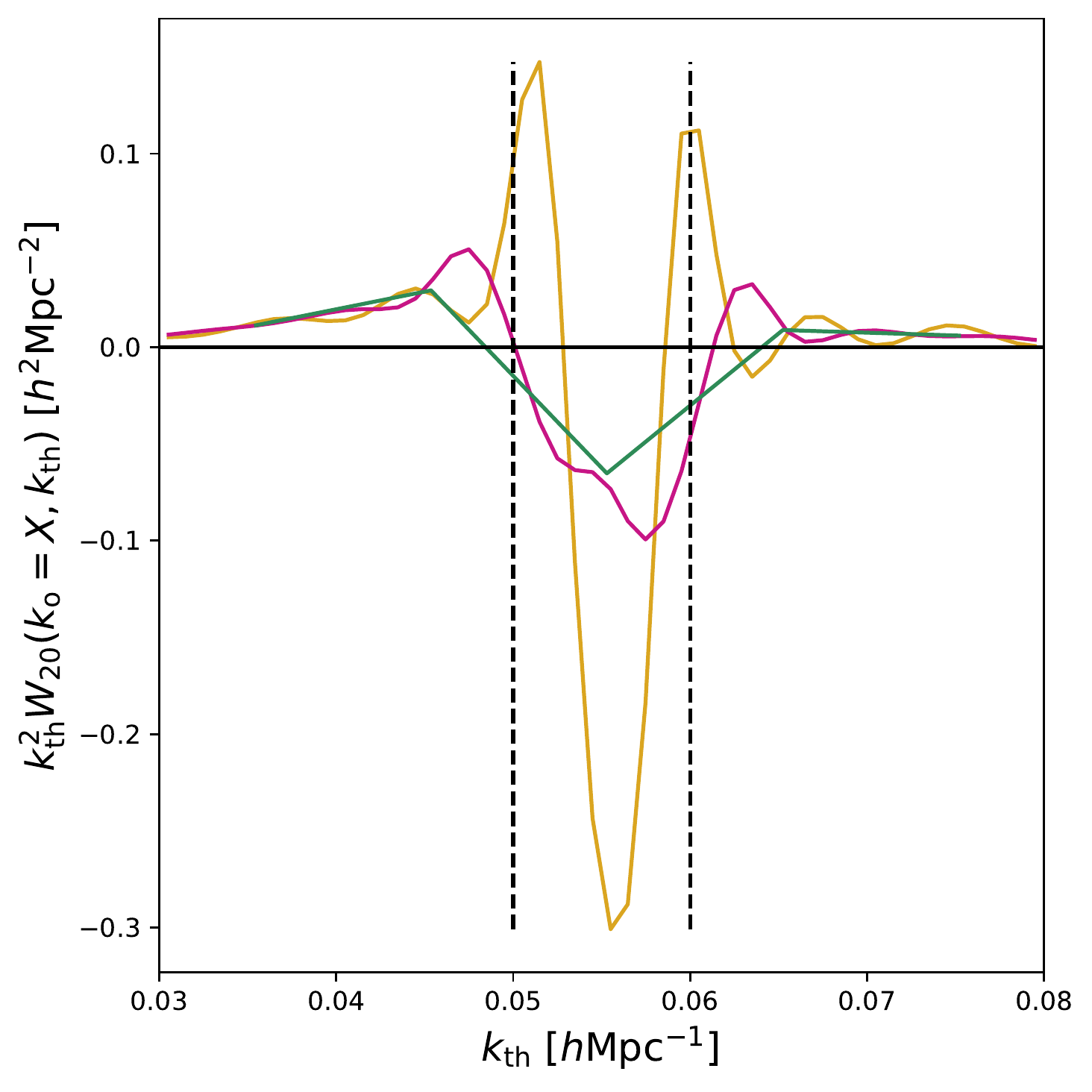}\\
    \includegraphics[width=0.32\textwidth]{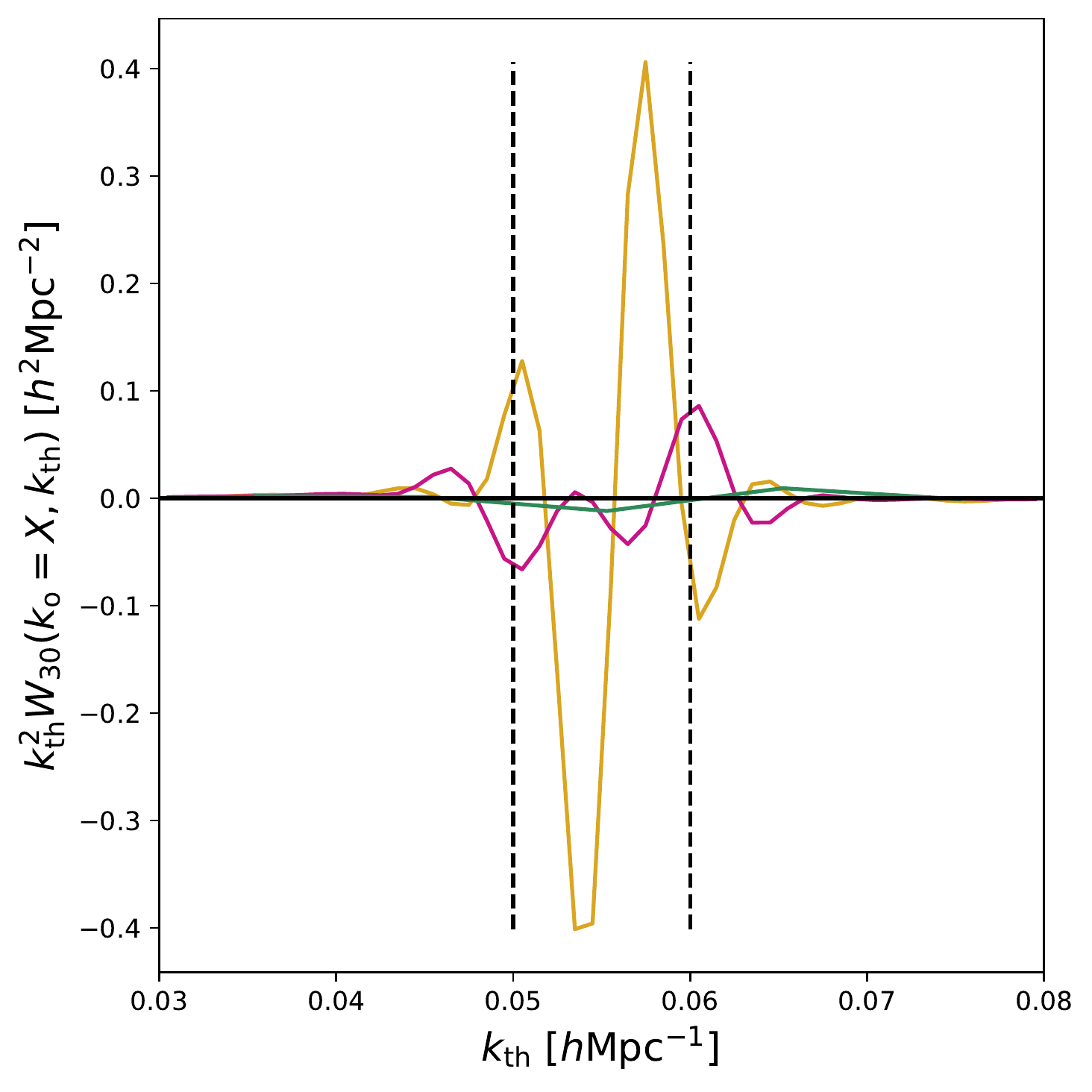}
    \includegraphics[width=0.32\textwidth]{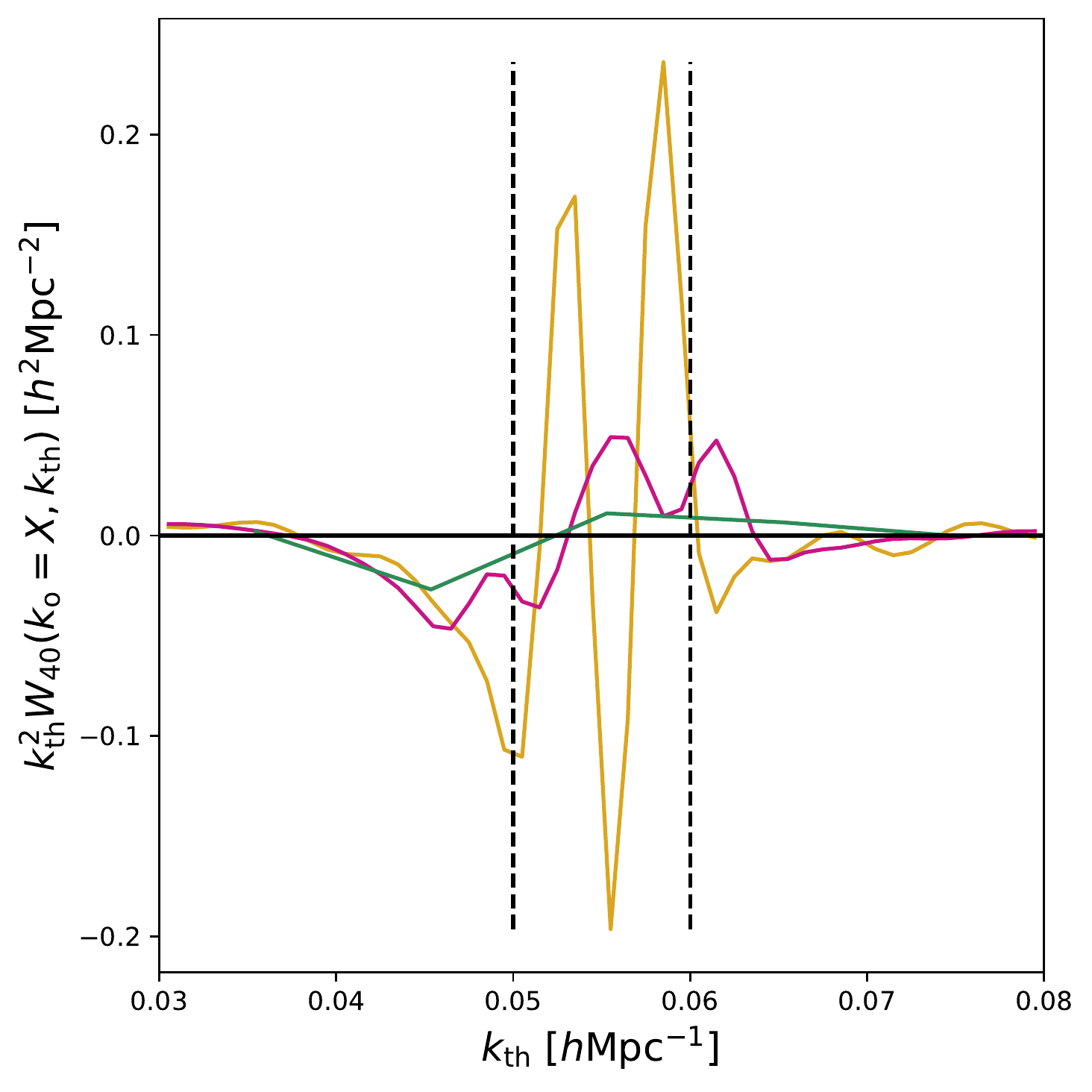}
    \caption{Comparison of the window function terms $W_{\ell0}$ for different binning schemes at $k_{\rm o}=0.055\kMpc$ for BOSS DR12 NGC in the low redshift bin (z1). The yellow line shows the window with bins of $\Delta k_{\rm o} = \Delta k_{\rm th} = 0.001\kMpc$. The red line shows the window with larger observational bins of $\Delta k_{\rm o} = 0.01\kMpc$, by averaging $10$ bins of the yellow line. The green line is averaged over both $k_{\rm o}$ and $k_{\rm th}$. The black dashed lines show the edges of the bin at $k_{\rm min} = 0.05\kMpc$ and $k_{\rm max} = 0.06\kMpc$. The relative heights of the three windows are scaled to make them comparable. The red line corresponds to our default choice.}
    \label{fig:binwindow}
\end{figure}

\begin{figure}[t]
    \centering
    \includegraphics[width=0.5\textwidth]{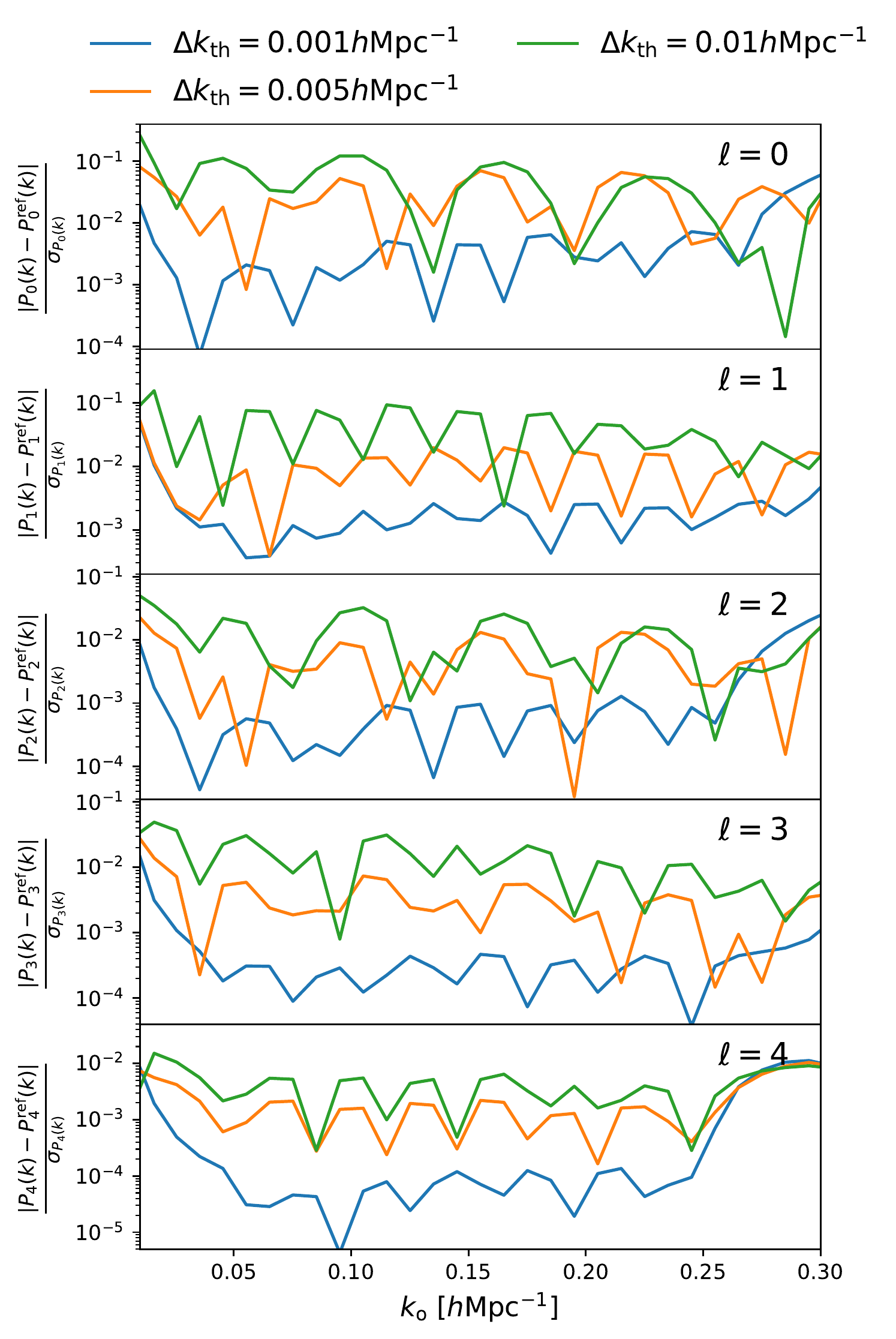}
    \caption{Comparison of convolved power spectrum multipole models based on 
different window function binning choices of $\Delta k_{\rm th}$ using BOSS DR12 NGC in the low redshift bin (z1). 
The window function transforms the theoretical binning ($\Delta k_{\rm th}$) 
into the binning used for the data ($\Delta k_{\rm o}$). The reference model is based on the current standard 
method~\citep{Wilson2015:1511.07799v2,Beutler2016:1607.03149v1}, which Hankel 
transforms the model into configuration-space, multiplies with the 
real-space window and Hankel transforms back into Fourier space. 
The window function uses $\Delta k_{\rm o}=0.01\kMpc$ in all cases. 
The window functions provided with this paper uses a default binning of $\Delta k_{\rm th} = 0.001\kMpc$, except of the case of deconvolution, where we use $\Delta k_{\rm th} = 0.01\kMpc$.
}
    \label{fig:binning_delta}
\end{figure}

All power spectrum multipole measurements provided with this paper use $k$-bins of $\Delta k_{\rm o} = 0.001\kMpc$. However, for practical purposes~\footnote{Mainly to reduce the size of the associated covariance matrix.} we use $10$ times larger bins of $\Delta k_{\rm o} = 0.01\kMpc$. The measured power spectra can easily be re-binned into larger bins using 
\begin{align}
    \bold{P}(k^i_{\rm o}) = \sum^{10}_{\text{bin }j\text{ within } \Delta k^i_{\rm o}}\frac{N^{\rm modes}_j}{N^{\rm modes}_i}\bold{P}(k^j_{\rm o})\, ,
    \label{eq:Pkbinaverage}
\end{align}
where $\Delta k^j_{\rm o} = \frac{1}{10}\Delta k_{\rm o}^i = 0.001\kMpc$ and $N^{\rm modes}_{j}$ is the number of observed k modes in bin $j$ with
\begin{equation}
    N^{\rm modes}_i = \sum^{10}_{\text{bin }j\text{ within } \Delta k^i_{\rm o}}N^{\rm modes}_j\, .
\end{equation}
One crucial point to highlight here is that the fundamental mode for all galaxy datasets included in this paper is smaller than our choice for $\Delta k_{\rm o}$. This means we need to bin average our theoretical model in accordance with the average of \eq{eq:Pkbinaverage}. For that reason we measure the window function in bins of $\Delta k_{\rm o} = 0.001\kMpc$ and average as 
\begin{align}
    \bold{W}(k^i_{\rm o},k_{\rm th}) &= \int dk\,k^2\bold{W}(k,k_{\rm th})\Theta(k^i_{\rm o}, k)\\
    &\approx \sum^{10}_{\text{bin }j\text{ within } \Delta k^i_{\rm o}}\frac{N^{\rm modes}_j}{N^{\rm modes}_i}\bold{W}(k^j_{\rm o},k_{\rm th})\, .
    \label{eq:binaverage}
\end{align}
Figure~\ref{fig:binwindow} shows the window function contributions to the monopole, $W_{\ell 0}$, with $\Delta k_{\rm o} = 0.001\kMpc$ (yellow line) and $\Delta k_{\rm o} = 0.01\kMpc$ (red line) for the low redshift (z1) sample of BOSS DR12 NGC. 

\subsubsection{The theoretical binning $\Delta k_{\rm th}$}

The multiplication of the power spectrum model with the window function matrix effectively calculates a convolution. The question we want to address here is, what would be the required resolution for this convolution, meaning how sensitive is the convolved power spectrum to the choice of $\Delta k_{\rm th}$? In \fig{fig:binning_delta} we show convolved power spectrum multipoles using our default binning of $\Delta k_{\rm o}=0.01\kMpc$ but different $\Delta k_{\rm th}$. The reference model is based on the current standard method, where the power spectrum model is (1) Hankel transformed into configuration-space, (2) multiplied by the configuration-space window and (3) Hankel transformed back into Fourier space~\citep{Wilson2015:1511.07799v2,Beutler2016:1607.03149v1}. The result is than bin-averaged to $\Delta k_{\rm o}=0.01\kMpc$.
The differences between our default choice of $\Delta k_{\rm th} = 0.001\kMpc$ and the reference model are $< 1\%$ of the measurement uncertainties for the monopole on most scales and even smaller for all other multipoles.

Finally we note that the above binning tests for the power spectrum and window function are based on a $\Lambda$CDM model (with the default cosmology given at the end of the introduction). Our default binning choices might not be optimal for non-standard analysis that searches for variations in the power 
spectrum smaller than the chosen bin width. Examples for such cases are 
primordial features~\citep{Beutler2019:1906.08758v1} or primordial 
non-Gaussianity~\citep{Castorina2019:1904.08859v1}. However, our binning 
choice should be perfectly suitable for the standard RSD and BAO analysis.

\section{Wide-angle effects as matrix multiplication}
\label{sec:wamatrix}

The odd power spectrum multipoles sourced by wide-angle effects are linearly related to the even multipoles through~\eq{eq:dipole_analytic} and (\ref{eq:octopole_analytic}). This implies that we can define a linear transformation between `flat-sky' statistics (no wide-angle effects) and `curved-sky' statistics (including wide-angle effects) as
\begin{equation}
    \bold{M}\Ptruefs = \Ptrue\, ,
    \label{eq:Mtransform}
\end{equation} 
where $\Ptruefs = (P_0,P_2,P_4)$ is a vector of the predicted even multipoles and $\bold{M}$ is a $5N_{k_{\rm th}}\times 3N_{k_{\rm th}}$ matrix, which transforms this vector into a new vector $\Ptrue$, which contains $5$ multipoles, including the dipole and octopole. We can define
\begin{equation}
    \bold{M} = 
    \begin{pmatrix}
        \mathcal{I} & 0 & 0\\
        0 & K^{2\rightarrow 1} & 0\\
        0 & \mathcal{I} & 0\\
        0 & K^{2\rightarrow 3} & K^{4\rightarrow 3}\\
        0 & 0 & \mathcal{I}\\
    \end{pmatrix} \, ,
    \label{eq:matrixM}
\end{equation}
where $\mathcal{I}$ is the identity matrix of size $N_{\rm th}\times N_{\rm th}$ and
\begin{align}
    K^{2\rightarrow 1}_{lm} &= -i\frac{3}{5d}\left[\frac{3}{\kth{l}}\Theta(\kth{l},\kth{m}) + \partial_{\kth{m}}\Theta(\kth{l},\kth{m})\right]\, ,\notag\\
    &= -i\frac{3}{5d}\left[\frac{3}{\kth{l}}\Theta(\kth{l},\kth{m}) + \frac{\Theta(\kth{l},\kth{m}-\Delta \kth{})-\Theta(\kth{l},\kth{m}+\Delta \kth{})}{2\Delta \kth{}}\right]\, ,\label{eq:K21}\\
    K^{2\rightarrow 3}_{lm} &= -i\frac{3}{5d}\left[\frac{2}{\kth{l}}\Theta(\kth{l},\kth{m}) - \frac{\Theta(\kth{l},\kth{m}-\Delta \kth{})-\Theta(\kth{l},\kth{m}+\Delta \kth{})}{2\Delta \kth{}}\right]\, ,\label{eq:K23}\\
    K^{4\rightarrow 3}_{lm} &= -i\frac{10}{9d}\left[\frac{5}{\kth{l}}\Theta(\kth{l},\kth{m}) + \frac{\Theta(\kth{l},\kth{m}-\Delta \kth{})-\Theta(\kth{l},\kth{m}+\Delta \kth{})}{2\Delta \kth{}}\right]\, ,\label{eq:K43}
\end{align}
with $\Theta$ defined in \eq{eq:step}. As shown in these equations, we can implement derivatives within a matrix multiplication by including off-diagonal terms. Here we use two-sided derivatives except for the first and last bin in the data vector, where we use forward and backwards derivatives instead.
The transformation matrix $\bold{M}$ does depend on the order of wide-angle effects $n$, since at $n=2$ there are new even multipoles sourced by wide-angle effects. Following the rest of this paper, we only include terms at $n\leq 1$. The plot on the right in \fig{fig:Wkokt} shows the matrix $\bold{M}$ for the low redshift bin of BOSS DR12 NGC. The only survey specific parameter in this matrix is the amplitude of the LOS vector $d$. The code to calculate the matrix M is publicly available~\footnote{\url{https://github.com/fbeutler/pk_tools/blob/master/wide_angle_tools.py}}.

\section{Deconvolution as matrix multiplication}
\label{sec:deconvolution}

\begin{figure}[t]
    \centering
    \includegraphics[width=0.48\textwidth]{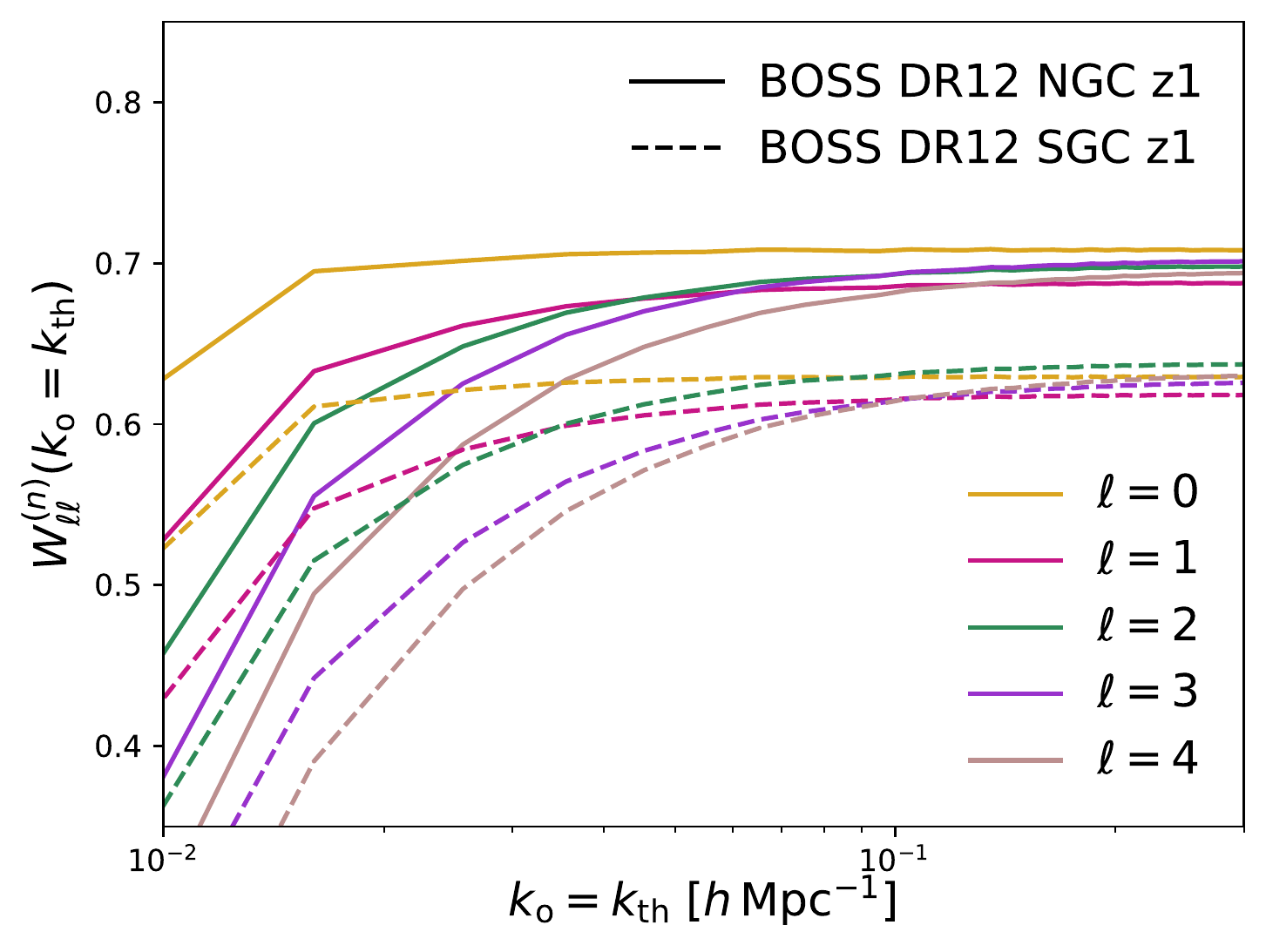}
    \includegraphics[width=0.48\textwidth]{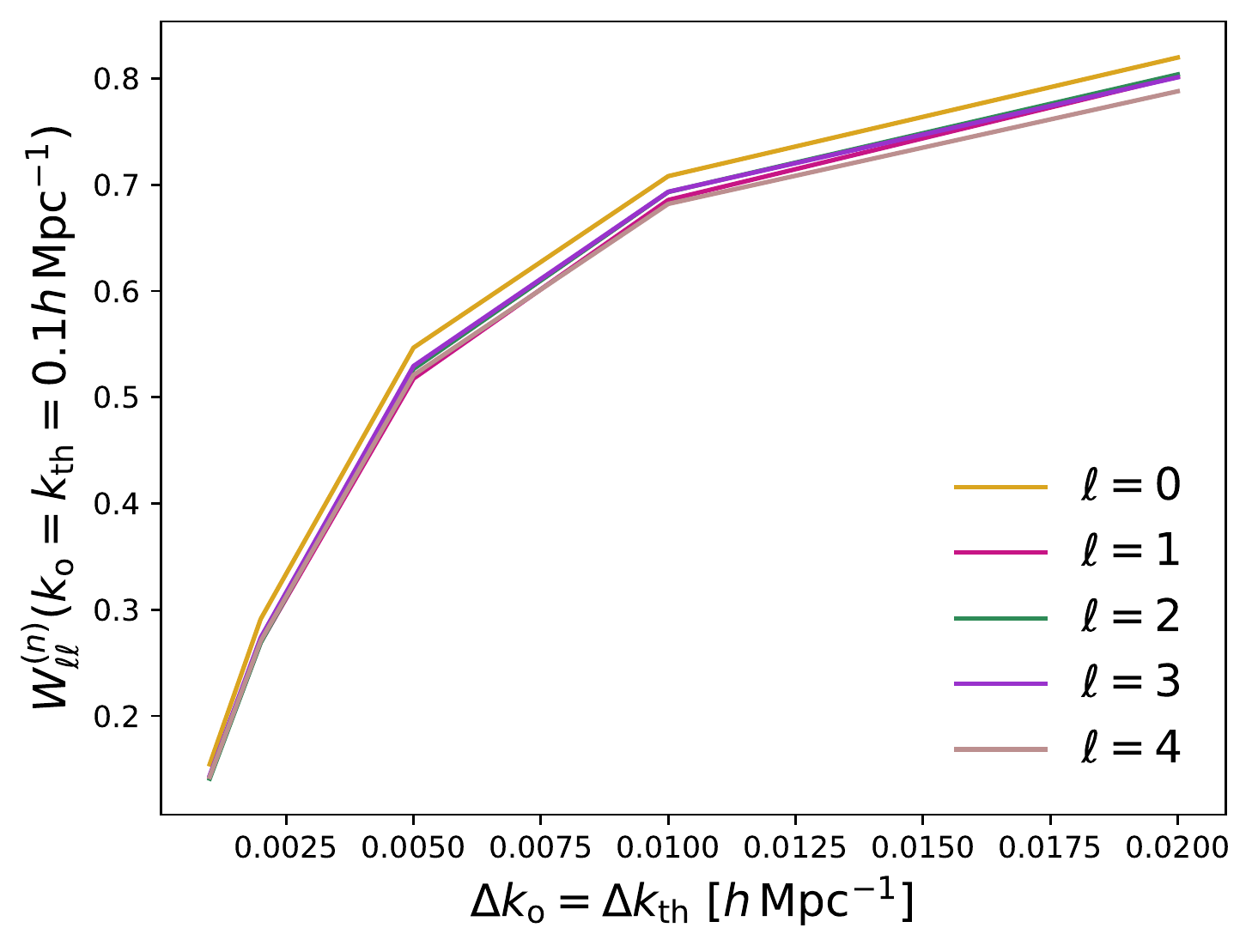}
    \caption{Left: Diagonal of the window function 
$W^{(n)}_{\ell\ell}(k_{\rm o}=k_{\rm th})$ for the low redshift bin of 
BOSS DR12 NGC (solid lines) and SGC (dashed lines). The differently colored 
lines correspond to the different multipoles. All lines in the plot on the 
left use $\Delta k_{\rm o} = \Delta k_{\rm th} = 0.01\kMpc$. The SGC has 
smaller values for $W^{(n)}_{\ell\ell}(k_{\rm o}=k_{\rm th})$ because its 
fundamental mode is larger. 
Right: The window function 
$W^{(n)}_{\ell\ell}(k_{\rm o}=k_{\rm th}=0.1\kMpc)$ as a function of the 
size of the bandpower bin width $\Delta k$. The quantity plotted on the 
y-axis of these two figures scales the bandpower uncertainty for the 
deconvolved power spectrum multipoles as shown in \eq{eq:Ik}.
\pvmhid{diag_window.py}
}
    \label{fig:Ik}
\end{figure}

\begin{figure}[t]
    \centering
    \includegraphics[width=0.48\textwidth]{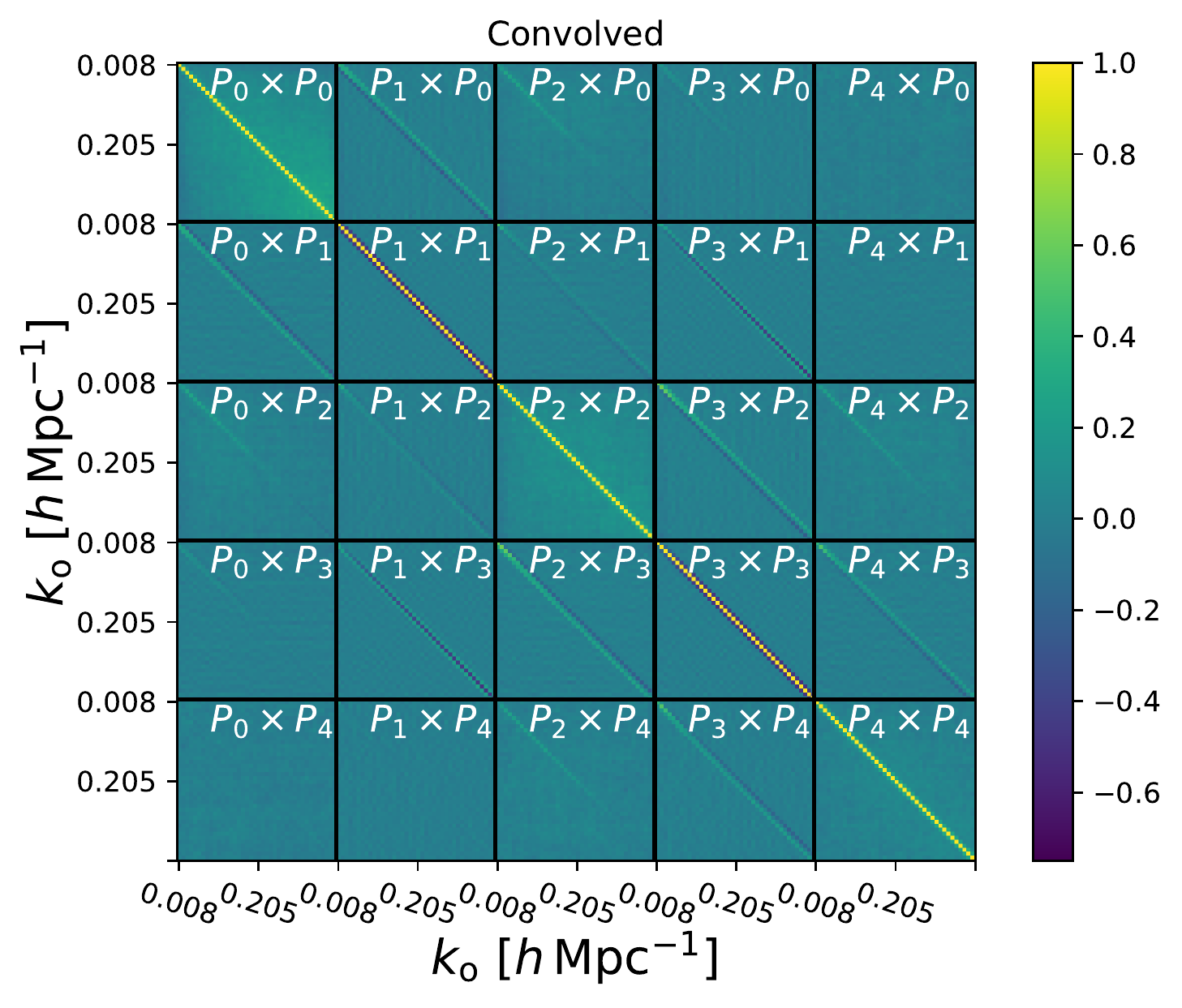}
    \includegraphics[width=0.48\textwidth]{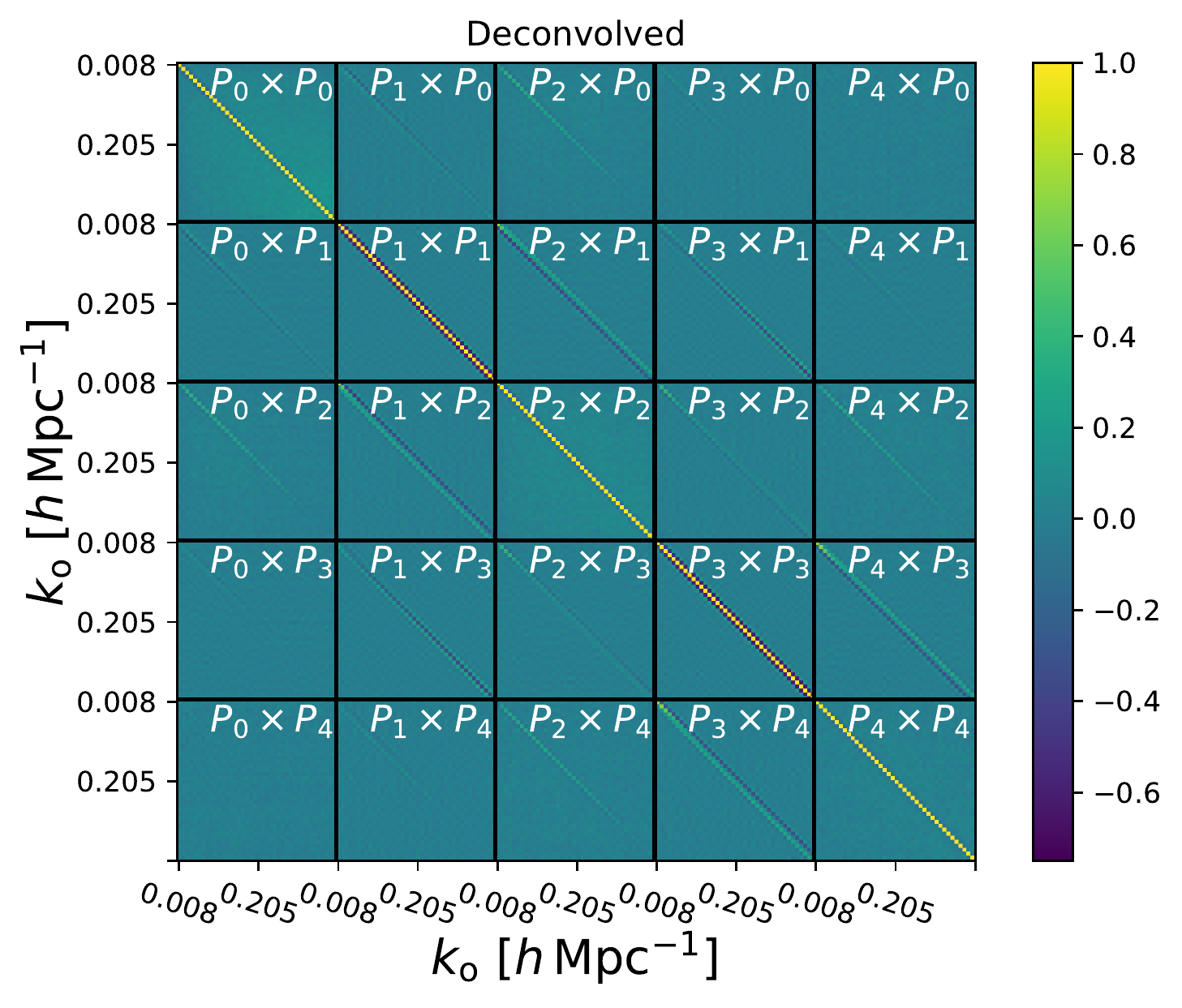}
    \caption{Left: Correlation matrix for the 5 power spectrum multipoles of 
BOSS DR12 NGC in the low redshift bin. These matrices are based on the $2048$ 
BOSS DR12 MultiDark-Patchy (MD-Patchy) mock catalogs with a binning of 
$\Delta k_{\rm o} = 0.01\kMpc$. Right: The same correlation matrix using the 
deconvolved power spectrum multipoles. The wide-angle expansion used here includes all terms with $n < 2$.}
    \label{fig:cov}
\end{figure}

\begin{figure}[t]
    \centering
    \includegraphics[width=0.6\textwidth]{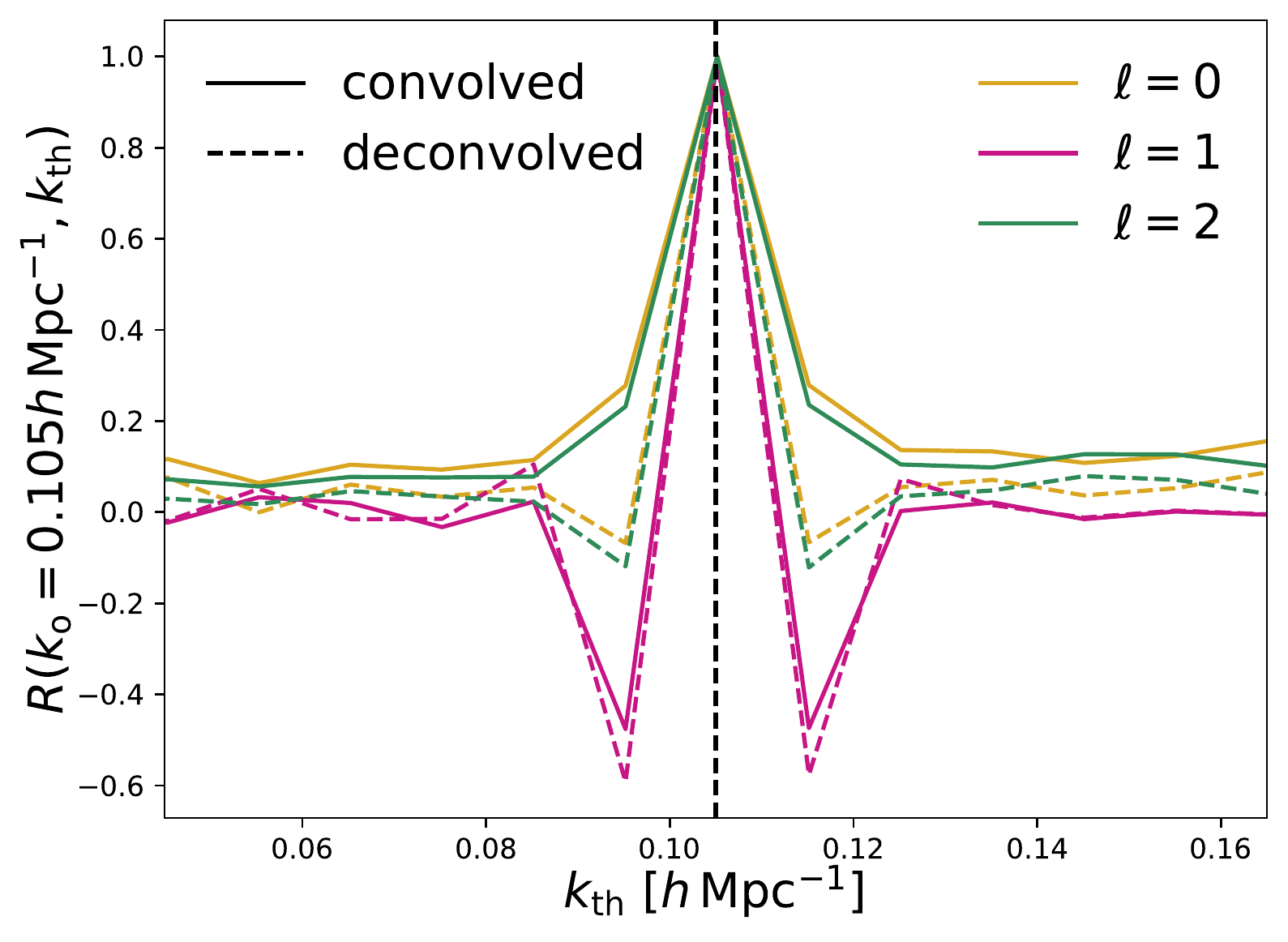}
    \caption{Slice through the correlation matrices shown in \fig{fig:cov} at $k_{\rm o} = 0.105\kMpc$ for the convolved (solid lines) and deconvolved (dashed lines) cases, respectively. We focus on the diagonal components of the monopole (yellow), dipole (pink) and quadrupole (green).}
    \label{fig:Cslice}
\end{figure}

Usually the aim of a power spectrum analysis is to obtain the likelihood
\begin{equation}
    \mathcal{L} \propto \exp\left[-\frac{1}{2}(\Pconvo - \bold{W}\Ptrue)^{\rm T}\invCconv(\Pconvo - \bold{W}\Ptrue)\right]\, ,
    \label{eq:likelihood}
\end{equation}
where $\invCconv$ is the inverse covariance matrix of the (convolved) power spectrum multipoles, $\Pconvo$ represents the measured power spectrum and $\Ptrue$ is the unconvolved power spectrum model. 

Using~\eq{eq:likelihood} together with the maximum likelihood condition 
$\partial\chi^2/\partial \Ptrue=0$ implies that the power spectrum before 
convolution with the window function is related to the convolved power 
spectrum by 
\begin{align}
     (\bold{W}^T\invCconv\bold{W})^{-1}\bold{W}^T\invCconv\Pconvo 
&\equiv \Pdeconv\, ,
    \label{eq:deconv}
\end{align}
where 
\begin{equation}
    (\bold{W}^T\invCconv\bold{W})^{-1}\bold{W}^T\invCconv = \bold{W}^{-1}
\end{equation} 
if $\bold{W}$ is a square matrix and non-singular. Note that the deconvolution process implied in \eq{eq:deconv} cannot recover the true underlying power spectrum (which can be calculated in theory), since the window function did erase all information below the fundamental mode.

The covariance matrix of the deconvolved power spectrum is given by
\begin{equation}
    \invCdeconv = \bold{W}^T\invCconv\bold{W}\, .
    \label{eq:cov_true}
\end{equation}
The correlation matrix, $R = C_{ij}/\sqrt{C_{ii}C_{jj}}$, for BOSS DR12 NGC in the low redshift bin is plotted in \fig{fig:cov} before (left) and after (right) deconvolution.
Deconvolution often leads to anti-correlated bins~\citep{Hamilton1999:astro-ph/9905192v4}, which can be intuitively explained by imagining a power spectrum model with very fine $k$ binning ($\Delta k \ll$ fundamental mode $k_f$). If one moves the power spectrum in one bin up, but compensates by moving the power spectrum in the adjacent bin down, the convolved version of this power spectrum will look very similar to the original power spectrum, since the window function averages neighbouring bins. Hence such a mode of variation is not well constrained in a deconvolved estimate, leading to anti-correlation.

Generally however, there is a decrease in the correlation between bandpowers as clearly visible in \fig{fig:Cslice}, which implies an increase in the variance of modes within that bin, given by 
\begin{equation}
    \bold{\sigma}_{\bold{P}_{\ell}^{\rm deconv}}(k) \propto \frac{1}{W^{(n)}_{\ell \ell}(k,k)}\, .
    \label{eq:Ik}
\end{equation}
Figure~\ref{fig:Ik} shows the behaviour of $W^{(n)}_{\ell \ell}(k,k)$ as a 
function of $k$ for different $\Delta k$ in the low redshift bin of BOSS DR12. 
This clearly shows that deconvolution can significantly increase the 
uncertainties in the power spectrum multipoles if $\Delta k$ is small. However, this 
increase in the uncertainties does not reflect any loss of information, but 
only shows that for smaller $\Delta k$ the window function has a larger 
impact. The larger uncertainty is compensated by having less correlation 
(or anti-correlation) between the bandpowers. The information content of 
$\Pdeconv$ and $\Pconvo$ in terms of the likelihood is identical, since we 
derived $\Pdeconv$ in \eq{eq:deconv} using the likelihood in 
\eq{eq:likelihood} (assuming $\Pdeconv$ is estimated in at least as many bands 
as $\Pconvo$). 

It should be noted, however, that our deconvolution procedure is based on a Gaussian likelihood (as given in eq.~\ref{eq:likelihood}), which might not be true on the largest scales of the survey (see e.g.~\citep{Hahn2018:1803.06348v1}). This could bias large-scale signals like primordial non-Gaussianity. Alternative to our procedure one could derive the deconvolved bandpowers using an MCMC approach without the need to assume a Gaussian likelihood or include a Gaussianisation step as e.g. suggested in~\citep{Wang2018:1811.08155v2}.

\subsection{Wide-angle compression}
\label{sec:compression}

\begin{figure}[t]
    \centering
    \includegraphics[width=0.49\textwidth]{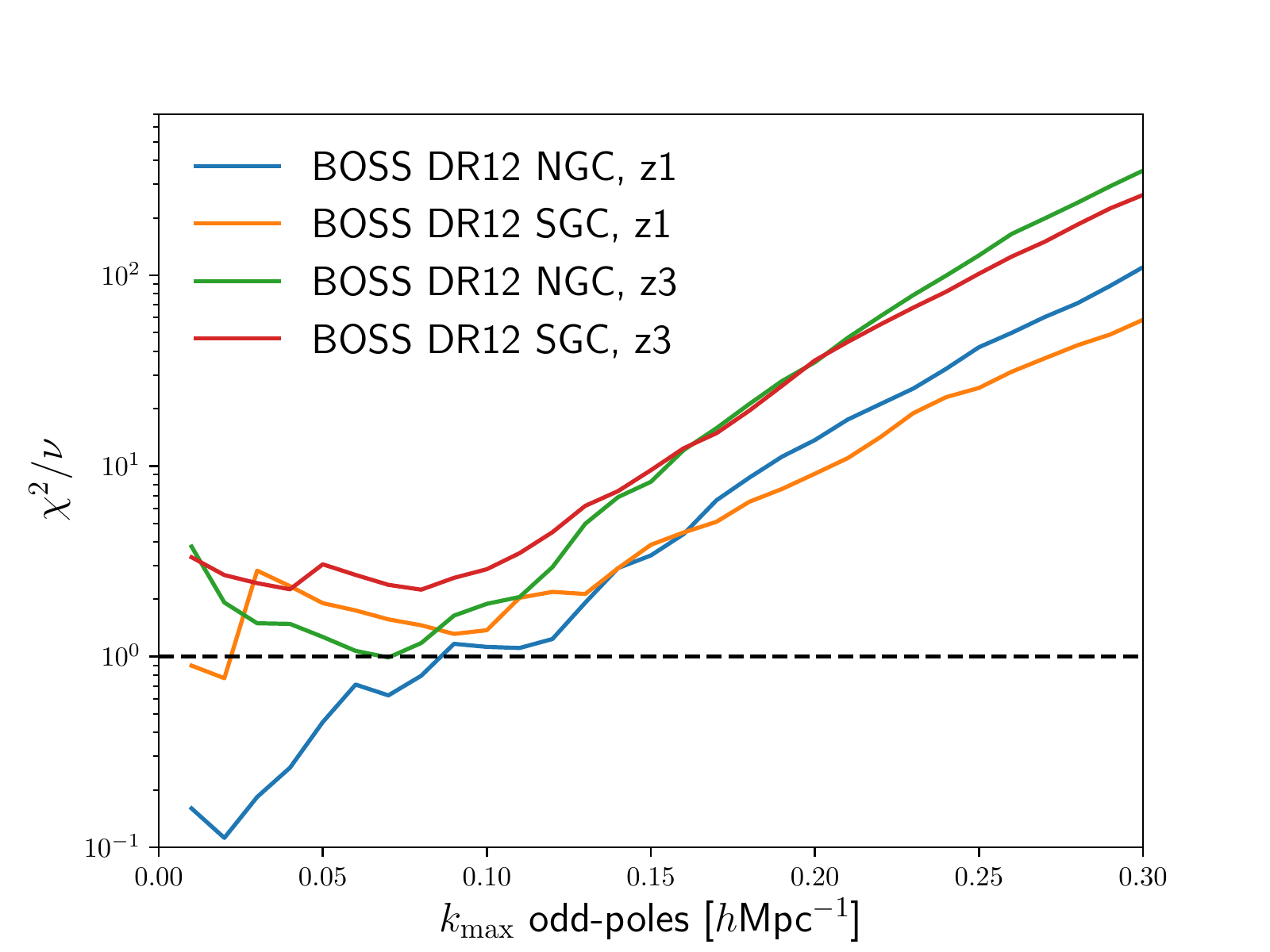}
    \includegraphics[width=0.49\textwidth]{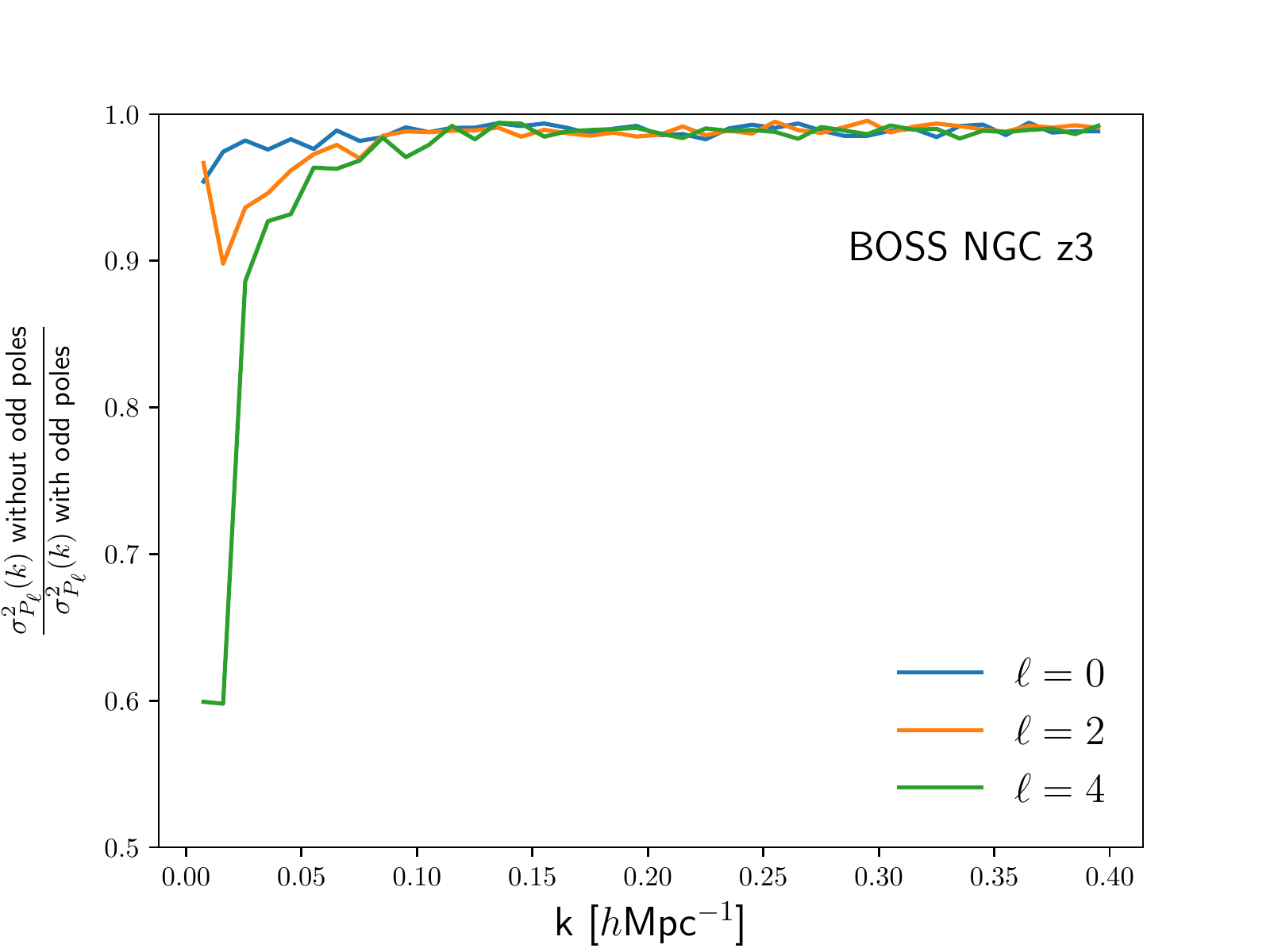}
    \caption{Left: The reduced $\chi^2$ for fits of band-parameterized even poles
($\Ptruefs$)
to all the measured even poles, and odd poles up to the maximum $k$ on the 
horizontal axis. 
Right: Diagonal covariance for the inferred flat sky even
poles (see eq.~\ref{eq:cov_compression}), including the constraint from $k<0.1\kMpc$ odd poles, 
divided by the same covariance with no odd pole constraint. 
}
\label{fig:decontoeven}
\end{figure}

In \eq{eq:deconv} we can replace $\vW$ with $\vtilW\equiv \vW \vM$ to
compress our vector of $5$ multipoles to a 
vector of $3$ even multipoles, i.e., 
\begin{align}
\Ptruefs &=   (\bold{\tilW}^T\invCconv\bold{\tilW})^{-1}\bold{\tilW}^T
\invCconv\Pconvo
\, ,
    \label{eq:compression}
\end{align}
while the corresponding covariance matrix is 
\begin{equation}
    \bold{C}_{\rm true,flat\text{-}sky}^{-1} = \vtilW^T\invCconv\vtilW\, .
    \label{eq:cov_compression}
\end{equation}
This is equivalent to a minimum $\chi^2$ fit of the band-parameterized model
for $\Ptruefs$ to all multipoles.
This is an 
over-constrained fit with the degrees of freedom equal to the number of bins in the
odd multipoles~\footnote{Assuming the theory is band-parameterized using the
same set of $k$ bins as the measurement, i.e., the number of free 
parameters is equal to the number of bins in the even power spectrum multipoles so that they 
cancel in the calculation of the degrees of freedom.}. Because the 
prediction of the odd multipoles from $\Ptruefs$ is model-independent, the reduced $\chi^2$ ($\chi^2/\nu$) for this fit is a 
model-independent test of the various 
quantities we report ($\vC$, $\vM$, etc.). Unfortunately, we find that
when we try to fit all multipoles on all scales this way, $\chi^2$ is 
unacceptably bad, i.e., the measured points are not consistent with 
{\it any model}, given our $\vM$, $\vW$, and $\vC$.  
As shown in figure \ref{fig:decontoeven} (left), this problem is 
driven by the high $k$ bins in the odd multipoles. 

Figure~\ref{fig:decontoeven} shows that $\chi^2/\nu$ is near unity if we drop the odd multipoles 
with $k\gtrsim 0.1\kMpc$, but quickly grows when including higher
$k$. This is not a straightforward thing to understand, or even see, 
by looking at the measured points, covariance matrix, etc. It appears to be
driven by a few special linear combinations of even and odd multipoles which 
have very low variance in the mocks used to compute the covariance matrix. 
These are then not well-predicted by the theory as multiplied by $\vM$ and
$\vW$. E.g., for BOSS DR12 NGC, z3, $k<0.2\kMpc$, we find $\chi^2=1400$, of
which $1127$ is contributed by the single worst eigenvector of the covariance matrix. We were unable to find any flaw in the calculations that would fix
this. Since we expect the wide-angle effects to be primarily important
at low-$k$, our solution is to simply drop the odd multipoles at $k>0.1\kMpc$ from 
fits. The right panel of figure \ref{fig:decontoeven} shows an example of
the impact of
the lower $k$ odd multipoles on the inferred deconvolved even multipole variance. As expected
the impact of the odd multipoles is primarily at very low $k$. 

\section{Datasets}
\label{sec:datasets}

\begin{table}
    \begin{center}
        \hspace*{-1.5cm}\begin{tabular}{lllllllll}
            Survey/Sample & $N_{\rm gal}$ & [$z_{\rm min}$ $z_{\rm max}$] & $z_{\rm eff}$ & $V_{\rm eff}$ & $k_f$ ($50\%$) & $k_f$ ($95\%$) & $N_{\rm m}$ & 
$A/A_{\rm old}$\\
             &  &  &  & [Gpc$^3$] & [$h$Mpc$^{-1}$] & [$h$Mpc$^{-1}$] & &\\
            \hline
            6dFGS DR3 & $75\,117$ & [$0.01$, $0.2$] & $0.096$ & $0.12$ & $0.00514$ & $0.00142$ & $600$ & 1.0168\\
            BOSS DR12 NGC z1 & $429\,182$ & [$0.2$, $0.5$] & 0.38 & $2.6$ & $0.00222$ & $0.00062$ & $2048$ & 0.9032\\
            BOSS DR12 SGC z1 & $174\,820$ & [$0.2$, $0.5$] & 0.38 & $1.0$ & $0.00310$ & $0.00086$ & $2048$ & 0.8896\\
            BOSS DR12 NGC z3 & $435\,742$ & [$0.5$, $0.75$] & 0.61 & $2.8$ & $0.00154$ & $0.00042$ & $2048$ & 0.9104\\
            BOSS DR12 SGC z3 & $158\,262$ & [$0.5$, $0.75$] & 0.61 & $1.0$ & $0.00226$ & $0.00062$ & $2048$ & 0.9016\\
            eBOSS DR16 QSO NGC & $218\,209$ & [$0.8$, $2.2$] & $1.52$ & $0.35$ & $0.00094$ & $0.00026$ & $1000$ & 0.9302\\
            eBOSS DR16 QSO SGC & $125\,499$ & [$0.8$, $2.2$] & $1.52$ & $0.18$ & $0.00114$ & $0.00034$ & $1000$ & 0.8974
        \end{tabular}\hspace*{-1.5cm}
        \caption{Properties of the different galaxy samples studied in this 
paper. 
The effective redshift is defined as $z_{\rm eff} = \sum_i w_iz_i/\sum_i w_i$, 
where $w_i$ includes any completeness weight as well as the FKP 
weight~\citep{Feldman1993:astro-ph/9304022v1}. To calculate the FKP weight as 
well as the effective volume, we assume 
$P_0 = 10\,000h^{-3}$Mpc$^3$ for BOSS and 6dFGS and 
$P_0 = 6\,000h^{-3}$Mpc$^3$ for eBOSS. The effective volume in this table 
might differ from other references due to (1) a different fiducial cosmology 
and (2) only the raw (unweighted) number density of galaxies is used in our 
calculation. $k_f$ represents the fundamental mode, which is given by the 
wavenumber at which the monopole window $Q^{(0)}_{0}(k)$ has $50\%$ of the 
height of $Q^{(0)}_{0}(k\rightarrow 0)$ (column 6) or $95\%$ (column 7). 
$N_m$ represents the number of mock realizations available for each sample, and 
the last column gives the ratio 
between our $A$ definition, enforcing $Q_0(s\rightarrow 0)=1$, and the 
normalization used in many previous papers (see 
section~\ref{sec:window}). For BOSS we also have catalogs after applying 
density field reconstruction. However, only $1000$ of the $2048$ BOSS mock 
catalogs have been processed in this way~\protect\footnotemark. More 
information about the different samples can be found in the 
relevant data release papers of 
6dFGS~\citep{Jones2009:0903.5451v1}, 
BOSS~\cite{Reid2015:1509.06529v2} and eBOSS~\citep{Ata2017:1705.06373v2}. 
\label{tab:survey_properties}
}
    \end{center}
\end{table}

\footnotetext{We actually have $997$ mock catalogs for the NGC in the low and high redshift bins and $1000$ mock catalogs for the SGC in both redshift bins. Some NGC mock catalogs have been excluded due to issues while processing the catalogs.}

\begin{figure}[t]
    \centering
    \includegraphics[width=1.\textwidth]{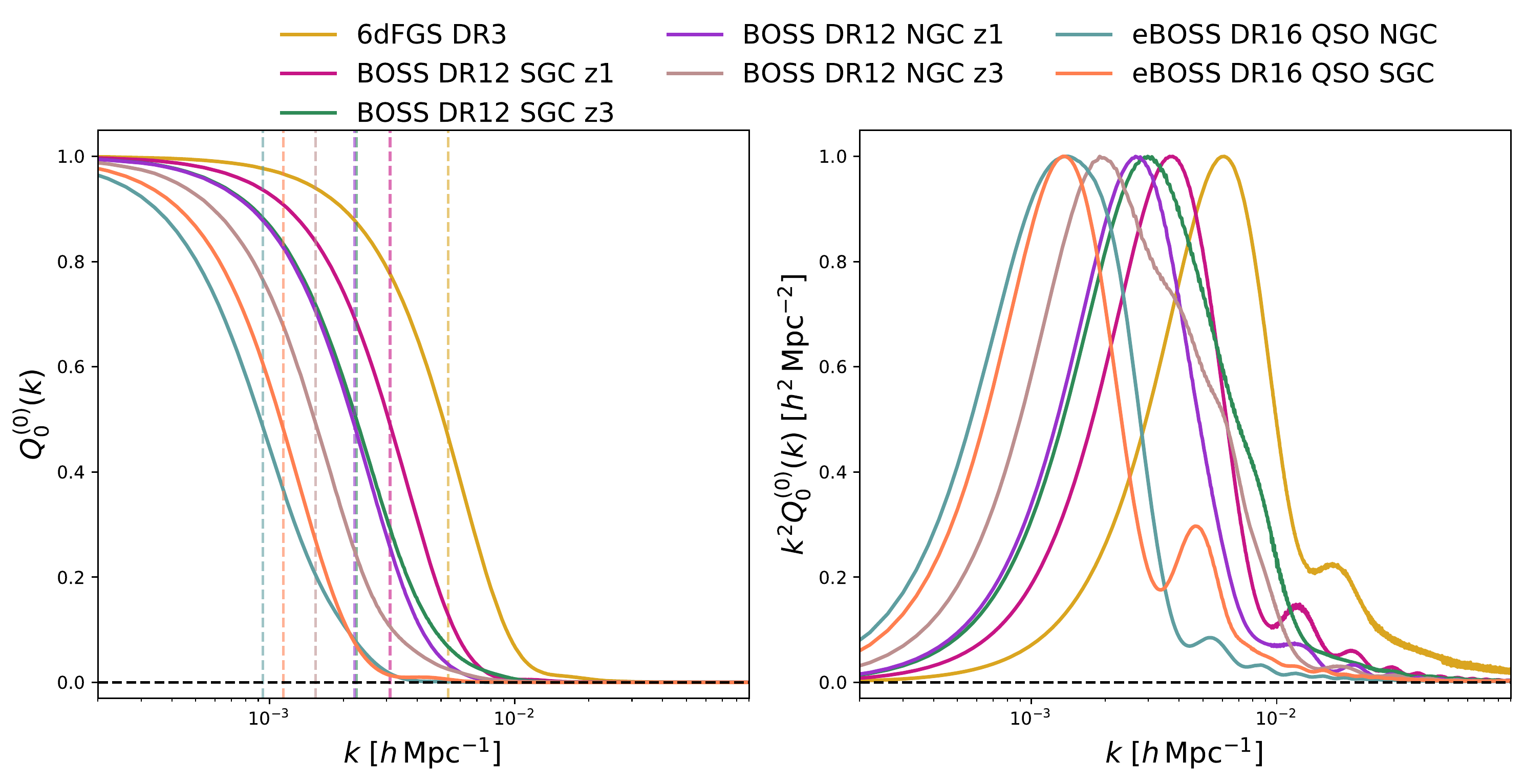}
    \caption{Comparison of the monopole window function for 6dFGS DR3, 
BOSS DR12 and eBOSS DR16. The dashed lines in the plot on the left indicate 
the fundamental mode defined as the $50\%$ height of the monopole. The Figure 
on the right shows the monopole weighted by $k^2$, which is the quantity which 
goes in the window function in \eq{eq:intW} at $n=0$. The double peaked shape 
of the eBOSS DR16 SGC window is probably caused by the narrow band geometry 
and/or the split of the SGC into two separate regions (see figure 2 
in~\citep{Ata2017:1705.06373v2}). The curves in the plot on the right hand 
side are normalized so that the maximum of the curve is at unity (which 
allows comparison between the different surveys).
}
    \label{fig:kspace_window}
\end{figure}

Here we will introduce the three galaxy redshift survey datasets we analyze in this paper, namely the 6dFGS DR3 sample, the BOSS DR12 sample and the eBOSS DR16 quasar
sample.

\subsection{6dFGS DR3}

The 6-degree Field Galaxy Survey (6dFGS~\citep{Jones2009:0903.5451v1}) is a K-band selected, magnitude limited ($K \leq 12.9$) galaxy survey, based on the 2MASS Extended Source Catalog (2MASS XSC;~\citep{Jarrett2000:astro-ph/0004318v1}).
6dFGS covers nearly the entire southern sky and is the lowest redshift sample included in this paper. The survey made use of the Six-Degree Field (6dF) multi-fibre instrument on the UK Schmidt telescope at the Siding Spring Observatory. The three data release papers~\citep{Jones2004:astro-ph/0403501v1,Jones2006:astro-ph/0603609v2,Jones2009:0903.5451v1} describe 6dFGS in full detail, including comparisons between 6dFGS, 2dFGRS and SDSS. Here we use the final K-band selected sample, which contains  $75\,117$ galaxies and has been used in several galaxy clustering studies~\citep{Beutler2012:1212.3610v1,Beutler2012:1204.4725v1,Beutler2011:1106.3366v1,Scrimgeour2015:1511.06930v1,Johnson2014:1404.3799v2,Carter2018:1803.01746v1} (see Table~\ref{tab:survey_properties} for more details). 

\paragraph{Mock catalogs:} The 6dFGS mock catalogs are based on COLA simulations~\citep{Koda2015:1507.05329v1} with $1728^3$ particles in boxes with $1.2h^{-1}$Gpc on each side. The simulations use $20$ time steps down to $z=0$ and a mass resolution of $2.8\times 10^{10}h^{-1}M_{\odot}$. A friends-of-friends (FoF) finder is used to locate halos with a minimum of $32$ dark matter particles per halo. The cosmology used in these simulations is $\Omega_m = 0.3$, $\Omega_b = 0.0478$, $h = 0.68$, $\sigma_8 = 0.82$ and $n_s = 0.96$. The derived halo catalogs are populated with galaxies using an HOD measured on the dataset itself~\citep{Beutler2012:1212.3610v1} (for more details see section 3 of~\citep{Carter2018:1803.01746v1}).\\

The 6dFGS sample has by far the smallest volume of all the datasets included in this analysis (see table~\ref{tab:survey_properties} and \fig{fig:kspace_window}). However, it does occupy a unique redshift range, which is not covered by BOSS or eBOSS. 

\subsection{BOSS DR12}
\label{sec:BOSS}

The Baryon Oscillation Spectroscopic Survey (BOSS) was part of SDSS-III~\citep{Eisenstein2011:1101.1529v2,Dawson2012:1208.0022v3} and measured spectroscopic redshifts of $1\,198\,006$ million galaxies~\citep{Reid2015:1509.06529v2}.
The survey covers $10\,252\,$deg$^2$ divided in two patches on the sky, the North Galactic Cap (NGC) and the South Galactic Cap (SGC), over a redshift range of $0.2$ - $0.75$. Here we split this redshift range into two redshift bins defined by $0.2 < z < 0.5$ and $0.5 < z < 0.75$ with the effective redshifts $z_{\rm eff} = 0.38$ and $0.61$, respectively.
We also include the different incompleteness weights as
\begin{equation}
    w_c = (w_{\rm rf} + w_{\rm fc}-1)w_{\rm sys}\, ,
\end{equation}
which account for redshift failures ($w_{\rm rf}$), fibre collisions ($w_{\rm fc}$) and photometric systematics related to the observational seeing conditions and correlations with stellar density ($w_{\rm sys}$)~\citep{Anderson2013:1312.4877v2,Ross2016:1607.03145v2}.

\paragraph{Mock catalogs:} The BOSS collaboration provided mock catalogs for the final BOSS DR12 dataset (MD-Patchy mock catalogs~\citep{Kitaura2015:1509.06400v3}). These catalogs have been produced using approximate gravity solvers and analytical-statistical biasing models calibrated to a reference sample from the BigMultiDark simulations~\citep{Klypin2014:1411.4001v2}. The BigMultiDark simulation is based on gadget-2~\citep{Springel2005:astro-ph/0505010v1} with $3840^3$ particles
in a volume of ($2.5 h^{-1}$Gpc)$^3$ assuming a $\Lambda$CDM cosmology with $\Omega_m = 0.307115$, $\Omega_b = 0.048206$, $\sigma_8 = 0.8288$, $n_s = 0.9611$ and $h = 0.6777$. The mock catalogs use halo abundance
matching to reproduce the observed BOSS two- and three-point clustering measurements~\citep{Rodriguez-Torres2015:1509.06404v3}. This technique is applied as a function of redshift to reproduce the BOSS DR12 redshift evolution.\\

We already saw the BOSS DR12 NGC window function $W^{(n)}_{\ell\ell'}$ of the low redshift bin (z1) in \fig{fig:wll0_NGC_z1} and \ref{fig:wll1_NGC_z1}. We also include the monopole window function, $Q^{(0)}_0(k)$, in \fig{fig:kspace_window}. The BOSS DR12 sample represents the largest galaxy redshift dataset 
(in terms of effective volume)
\citep{Alam2016:1607.03155v1, Ross2016:1607.03145v2, 
Grieb2016:1607.03143v2, Sanchez2016:1607.03147v1, Satpathy2016:1607.03148v2, 
Beutler2016:1607.03150v1, Beutler2016:1607.03149v1}.

\subsection{eBOSS DR16 QSO}

The extended Baryon Oscillation Spectroscopic Survey (\citep{Dawson2015:1508.04473v2}, eBOSS) is part of SDSS-IV~\citep{Blanton2017:1703.00052v2} and relies on the same optical spectrographs~\citep{Smee2012:1208.2233v2} as the SDSS-III BOSS survey. In addition to observing luminous red galaxies (LRGs) and emission line galaxies (ELGs), eBOSS collected redshifts for $\sim 500\,000$ quasars. While the Quasar density is comparatively low, this sample has the distinction of covering the largest cosmic volume, leading to the smallest fundamental mode of all samples discussed in this paper (see figure~\ref{fig:kspace_window}). Here we focus on the eBOSS quasar sample to avoid any overlap with the other samples. 

The eBOSS targets~\citep{Abolfathi2017:1707.09322v3,Paris2017:1712.05029v2} are selected from the  DR7~\citep{Abazajian2008:0812.0649v2} and DR8~\citep{collaboration2011:1101.1559v2} photometric catalogs as well as the Wide Field Infrared Survey Explorer (WISE,~\citep{Wright2010:1008.0031v2}), as described in~\citep{Myers2015:1508.04472v2}.
Just like BOSS, the eBOSS quasars are split in two angular regions, the North Galactic Cap (NGC) and South Galactic Cap (SGC). The effective areas of these regions are $2924\,$deg$^2$ and $1884\,$deg$^2$, respectively (see Table~\ref{tab:survey_properties} for more details). Each eBOSS object has a completeness weight including corrections for fibre collisions, redshift failures~\citep{Paris2016:1608.06483v1} and photometric systematic effects~\citep{Laurent2017:1705.04718v1}:
\begin{equation}
    w_{\rm c} = w_{\rm cp}w_{\rm npz}w_{\rm sys}\, .
\end{equation} 
We refer to section 5.6 of~\citep{Ross2020:2007.09000v1} for details about these weights. The cosmology results for the DR16 quasar sample were presented in~\citep{Collaboration]2020:2007.08991v1,Neveux2020:2007.08999v1}~\footnote{\url{https://www.sdss.org/dr16/}}, which reported a BAO distance measurement in the range $0.8 < z < 2.2$.

\paragraph{Mock datasets:} We make use of a set of $1000$ mock catalogs to 
estimate the covariance matrix of the eBOSS DR16 quasar power spectrum. The 
mocks are based on the Extended Zel'dovich (EZ) approximate N-body 
simulation
scheme~\citep{Chuang2014:1409.1124v2}. These mocks rely on the 
Zel'dovich approximation to generate a density field, while including  
nonlinear and halo biasing effects through the use of free parameters. 
These free parameters are tuned to produce two-point and three-point 
clustering of a desired data set. 
The EZ mock catalogs account for the redshift evolution of the eBOSS quasars by constructing a light-cone out of $7$ redshift shells, generated from periodic boxes of side length $L = 5000 h^{-1}$Mpc at different redshifts. 
These mock catalogs also mimic the fibre collisions and redshift failures, so that each object has an associated $w_{\rm foc}$ and $w_{\rm cp}$. 
The cosmology of the EZmocks is $\Lambda$CDM with $\Omega_m = 0.307115$, $\Omega_b = 0.048206$, $h = 0.6777$, $\sigma_8 = 0.8255$, and $n_s = 0.9611$ (same cosmology as the BOSS DR12 MD-Patchy mocks discussed above).\\

Figure~\ref{fig:kspace_window} shows the window function monopole for the different samples where the NGC sample of eBOSS DR16 indicates the smallest fundamental mode of all samples discussed in this paper. The fact that the eBOSS quasar sample probes the largest scales currently accessible with galaxy redshift surveys, makes it a valuable tool to test primordial non-Gaussianity~\citep{Castorina2019:1904.08859v1}.  However, the high shot noise level does mean that the effective volume of the eBOSS DR16 QSO sample at most wavenumbers is below BOSS DR12.

\section{Data analysis}
\label{sec:analysis}

Here we will apply the deconvolution formalism developed in section~\ref{sec:deconvolution} to the three datasets introduced in the previous section.

\subsection{Setup of power spectrum measurements}
\label{sec:pksetup}

All power spectra discussed in this section are measured using the estimator of~\citep{Bianchi2015:1505.05341v2} and \citep{Scoccimarro2015:1506.02729v2} including the odd multipoles as discussed in appendix E of~\citep{Beutler2018:1810.05051v3}. For all power spectrum measurements we use a cubic grid with $L= 2000$, $3500\,h^{-1}$Mpc and $N=600$, $700$ for 6dFGS and BOSS, respectively. For the eBOSS QSO sample we use $L=5400$, $6600\,h^{-1}$Mpc with $N=1040$, $1270$ for the SGC and NGC, respectively. This setup ensures that for all three surveys the Nyquist frequency is $> 0.6\kMpc$. The galaxies are assigned to the grid using the triangular shape cloud procedure and we correct for the associated pixel window function~\citep{Jing2004:astro-ph/0409240v2}. To further reduce aliasing effects we included the interlacing procedure of~\citep{Sefusatti2015:1512.07295v2}. 
As defined in Section \ref{sec:window}, we use a different
normalization for the power spectra than other recent papers, which brings the
convolved power closer to the true/deconvolved power. 
For comparison with past work, Table~\ref{tab:survey_properties} gives the
ratio, of our definition of the normalisation to the one that we would compute 
using eq.~(13) 
of \cite{Beutler2016:1607.03150v1}.

\subsection{Deconvolution}

\begin{figure}[t]
    \centering
    \includegraphics[width=0.32\textwidth]{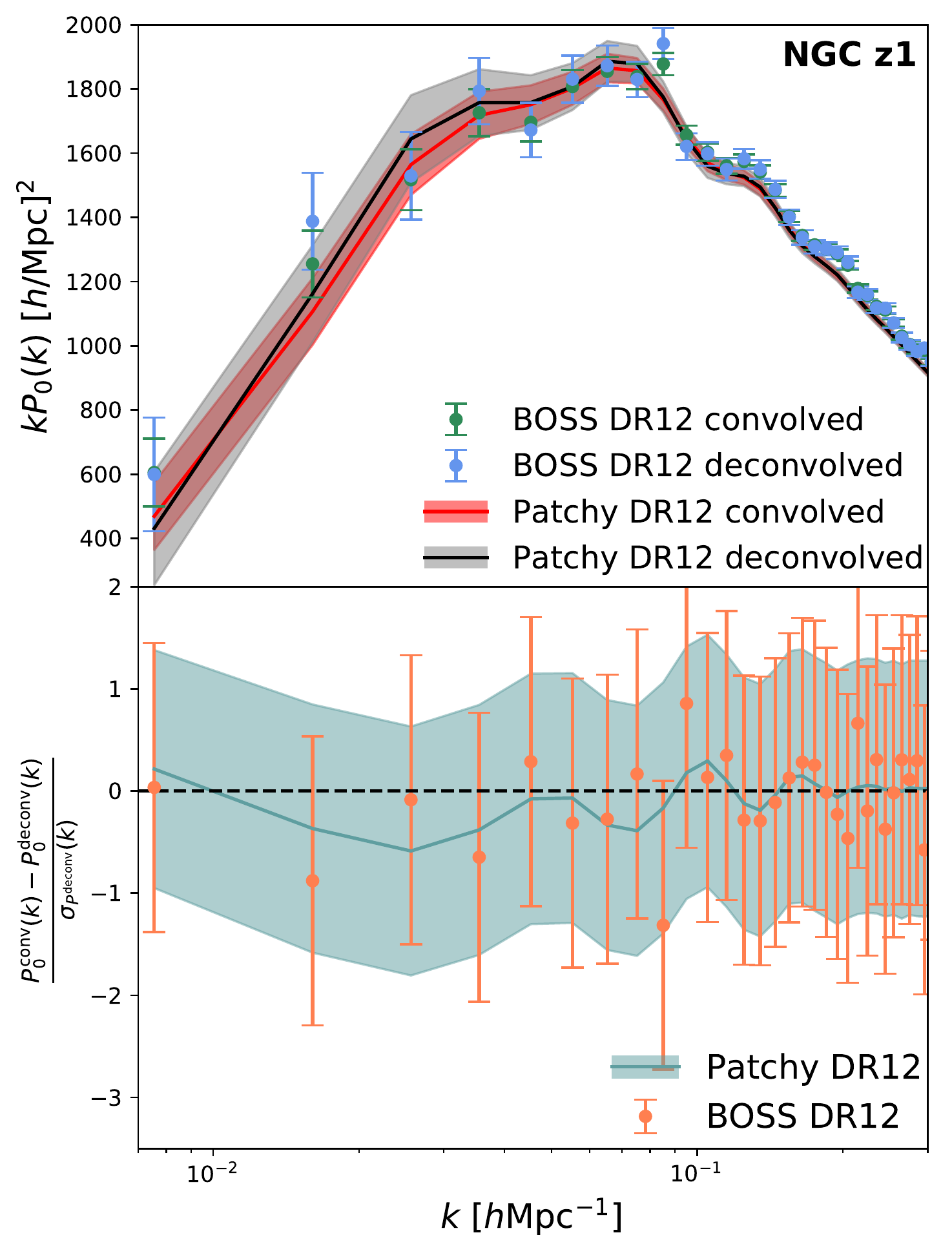}
    \includegraphics[width=0.32\textwidth]{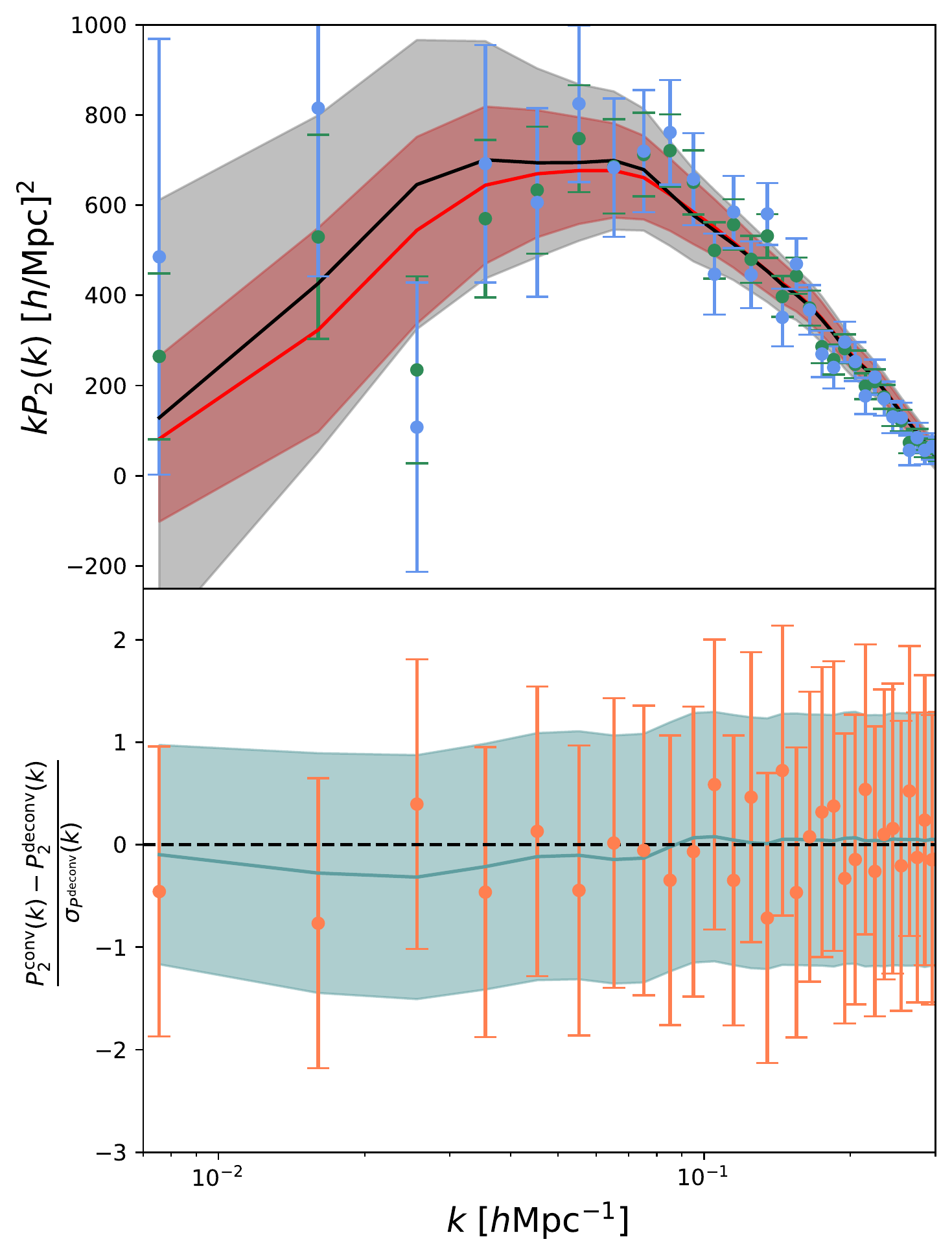}
    \includegraphics[width=0.32\textwidth]{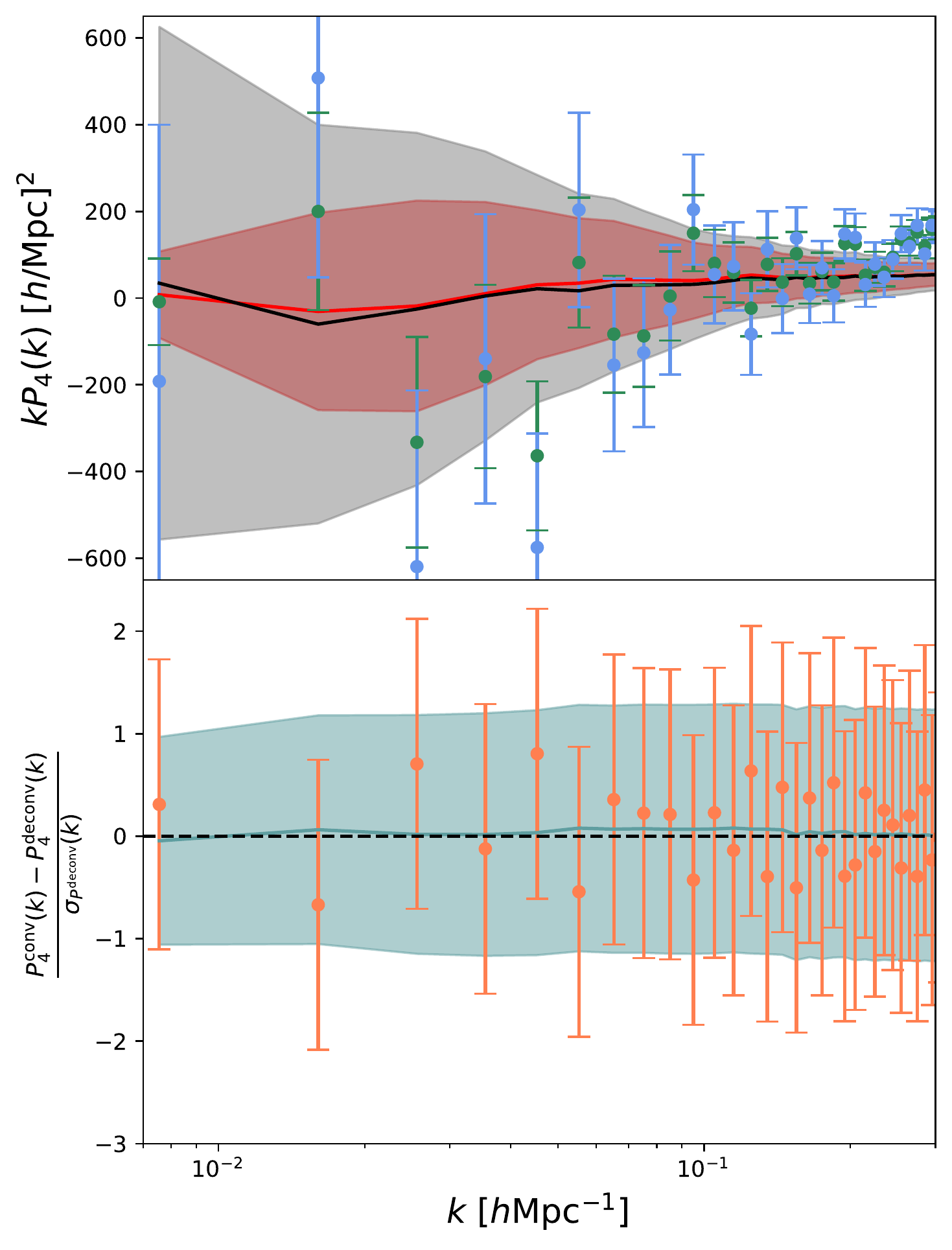}
    \caption{Comparison of the convolved and deconvolved even power spectrum multipoles for the BOSS DR12 low redshift bin (z1) in the NGC. The red line and shaded area shows the mean of the measured MD-Patchy DR12 power spectrum multipoles together with the standard deviation. The black line and shaded area shows the equivalent power spectra after deconvolution. The data points show the actual BOSS DR12 data. The lower panel shows the difference between the red and black lines in the upper panel, relative to the uncertainties post-deconvolution, which highlights the impact of the window function.}
    \label{fig:even_decon_patchy_NGC_z1}
\end{figure}

\begin{figure}[t]
    \centering
    \includegraphics[width=0.45\textwidth]{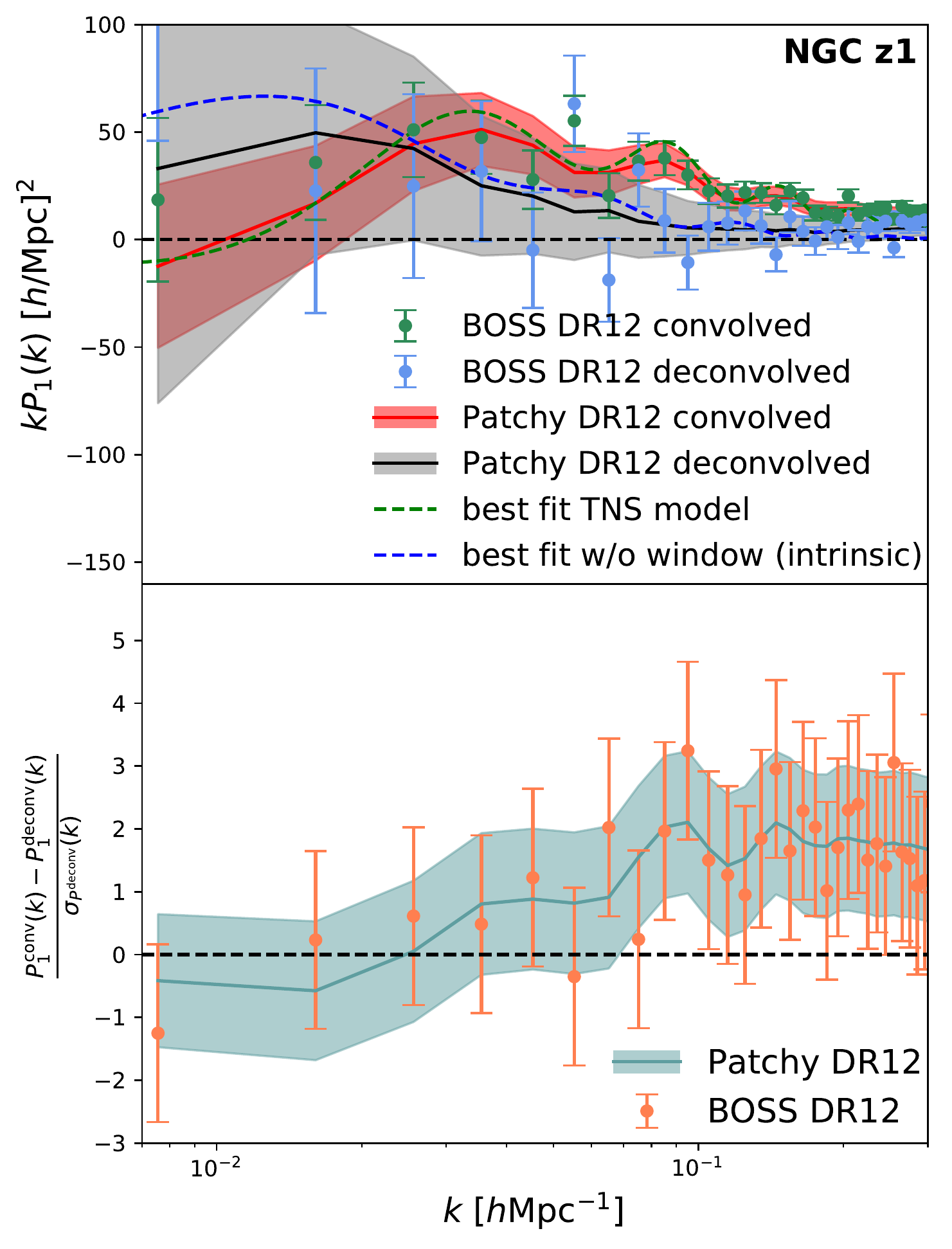}
    \includegraphics[width=0.45\textwidth]{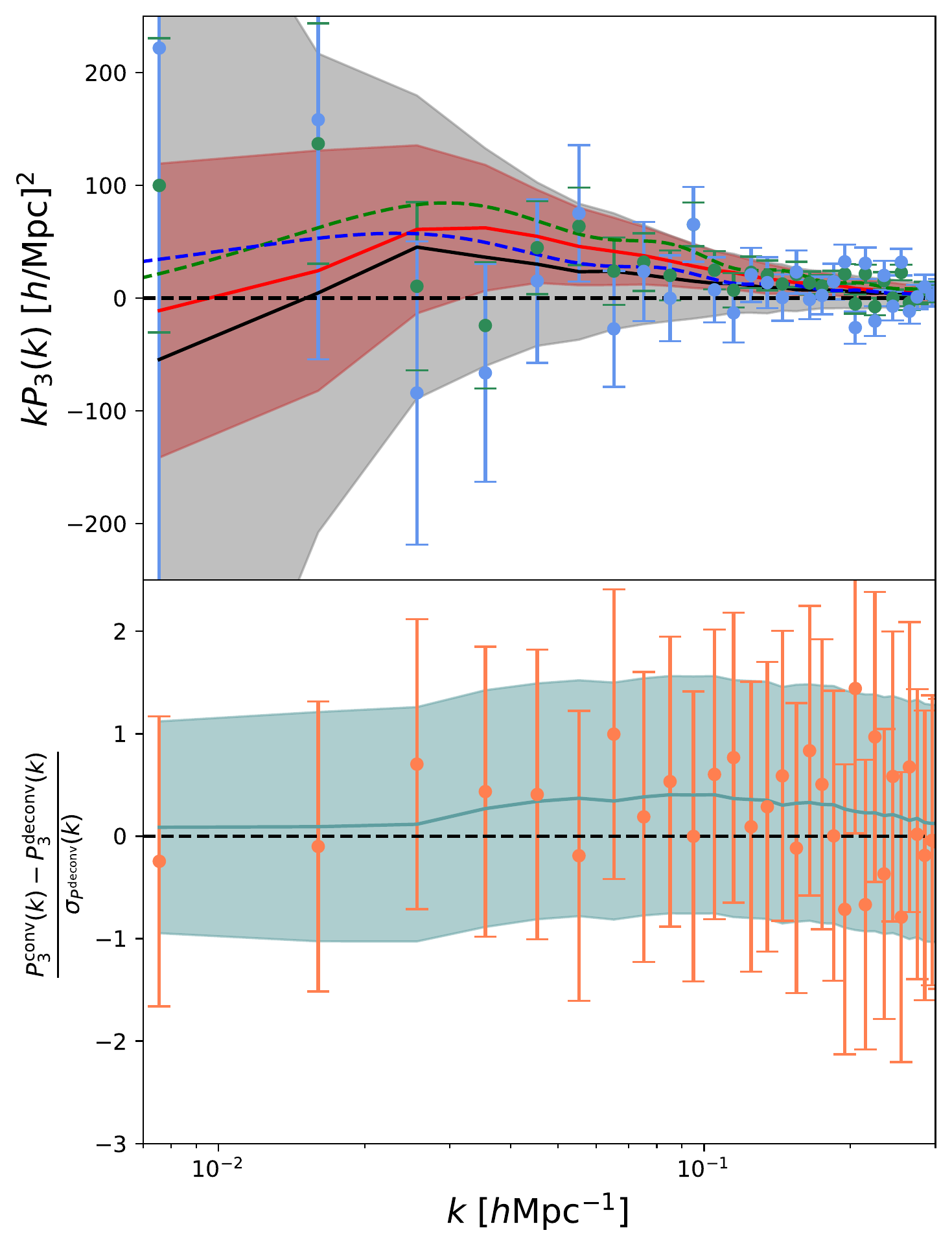}
    \caption{Comparison of the convolved and deconvolved odd power spectrum multipoles for the BOSS DR12 low redshift bin (z1) in the NGC. The red line and shaded area shows the mean of the measured MD-Patchy DR12 power spectrum multipoles together with the standard deviation. The black line and shaded area shows the equivalent power spectra after deconvolution. The data points show the actual BOSS DR12 data. The green dashed line shows the best fit as discussed in section 3.6 of \citep{ Beutler2018:1810.05051v3}. The dashed blue line shows the intrinsic odd multipoles introduced by wide-angle effects based on the same model as the green dashed line. The lower panel shows the difference between the red and black lines in the upper panel, relative to the uncertainties post-deconvolution, which highlights the impact of the window function.}
    \label{fig:odd_decon_patchy_NGC_z1}
\end{figure}

Figure~\ref{fig:even_decon_patchy_NGC_z1} and \ref{fig:odd_decon_patchy_NGC_z1} show a comparison of the convolved and deconvolved power spectrum multipoles of BOSS DR12 NGC in the low redshift bin. For these plots we combined $10$ measured bandpowers of $\Delta k_{\rm o} = 0.001\kMpc$ into larger bins of $\Delta k_{\rm o} = 0.01\kMpc$. Since deconvolution requires a square matrix in \eq{eq:deconv}, we also use $\Delta k_{\rm th} = 0.01\kMpc$. In \fig{fig:even_decon_patchy_NGC_z1} we can see the increase in the variance caused by deconvolution (comparison of the red and black shaded regions). At the same time deconvolution reduces the correlation between bandpowers. The odd multipoles in \fig{fig:odd_decon_patchy_NGC_z1} also include a best fitting model (green dashed lines) as well as the intrinsic dipole and octopole (blue dashed lines) as given by~\eq{eq:dipole_analytic} and \eq{eq:octopole_analytic}. These models are based on a fit to the even multipoles as discussed in section 3.6 of \citep{Beutler2018:1810.05051v3}. One can clearly see that deconvolving the dipole removes the window function contributions, which dominate the dipole on most scales and recovers the intrinsic dipole expected due to wide-angle effects. We included the corresponding results for 6dFGS and eBOSS in appendix~\ref{app:deconvolution}.

Figure~\ref{fig:decon_6dfgs} in appendix~\ref{app:deconvolution} shows the convolved and deconvolved power spectrum multipoles of 6dFGS.
From these plots we can see that the monopole power spectrum of the mock catalogs does not perfectly match the data power spectrum amplitude, while higher order multipoles agree well~\citep{Blake2018:1801.04969v2}. Even though 6dFGS is the lowest redshift sample, the wide-angle effects seem far less important compared to BOSS DR12, and the dipole is (by eye) consistent with zero. This agrees with~\citep{Beutler2011:1106.3366v1} where wide-angle effects are discussed in appendix C. Window function effects also seem to be much smaller than the statistical noise, which might be caused by the very compact geometry of the 6dFGS survey (see~\citep{Jones2009:0903.5451v1} for details about the angular and redshift distribution of 6dFGS).

Figure~\ref{fig:decon_eBOSS} in appendix~\ref{app:deconvolution} compares the power spectrum multipoles of the eBOSS DR16 quasar sample with the corresponding mock results before and after deconvolution. While the dipole and octopole moments are consistent with zero for each bandpower estimate, there is a clear systematic dipole signal~\footnote{Even though the Nyquist frequency is $k_{\rm Ny} = 0.6\kMpc$, the eBOSS power spectrum dipole below $k<0.3\kMpc$ does seem to be affected by aliasing, causing the rise in the dipole at high $k$. This should be taken into account when analyzing the eBOSS dataset.}.

\subsection{Deconvolution and BAO}

\begin{figure}[t]
    \centering
    \includegraphics[width=0.48\textwidth]{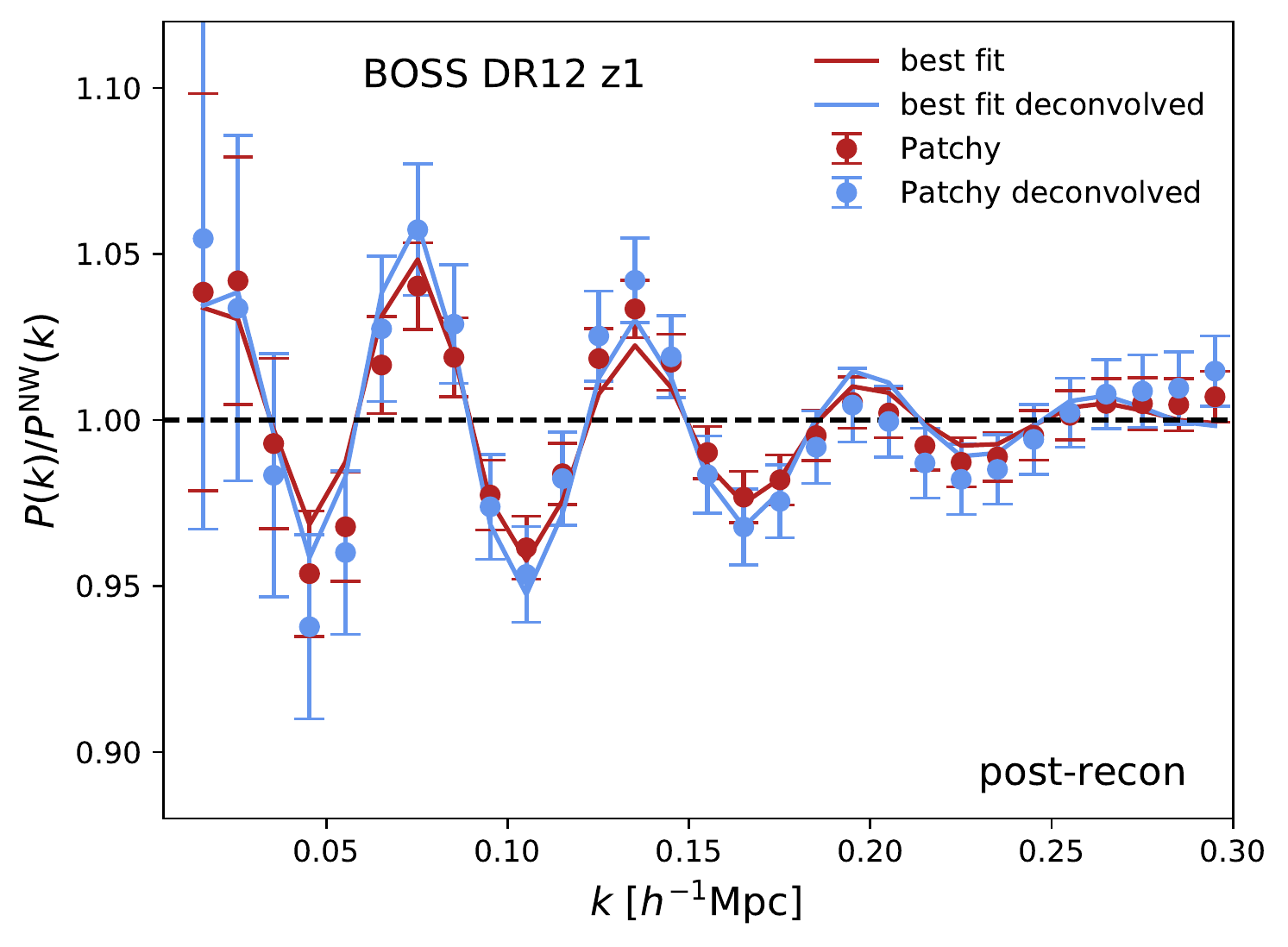}
    \includegraphics[width=0.48\textwidth]{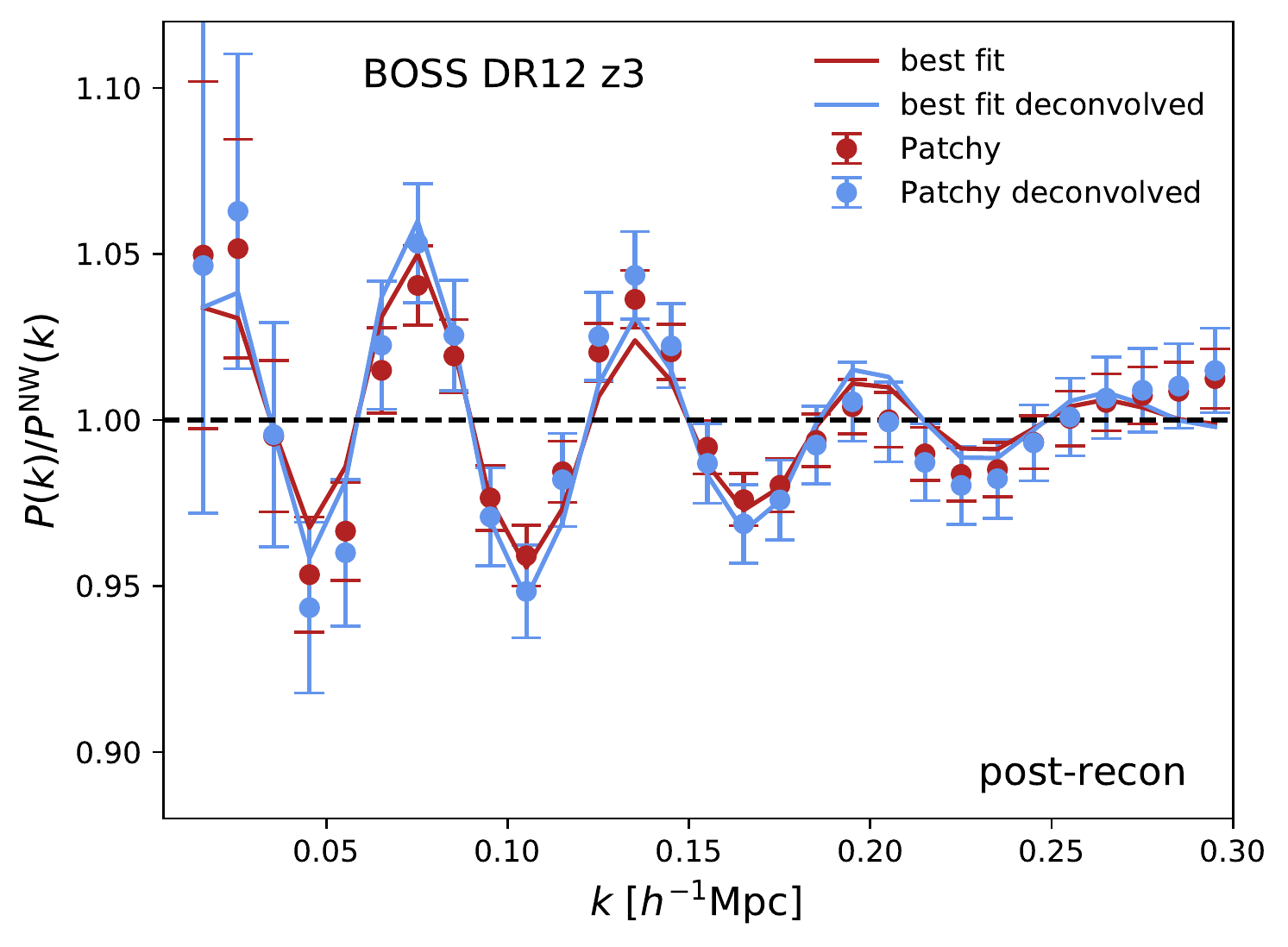}
    \caption{Comparison of the convolved and deconvolved BAO signal in the power spectrum of the low redshift bin (left) and high redshift bin (right) of the  BOSS DR12 MD-Patchy mock catalogs (we are plotting the mean of the $1000$ mock catalogs together with the variance). The red data points correspond to the convolved power spectrum with the best fitting model shown as the solid red line. The deconvolved measurements are shown by the blue data points together with the best fitting model (blue solid line). The data points are the weighted mean of the NGC and SGC. The errorbars correspond to the square root of the diagonal terms of the covariance matrix.}
    \label{fig:BAO_deconvolved}
\end{figure}

The convolution of the measured power spectrum with the survey window 
function could smear out a signal which exists at the same scale as the 
fundamental mode (or the $k_o$ band width, whichever is larger).
The BAO signal has a wavelength of $\sim 0.06\kMpc$, which is much larger than the fundamental mode of the surveys discussed in this paper (see table~\ref{tab:survey_properties}). 

Nevertheless, deconvolving the power spectrum does increase the BAO signature as clearly visible in~\fig{fig:BAO_deconvolved}. This figure shows the BAO signal in the low and high redshift bins of the BOSS DR12 MD-Patchy mock catalogs together with the best fitting models. The data points represent the mean and variance of $1000$ post-reconstruction MD-Patchy mock catalogs.

It is important to note that even though the BAO signal appears to be 
enhanced post-deconvolution, our deconvolution procedure is based 
on~\eq{eq:likelihood} and hence the convolved and deconvolved analysis 
should lead to the same likelihood and the same model parameters, if the
same cuts and theory binning are used. 
To demonstrate this point we perform an isotropic BAO analysis based on the 
mean of $1000$ BOSS DR12 MD-Patchy mock power spectrum monopoles 
(post-reconstruction). 
For the pre-deconvolution fit we build the monopole model following the 
current standard analysis pipeline described 
in~\cite{Beutler2016:1607.03149v1}. We also provide higher order multipoles 
for the quadrupole and hexadecapole based on the simple Kaiser model. 
We use the window function and wide-angle matrices ($\bold{W}$ and 
$\bold{M}$) and limit the likelihood evaluation to the monopole only 
(see the second example in appendix~\ref{app:userguide}). Our fitting range 
is $0.01 < k < 0.3\kMpc$ and we are 
jointly fitting the NGC and SGC. For the post-deconvolution fit we only fit 
the monopole without any window function or wide-angle matrix, again using 
the model of~\cite{Beutler2016:1607.03149v1}. Our fitting procedure is 
using the \code{Python}-based MCMC sampler
\code{zeus}~\cite{Karamanis2020:2002.06212v1,Karamanis2021:2105.03468v1}~
\footnote{https://zeus-mcmc.readthedocs.io/en/latest/}.

In the low redshift bin of BOSS DR12 we find $\alpha = 0.999\pm0.013$ 
before deconvolution and an identical value after deconvolution. 
Figure~\ref{fig:BAO_deconvolved} (left) 
compares the best fitting model and the mean of the MD-Patchy mocks (the plot 
shows the weighted average of the NGC and SGC). 
The equivalent values for the high redshift bin shown in 
\fig{fig:BAO_deconvolved} (right) are $\alpha = 1.003\pm0.012$ before 
deconvolution and $\alpha = 1.002\pm0.012$ after deconvolution. 
Any observed differences are consistent with noise in the MCMC chain, 
the slightly different cuts implied by using the monopole within $0.01<k<0.3 \kMpc$ 
in convolved vs. deconvolved space, and 
coarsening of the theory side of the window function used in the 
deconvolution. The window matrix used for
deconvolution has to be a square matrix with relatively large 
k-bins ($\Delta k = 0.01\kMpc$ in our case). Such large k-bins cannot 
capture small-scale features in the theory power spectrum (see 
\fig{fig:binwindow}). While these effects should not introduce any issues within a smooth $\Lambda$CDM power spectrum (as demonstrated here for the BAO case), it could become relevant for non-$\Lambda$CDM models especially if small scale features are present.

\section{Conclusion}
\label{sec:conclusion}

When analysing galaxy redshift surveys in Fourier space, one needs to account for the survey window function as well as wide-angle effects. In this paper we leverage recent new developments dealing with the survey window function and wide-angle effects, to lay out a simple power spectrum analysis framework based on matrix multiplications. The main results of this paper are:
\begin{itemize}
    \item[(1)] We derive a matrix to account for wide-angle effects in the power spectrum multipoles. We use a new analytic approach rather than the commonly used Hankel transforms.
    \item[(2)] We expand the window function matrix approach presented in~\citep{D'Amico2019:1909.05271v1} by including wide-angle effects.
    \item[(3)] We use this matrix-based analysis framework for the power spectrum multipoles to demonstrate two possible analysis pipelines, one using the standard path of convolving the model vector and one based on the deconvolution of the data vector.
    \item[(4)] We apply the deconvolution procedure to a set of existing galaxy redshift surveys, namely 6dFGS DR3, BOSS DR12 and eBOSS DR16. Using a BAO analysis we demonstrate that our deconvolution analysis framework leads to the same likelihood as the standard analysis. 
    \item[(5)] We provide the power spectrum multipoles as well as the window function matrices, wide-angle matrices and covariance matrices for 6dFGS DR3, BOSS DR12 and eBOSS DR16. In the appendix we also provide \code{Python}-based examples and a general user guide for a clustering analysis. These easy to use components hopefully simplify the analysis of these datasets and make them more accessible for the wider cosmology community.
\end{itemize}
The deconvolution framework outlined in this paper does not suffer from limitations inherent to other methods presented in the literature, such as the assumption of a global plane parallel approximation. Nevertheless, the inversion of the window function does require a square window matrix, which enforces large $\Delta k_{\rm th}$ bins, a limitation which is not present when convolving the model vector.

Our analysis focuses on the key science targets of galaxy redshift surveys such as RSD and BAO. Other observables such as primordial non-Gaussianity through the scale-dependent bias in the power spectrum, naturally requires to focus on the largest scales of the survey. The products provided with this paper are sub-optimal for such an observable. However, the formalism presented in this paper can easily be adapted to suit such an observable.

A similar matrix-based analysis approach could also be developed for higher order statistics, like the bispectrum. The extension of our analysis framework to higher order statistics will be addressed in future work.

\acknowledgments

The authors would like to thank Antonio Cuesta for help with the reconstructed BOSS catalogs and Richard Neveux, Arnaud De-Mattia and Hector Gil-Marin for helpful discussions regarding the window function normalisation. This project has received funding from the European Research Council (ERC) under the European Union's Horizon 2020 research and innovation programme (grant agreement 853291). FB is a Royal Society University Research Fellow.
PM was supported by the U.S. Department of Energy, Office of Science,
Office of High Energy Physics, under Contract no. DE-AC02-05CH11231.

This work has benefited from a variety of \code{Python}
packages including \code{numpy}~\cite{Walt2011:1102.1523v1}, \code{scipy}~\cite{Virtanen2019:1907.10121v1}, \code{matplotlib}~\citep{4160265}, \code{zeus}~\cite{Karamanis2020:2002.06212v1,Karamanis2021:2105.03468v1} and \code{hankl}~\citep{Karamanis2021}.

\bibliographystyle{JHEP}
\bibliography{deconvolve}{}

\newpage

\appendix

\section{User guide}
\label{app:userguide}

Together with this publication we provide:
\begin{enumerate}
    \item The window function $\textbf{W}$, in the form of $5N_{\rm o}\times 5N_{\rm th} = 200\times 2000$ and $5N_{\rm o}\times 5N_{\rm th} = 200\times 200$) matrices.
    \item The wide-angle transformation matrix $\textbf{M}$ ($5N^{\rm flat\text{-}sky}_{\rm th}\times 3N_{\rm th} = 2000\times 1200$).
    \item The covariance matrix $\textbf{C}_{\rm conv}$ derived from the mock catalogs ($5N_{\rm o}\times 5N_{\rm o} = 200\times 200$).
    \item The power spectrum multipole measurements for the mock catalogs and data in $\Delta k_{\rm o}=0.001\kMpc$ (note that to use them with the products above you need to re-bin to $\Delta k_{\rm o}=0.01\kMpc$, see \code{combine\_bins} parameter in \code{pk\_tools.read\_power()}). For BOSS DR12 we also provide the post-reconstruction power spectrum measurements.
\end{enumerate} 
All quantities assume $5$ multipoles in the $k$-range $0 < k < 0.4\kMpc$ in bins of $\Delta k_{\rm o} = 0.01\kMpc$ and $\Delta k_{\rm th} = 0.001\kMpc$. The only exception is the square window function ($200 \times 200$) meant to be used for deconvolution, which assumes $\Delta k_{\rm th} = \Delta k_{\rm o} = 0.01\kMpc$. All products listed above are available at \url{https://fbeutler.github.io/hub/deconv_paper.html} and \code{Python}-based modules to read these quantities are available at \url{https://github.com/fbeutler/pk_tools}.

We note that due to the Nyquist frequency in the measured power spectra, any analysis should be limited to $k_{\rm max} < 0.3\kMpc$. The purpose for the addition of the $k$-range $0.3 < k < 0.4\kMpc$ is mainly to allow a sensible window function contribution from outside the fitting range. We generally advice to exclude the first $k$-bin ($k_{\rm min} > 0.01\kMpc$), which could suffer from large scale systematics not studied in detail in this analysis. 

Here is an example of a likelihood analysis using \code{Python}:
\begin{python}
import numpy as np 
import pk_tools # see above
# Read data power spectrum (Delta k_o=0.01h/Mpc) as dictionary
pk_data_dict = pk_tools.read_power(pkfile, combine_bins=10)
# Turn dictionary into vector as [P_0,P_1,P_2,P_3,P_4]
kbins, pk_data_vector = pk_tools.dict_to_vec(pk_data_dict)
# Read covariance matrix
C = pk_tools.read_matrix(covfile)
Cinv = np.linalg.inv(C)
# Read window matrix (section 2, eq.~2.14)
W = pk_tools.read_matrix(Wfile)
# Read expansion matrix (section 3, eq.~3.2)
M = pk_tools.read_matrix(Mfile)
# Get your favourite power spectrum model as dictionary
pk_model_dict = your_favourite_model()
# Turn dictionary into vector as [P_0,P_2,P_4]
kbins_long, pk_model_vector = pk_tools.dict_to_vec(pk_model_dict, use_ell=[0,2,4])
# Expand pk model (true,flat-sky) -> (true) see eq.~3.1
expanded_model = np.matmul(M, pk_model_vector)
# Convolve with window (true) -> (conv) see eq.~2.18
convolved_model = np.matmul(W, expanded_model)
# Calculate chi2
diff = pk_data_vector - convolved_model
chi2 = np.dot(diff,np.dot(Cinv,diff))
\end{python}
Note that you can speed up these calculations by multiplying the matrices $\textbf{W}$ and $\textbf{M}$ beforehand, which reduces the required matrix multiplications in the likelihood evaluation to one.

\paragraph{How can I limit the k-range?} 
The power spectrum model should always have a k-range of $0 < k_{\rm th} < 0.4\kMpc$ in bins of $\Delta k_{\rm th} = 0.001\kMpc$. If your model cannot predict the power spectrum up to $k_{\rm max}=0.4\kMpc$, you should provide some ``sensible'' estimate, so that the window function contributions from those scales can be included. The likelihood analysis should be limit to $k_{\rm max} < 0.3\kMpc$, since the power spectra outside this k-range suffer from aliasing given that $k_{\rm Ny} \approx 0.6\kMpc$ for all measured power spectra. 
One can specify the k-range as
\begin{python}
kmin = 0.01
kmax = 0.3
# Assuming 40 bins in k_o with Delta k_o=0.01h/Mpc
krange = np.linspace(0.005, 0.395, num=40)
fit_selection = np.logical_and(kmin<krange,krange<kmax)
# Select fitting-range for data power spectrum and inverse C
fit_pk_data_vector = pk_data_vector[fit_selection]
fit_Cinv = np.linalg.inv(C[np.ix_(fit_selection, fit_selection)])
# Select fitting-range for the convolved model power spectrum
fit_model = convolved_model[fit_selection]
# ... and proceed to calculate $\chi^2$
\end{python}

\paragraph{How can I limit my analysis to the monopole?}
Since the window function couples the different multipoles you have to provide a model for all multipoles, so that the window function contributions can be calculated. For the likelihood itself one can limit the analysis to the monopole by 
\begin{python}
# Select multipoles to be included [P_0, P_1, P_2, P_3, P_4]
pole_selection = [True, False, False, False, False]
fit_selection = np.repeat(pole_selection, 40)
# Adjust data power spectrum and inverse C
fit_pk_data_vector = pk_data_vector[fit_selection]
fit_Cinv = np.linalg.inv(C[np.ix_(fit_selection, fit_selection)])
# Adjust convolved model power spectrum
fit_model = convolved_model[fit_selection]
# ... and proceed to calculate $\chi^2$
\end{python}

\paragraph{Speeding up the likelihood evaluation:} 
Often it is possible to achieve a speedup in the likelihood analysis if the convolution of the power spectrum model does not have to be performed in every model-data comparison. We can re-write the likelihood 
as~\citep{D'Amico2019:1909.05271v1}
\begin{align}
    \mathcal{L}&\sim \exp\left[-\frac{1}{2}(\Pconvo-\bold{W}\bold{M}\Ptruefs)^T\invCconv(\Pconvo-\bold{W}\bold{M}\Ptruefs)\right]\\
    \begin{split}
        &= \exp\left[-\frac{1}{2}\PconvoT\invCconv\Pconvo + \PtruefsT\bold{M}^{\rm T}\bold{W}^{\rm T}\invCconv\Pconvo\right. \\
        &\;\;\;\;\;\;\;\;\;\;\;\; \left. - \frac{1}{2}\PtruefsT\bold{M}^{\rm T}\bold{W}^{\rm T}\invCconv\bold{W}\bold{M}\Ptruefs\right]
    \end{split}\\
    \begin{split}
        &= \exp\left[
-\frac{1}{2}\PconvoT
               \invCconv\Pconvo + 
              \Ptruefs
               \bold{C}^{-1}_{\rm conv, W}\Pconvo\right.  \\
        &\;\;\;\;\;\;\;\;\;\;\;\; \left. - 
        \frac{1}{2}\PtruefsT
        \bold{C}^{-1}_{\rm conv, WW}\Ptruefs
\right]     
    \end{split}
\end{align}
with 
\begin{align}
    \bold{C}^{-1}_{\rm conv, WW} &= \bold{M}^{\rm T}\bold{W}^{\rm T}\invCconv\bold{W}\bold{M}\\
    \bold{C}^{-1}_{\rm conv, W} &= \bold{M}^{\rm T}\bold{W}^{\rm T}\invCconv\, .
\end{align}
An implementation of these equations could look like this:
\begin{python}
# Select multipoles to be included [P_0, P_1, P_2, P_3, P_4]
pole_selection = [True, False, False, False, False]
fit_selection = np.repeat(pole_selection, 40)
# Select the fitting range
Cinv = np.linalg.inv(C[np.ix_(fit_selection, fit_selection)])
pk_data_vector = pk_data_vector[fit_selection]
W = W[np.ix_(fit_selection, np.ones(2000) < 1.1)]
# Pre-calculate the two covariance matrices and the 
# data-data contribution
WCinv = np.matmul(np.transpose(W), Cinv)
MWCinv = np.matmul(np.transpose(M), WCinv)
MWCinvD = np.matmul(MWCinv, pk_data_vector)
MWCinvW = np.matmul(MWCinv, W)
MWCinvWM = np.matmul(MWCinvW, M)
data_term = np.dot(pk_data_vector, np.dot(Cinv, pk_data_vector))

## The following part needs to be run for every parameter evaluation 
# Get your favourite power spectrum model as disctionary
pk_model_dict = your_favourite_model()
# Turn dictionary into vector as [P_0,P_2,P_4]
kbins_long, pk_model_vector = pk_tools.dict_to_vec(pk_model_dict, use_ell=[0,2,4])
# Calculate chi2
chi2 = data_term
chi2 -= 2.*np.dot(pk_model_vector, MWCinvD) 
chi2 += np.dot(pk_model_vector, np.dot(MWCinvWM, pk_model_vector))
\end{python}
This implementation is about a factor of two times faster than the brute-force implementation shown in the first code example. Of course that is only significant if the window function convolution is dominating the likelihood evaluation. In many cases most of the time will be spend in \code{your\_favourite\_model()}.

\paragraph{How can I account for uncertainties in the covariance matrix?} When deriving the covariance matrix from a finite set of mock realisations, the resulting likelihood is no longer Gaussian, but follows a t-distribution. If assuming a Gaussian likelihood the parameter inference will be biased and this bias depends on the ratio of bins in the data vector and the number of mock realisations~\citep{Hartlap2006:astro-ph/0608064v2}. We can account for this by scaling the likelihood as
\begin{equation}
    \ln\mathcal{L} \propto -\chi^2\frac{N_{m}-N_{d}-2}{2(N_{m}-1)}\, ,
    \label{eq:hartlap}
\end{equation}
where $N_{m}$ is the number of mock realisations 
(given in table~\ref{tab:survey_properties}) and $N_{d}$ is the size of the 
data vector. Alternatively one can directly  account for the non-Gaussian 
likelihood as proposed in \citep{Sellentin2015:1511.05969v2}. For all cases 
discussed in this paper this approach agrees very well with \eq{eq:hartlap}.

Assuming your MCMC sampler expects $\log\mathcal{L}$ as a return value you could implement this equation as
\begin{python}
# Hartlap et al. (2007)
H = Nmocks - len(pk_data_vector) - 2
H /= (Nmocks - 1)
return -chi2*H/2.
\end{python}
If the final parameter uncertainty is derived from the likelihood itself, you need to account for a bias caused by the mock based covariance estimate. We can do that with a re-scaling of the parameter errors~\citep{Dodelson2013:1304.2593v2,Percival2013:1312.4841v1} by the square root of
\begin{equation}
    m_1 = \frac{1 + B(N_d - N_p)}{1 + A + B(N_p+1)}
\end{equation}
with 
\begin{align}
    A &= \frac{2}{(N_m - N_d - 1)(N_m-N_d-4)}\, ,\\
    B &= \frac{N_m - N_d - 2}{(N_m - N_d - 1)(N_m-N_d-4)}\, 
\end{align}
(we find that the approach of \citep{Sellentin2015:1511.05969v2}
does not take care of this factor). 

\paragraph{How can I deconvolve a power spectrum measurement?} To perform a deconvolution, all you have to do is to follow \eq{eq:deconv}. The difficulty here is that this equation only holds for square matrices $\bold{W}$. The window functions using $\Delta k_{\rm o} = \Delta k_{\rm th} = 0.01\kMpc$ are available.

\begin{python}
# Read data power spectrum (Delta k_o=0.01h/Mpc) as disctionary
pk_data_dict = pk_tools.read_power(pkfile, combine_bins=10)
# Turn dictionary into vector as [P_0,P_1,P_2,P_3,P_4]
kbins, pk_data_vector = pk_tools.dict_to_vec(pk_data_dict)
# Read window matrix (needs to be a square matrix)
W = pk_tools.read_matrix(Wfile)
# Invert window function 
Winv = np.linalg.inv(W)
# Deconvolution 
pk_data_vector_deconvolved = np.matmul(Winv, pk_data_vector)
\end{python}

\section{Window function pre-factors}
\label{app:Cls}

Focusing on the first $5$ multipoles ($0 \leq L \leq 4$) of the window function, including wide-angle terms up to second order ($n<3$), the weights for the individual contributions in \eq{eq:2Dwindow} are given by
\begin{equation}
    \begin{split}
        C^{(0)}_{0\ell L} = C^{(2)}_{0\ell L} &= \left(\begin{matrix}
        1 & 0 & 0 & 0 & 0\\
        0 & 0 & 0 & 0 & 0\\
        0 & 0 & \frac{1}{5} & 0 & 0\\
        0 & 0 & 0 & 0 & 0\\
        0 & 0 & 0 & 0 & \frac{1}{9}
        \end{matrix}\right)\;\;\;\;\;\;\;\;\;\;
        C^{(1)}_{0\ell L} = \left(\begin{matrix}
        0 & 0 & 0 & 0 & 0\\
        0 & \frac{1}{3} & 0 & 0 & 0\\
        0 & 0 & 0 & 0 & 0\\
        0 & 0 & 0 & \frac{1}{7} & 0\\
        0 & 0 & 0 & 0 & 0
        \end{matrix}\right)\\
        C^{(0)}_{1\ell L} = C^{(2)}_{1\ell L} &= \left(\begin{matrix}
        0 & 1 & 0 & 0 & 0\\
        0 & 0 & 0 & 0 & 0\\
        0 & \frac{2}{5} & 0 & \frac{9}{35} & 0\\
        0 & 0 & 0 & 0 & 0\\
        0 & 0 & 0 & \frac{4}{21} & 0
        \end{matrix}\right)\;\;\;\;\;\;\;\;
        C^{(1)}_{1\ell L} = \left(\begin{matrix}
        0 & 0 & 0 & 0 & 0\\
        1 & 0 & \frac{2}{5} & 0 & 0\\
        0 & 0 & 0 & 0 & 0\\
        0 & 0 & \frac{9}{35} & 0 & \frac{4}{21}\\
        0 & 0 & 0 & 0 & 0
        \end{matrix}\right)\\
        C^{(0)}_{2\ell L} = C^{(2)}_{2\ell L} &= \left(\begin{matrix}
        0 & 0 & 1 & 0 & 0\\
        0 & 0 & 0 & 0 & 0\\
        1 & 0 & \frac{2}{7} & 0 & \frac{2}{7}\\ 
        0 & 0 & 0 & 0 & 0\\
        0 & 0 & \frac{2}{7} & 0 & \frac{100}{693}\end{matrix}\right)\;\;\;\;\;\;\;
        C^{(1)}_{2\ell L} = \left(\begin{matrix}
        0 & 0 & 0 & 0 & 0\\
        0 & \frac{2}{3} & 0 & \frac{3}{7} & 0\\
        0 & 0 & 0 & 0 & 0\\ 
        0 & \frac{3}{7} & 0 & \frac{4}{21} & 0\\
        0 & 0 & 0 & 0 & 0 \end{matrix}\right)
    \end{split}
\end{equation}
and 
\begin{equation}
    \begin{split}
        C^{(0)}_{3\ell L} = C^{(2)}_{3\ell L} &= \left(\begin{matrix}
        0 & 0 & 0 & 1 & 0\\
        0 & 0 & 0 & 0 & 0\\
        0 & \frac{3}{5} & 0 & \frac{4}{15} & 0\\
        0 & 0 & 0 & 0 & 0\\
        0 & \frac{4}{9} & 0 & \frac{2}{11} & 0\end{matrix}\right)\;\;\;\;\;\;\;\;
        C^{(1)}_{3\ell L} = \left(\begin{matrix}
        0 & 0 & 0 & 0 & 0\\
        0 & 0 & \frac{3}{5} & 0 & \frac{4}{9}\\
        0 & 0 & 0 & 0 & 0\\
        1 & 0 & \frac{4}{15} & 0 & \frac{2}{11}\\
        0 & 0 & 0 & 0 & 0 \end{matrix}\right)\\
        C^{(0)}_{4\ell L} = C^{(2)}_{4\ell L} &= \left(\begin{matrix}
        0 & 0 & 0 & 0 & 1\\
        0 & 0 & 0 & 0 & 0\\
        0 & 0 & \frac{18}{35} & 0 & \frac{20}{77}\\ 
        0 & 0 & 0 & 0 & 0\\
        1 & 0 & \frac{20}{77} & 0 & \frac{162}{1001}\end{matrix}\right)\;\;\;\;\;
        C^{(1)}_{4\ell L} = \left(\begin{matrix}
        0 & 0 & 0 & 0 & 0\\
        0 & 0 & 0 & \frac{4}{7} & 0\\
        0 & 0 & 0 & 0 & 0\\ 
        0 & \frac{4}{7} & 0 & \frac{18}{77} & 0\\
        0 & 0 & 0 & 0 & 0 \end{matrix}\right),
    \end{split}
\end{equation}
where the wide-angle correction terms at $n=0$ and $n=2$ have the same shape.

\section{Derivation of the 2D window function} \label{app:deriveW}

In this section we derive the equation for the 2D window function $W^{(n)}_{\ell\ell}(k,k')$. We first derive this equation for $n=0$ showing consistency with~\citep{Beutler2013:1312.4611v2}, where this equation first appeared. We than include the wide-angle correction terms following~\citep{Beutler2018:1810.05051v3}, which leads to our~\eq{eq:2Dwindow}.

\subsection{Excluding wide-angle terms and consistency with ~\citep{Beutler2013:1312.4611v2}}

Here we show the relation between eq.~(33) of~\citep{Beutler2013:1312.4611v2} and \eq{eq:2Dwindow_nowa}. The convolution of the power spectrum can be written as (using a LOS of $\vbunit{d}=\vbunit{s}_1$ and following eq.~B.1 of~\citep{Beutler2013:1312.4611v2})
\begin{align}
    P_{\ell}^{\rm conv}(k) &= \frac{2\ell + 1}{2}\int d\mu\int\frac{d\phi}{2\pi}\int \frac{d\vb{k}'}{(2\pi)^3}\,P^{\rm true}(\vb{k}')|W(\vb{k} - \vb{k}')|^2\mathcal{L}_{\ell}(\vbunit{k}'\cdot\vbunit{s}_1)\\
    \begin{split}
        &=\frac{2\ell + 1}{2}\int d\mu\int\frac{d\phi}{2\pi}\int d\mu'\int \frac{d\phi'}{4\pi}\int \frac{dk'\,k'^2}{2\pi^2} P^{\rm true}(k',\mu')\times\\
        &\;\;\;\;\sum^{N_{\rm ran}}_{i,j,i=j}w_{\rm FKP}(x_i)w_{\rm FKP}(x_j)e^{i\vb{k}\cdot\vb{s}}e^{-i\vb{k}'\cdot\vb{s}}\mathcal{L}_{\ell}(\vbunit{k}\cdot\vbunit{s}_1)\, .
    \end{split}
\end{align}
Using multipole expansion as well as 
\begin{align}
    e^{iks\mu} &= \sum_Li^L(2L+1)j_L(ks)\mathcal{L}_L(\mu)\, ,\\
    \mathcal{L}_{\ell}(\vbunit{s}_1\cdot\vbunit{s})\delta_{\ell\ell'} &= \frac{2\ell+1}{2}\int d\mu\int\frac{d\phi}{2\pi}\mathcal{L}_{\ell}(\vbunit{k}\cdot\vbunit{s})\mathcal{L}_{\ell'}(\vbunit{k}\cdot\vbunit{s}_1)\, ,
\end{align}
we get~\footnote{Note that our Fourier transform convention is
$P(\vk) \equiv \int d\vx~e^{-i \vk\cdot \vx} \xi(\vx)$.}
\begin{equation}
    \begin{split}
        P_{\ell}^{\rm conv}(k) &= (-i)^Li^{\ell}(2\ell+1)\int \frac{dk'\, k'^2}{2\pi^2} \sum_LP^{\rm true}_L(k')j_L(k's)j_{\ell}(ks)\times\\
        &\;\;\;\;\sum^{N_{\rm ran}}_{i,j,i=j}w_{\rm FKP}(x_i)w_{\rm FKP}(x_j)\mathcal{L}_{\ell}(\vbunit{s}_1\cdot\vbunit{s})\mathcal{L}_{L}(\vbunit{s}_1\cdot\vbunit{s})\, .
    \end{split}
\end{equation}
Now using 
\begin{equation}
    \begin{split}
        \sum^{N_{\rm ran}}_{i,j,i=j}w_{\rm FKP}(x_i)w_{\rm FKP}(x_j) &= \int d^3s_1\int d^3s_2W(\vb{s_1})W(\vb{s_2})\\
        &= \int ds\;s^2\int d\Omega_s\int d^3s_1W(\vb{s}_1)W(\vb{s}+\vb{s}_1)
    \end{split}
\end{equation}
and the definition (see eq.~2.21 of \citep{Beutler2018:1810.05051v3})
\begin{equation}
    Q_L(s) = (2L+1)\int \frac{d\Omega_s}{4\pi}\int d^3s_1 W(\vb{s}_1)W(\vb{s}+\vb{s}_1)\mathcal{L}_{L}(\vbunit{s}\cdot\vbunit{s}_1)
\end{equation}
as well as 
\begin{align}
    \mathcal{L}_{\ell}(\vbunit{s}_1\cdot\vbunit{s})\mathcal{L}_{L}(\vbunit{s}_1\cdot\vbunit{s}) &= \sum^{\ell+L}_{p=|\ell-L|}(2p+1)\left(\begin{matrix} \ell & L & p\cr 0 & 0 & 0\end{matrix}\right)^2\mathcal{L}_p(\vbunit{s}_1\cdot\vbunit{s})\, ,
\end{align}
results in
\begin{align}
    \begin{split}
        P_{\ell}^{\rm conv}(k) &= 4\pi (-i)^Li^{\ell}\int \frac{dk'\, k'^2}{2\pi^2} \sum_LP^{\rm true}_L(k')\times\\
        &\;\;\;\;\int ds\,s^2j_L(k's)j_{\ell}(ks)(2\ell+1) \sum^{\ell+L}_{p=|\ell-L|}\left(\begin{matrix} \ell & L & p\cr 0 & 0 & 0\end{matrix}\right)^2Q_p(s)
    \end{split}\\
    &= \int dk'\,k'^2 \sum_L W_{\ell L}(k,k')P_{L}^{\rm true}(k')\, ,
\end{align}
where
\begin{equation}
    W_{\ell L}(k,k') = (-i)^Li^{\ell} \frac{2}{\pi}\int ds\,s^2j_L(k's)j_{\ell}(ks)A_{\ell L}(s)
\end{equation}
with
\begin{align}
    A_{\ell L}(s) &= (2\ell+1)\sum^{\ell+L}_{p=|\ell-L|}\left(\begin{matrix} \ell & L & p\cr 0 & 0 & 0\end{matrix}\right)^2Q_p(s)\label{eq:Cldef}\\
    &=\sum^{\ell+L}_{p=|\ell-L|} C_{\ell L p}Q_p(s)
\end{align}
and the factors $C_{\ell L p}=C^{(0)}_{\ell L p}$ are given in appendix~\ref{app:Cls}.

\subsection{Including wide-angle terms and consistency with ~\citep{Beutler2018:1810.05051v3}}
\label{app:W_with_wa}

The convolution of the power spectrum multipoles including the wide-angle correction terms has first been derived in~\citep{Beutler2018:1810.05051v3} and is given by~\footnote{We added a factor of $1/4\pi$ to account for the difference in the window function definition (see eq.~2.21 in~\citep{Beutler2018:1810.05051v3} and eq.~\ref{eq:window} in this paper.)}
\begin{align}
    P^{\rm conv}_{\ell}(k) &=
    (-i)^{\ell}\frac{(2\ell+1)}{4\pi}\sum_{L}\sum^{\ell+L}_{p=|\ell-L|}\left(\begin{matrix} L & p & \ell\cr 0 & 0 & 0\end{matrix}\right)^2\int ds \sum_n s^{n+2}j_{\ell}(ks)\xi^{(n)}_{L}(s)Q_{p}^{(n)}(s)\, .
\end{align}
Now using 
\begin{equation}
    \xi^{(n)}_{L} = i^{L}\int\frac{dk \,k^2}{2\pi^2}(ks)^{-n}P^{(n)}_{L}(k)j_{L}(ks)
\end{equation}
we get
\begin{align}
    \begin{split}
        P^{\rm conv}_{\ell}(k) &=(-i)^{\ell}\frac{2}{\pi}\int ds\sum_{L}C^{(n)}_{\ell L}(s) \sum_n  s^{n+2}j_{\ell}(ks)\times\\
        &\;\;\;\;\;\;i^{L}\int k'^2dk'(k's)^{-n}P^{(n), \rm true}_{L}(k')j_{L}(k's)
    \end{split}\\
    &=\int dk'\,k'^{2-n} \sum_{L,n}W^{(n)}_{\ell L}(k,k')P^{(n), \rm true}_{L}(k')\, ,
\end{align}
where
\begin{equation}
    W^{(n)}_{\ell L}(k,k') = (-i)^{\ell}i^{L}\frac{2}{\pi}\int ds\, s^2j_{L}(k's)j_{\ell}(ks)A^{(n)}_{\ell L}(s)
\end{equation}
and 
\begin{align}
    A^{(n)}_{\ell L}(s) &= (2\ell+1)\sum^{\ell+L}_{p=|\ell-L|}\left(\begin{matrix} L & p & \ell\cr 0 & 0 & 0\end{matrix}\right)^2Q^{(n)}_{p}(s)\\
    &=\sum^{\ell+L}_{p=|\ell-L|} C^{(n)}_{\ell L p}Q^{(n)}_p(s)\, .
\end{align}
The factors $C^{(n)}_{\ell L p}$ are given in appendix~\ref{app:Cls}. The equation above is consistent with \eq{eq:Cldef} since 
\begin{equation}
    \left(\begin{matrix} \ell_1 & \ell_2 & \ell_3\cr 0 & 0 & 0\end{matrix}\right)^2 = \left(\begin{matrix} \ell_2 & \ell_3 & \ell_1\cr 0 & 0 & 0\end{matrix}\right)^2\, .
\end{equation}

\section{Analytic calculation of the odd power spectrum multipoles}
\label{app:dipole_der}

Here we derive \eq{eq:dipole_analytic} and \eq{eq:octopole_analytic} used in section~\ref{sec:wa}. The dipole power spectrum is given by
\begin{equation}
    P^{(1)}_1(k) = -ik\frac{6}{5\pi}\int dk'k'^2P^{(0)}_2(k')\int s^3ds\,j_2(k's)j_1(ks)\, .
\end{equation}
Following eq.~(F.1), (F.6) and (F.10) of ~\citep{Reimberg2015:1506.06596v2} we can write 
\begin{equation}
     \frac{2}{\pi}\int s^3ds\,j_2(k's)j_1(ks) = -k'\partial_{ k'}\left[\frac{1}{k'^3}\delta_D(k'-k)\right]\, ,
\end{equation}
where $\delta_D$ is the Dirac delta function. Using 
\begin{equation}
    I_{11-1}(k,k') = \frac{1}{k^2}\delta_D(k'-k)
\end{equation}
to replace the integral with a derivative, results in
\begin{align}
    P^{(1)}_1(k)
    &= ik\frac{3}{5}\int dk'k'^3P^{(0)}_2( k')\partial_{ k'}\left[\frac{1}{k'^3}\delta_D(k'-k)\right]\\
    &= ik\frac{3}{5}\int dk'k'^3P^{(0)}_2( k')\left[-\frac{3}{k'^4}\delta_D(k'-k) + \frac{1}{k'^3}\partial_{k'}\delta_D(k'-k)\right]\, .
\end{align}
Now we can use 
\begin{equation}
    \int dx f(x)\partial_{x}\delta_D(x-a) = -\partial_a f(a)\, ,
\end{equation}
yielding
\begin{align}
    P^{(1)}_1(k) &= -i\frac{3}{5}\left[3P^{(0)}_2(k) + k\partial_{k}P^{(0)}_2(k)\right]\\
    &= -if\left(\frac{4}{5}b_1+\frac{12}{35}f\right)\left[3P_m(k) + k\partial_{k}P_m(k)\right]\, ,
\end{align}
where the second line assumes linear theory. 
This final equation is numerically much easier to evaluate compared to the 
equation we started with.

Equivalently we can derive the octopole
\begin{align}
    P^{(1)}_3(k) &= -i\left[\frac{3}{5}\left(2P^{(0)}_2(k) - k\partial_{k}P^{(0)}_2(k)\right) + \frac{10}{9}\left(5P^{(0)}_{4}(k) + k\partial_{k}P^{(0)}_4(k)\right)\right]\\
    &= -i4f\left[\frac{1}{5}\left(b_1 + \frac{3}{7}f\right)\left(2P_m(k) - k\partial_{k}P_m(k)\right) + \frac{4}{63}f\left(5P_m(k) + k\partial_{k}P_m(k)\right)\right]\, .
\end{align}

\section{Deconvolution results for all datasets}
\label{app:deconvolution}

\begin{figure}[h]
    \centering
    \includegraphics[width=0.19\textwidth]{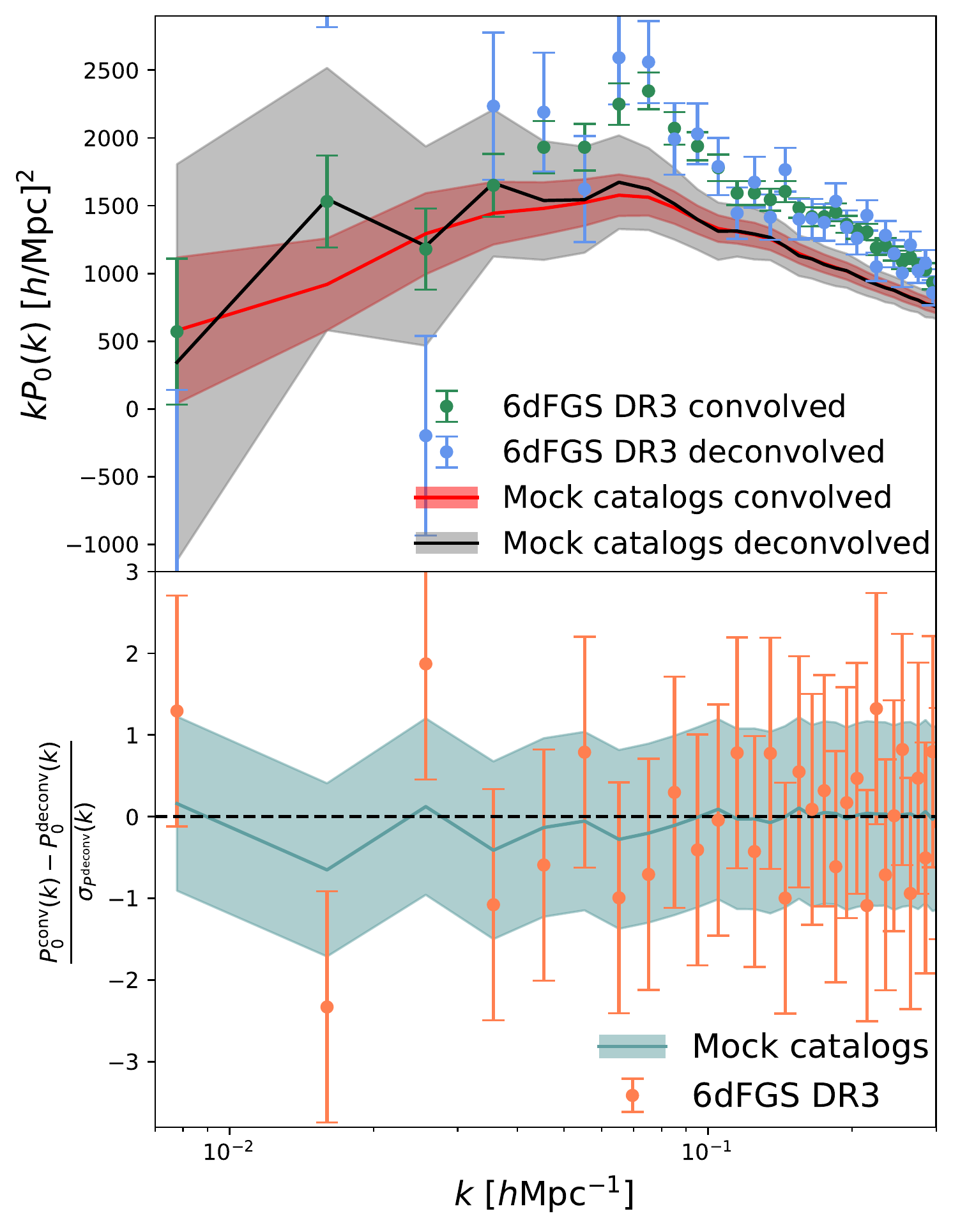}
    \includegraphics[width=0.19\textwidth]{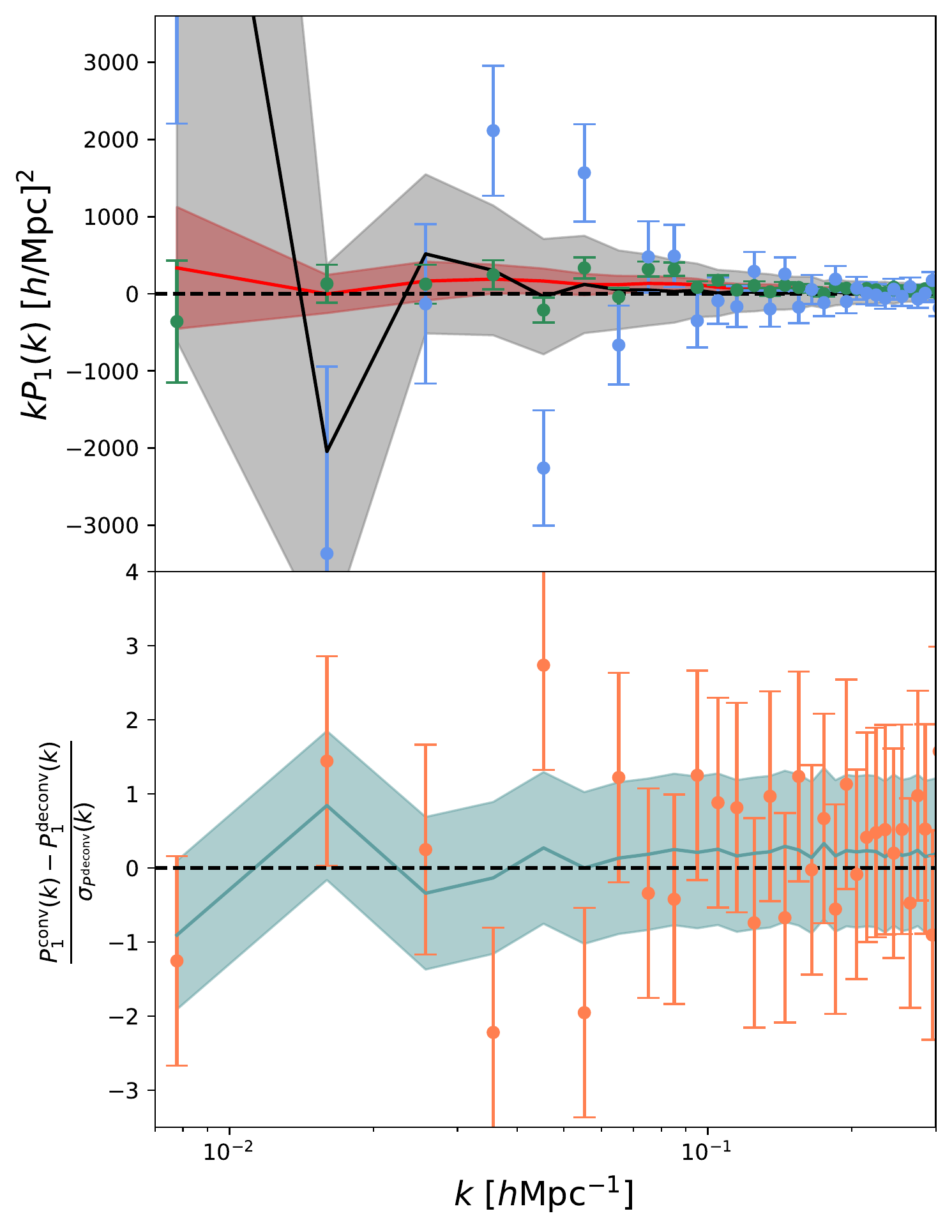}
    \includegraphics[width=0.19\textwidth]{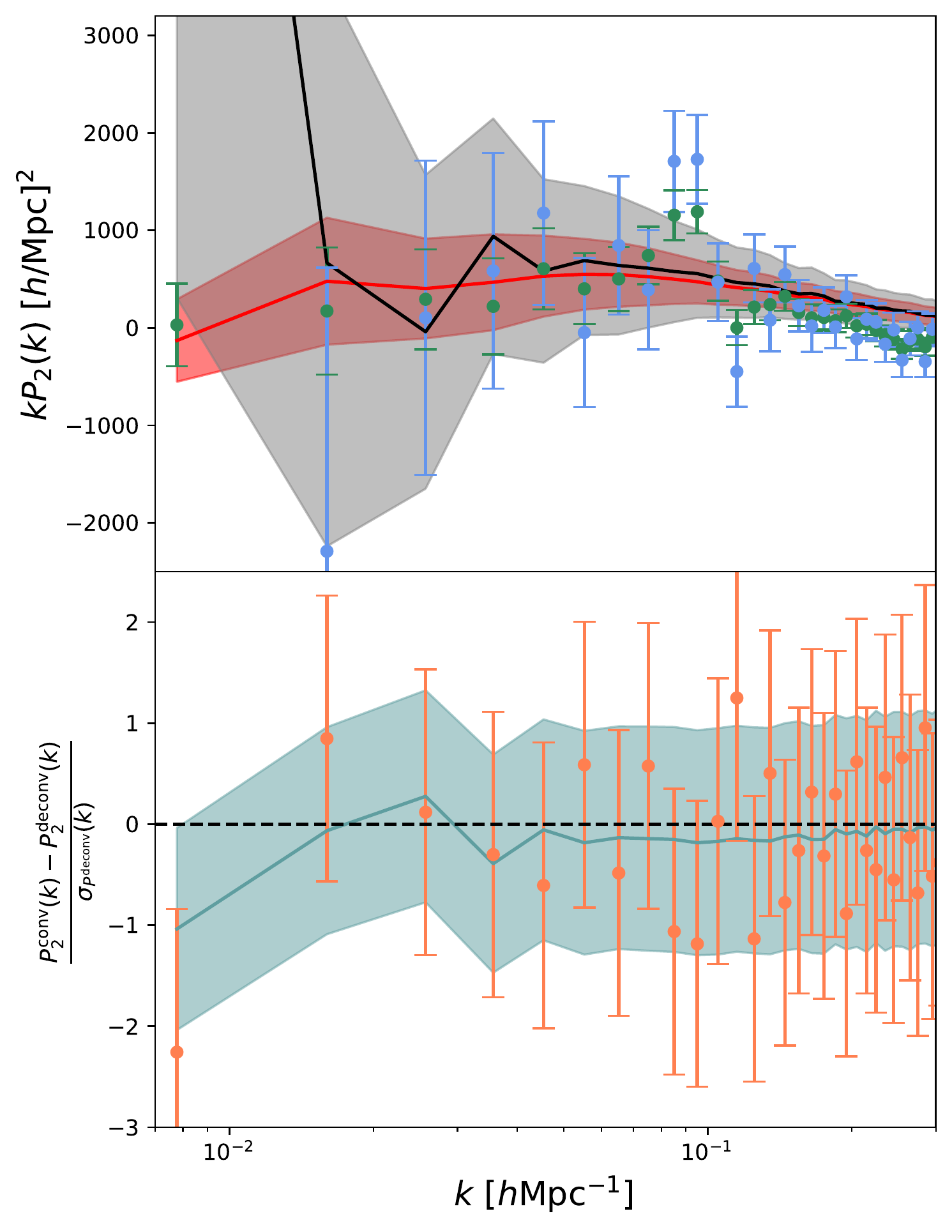}
    \includegraphics[width=0.19\textwidth]{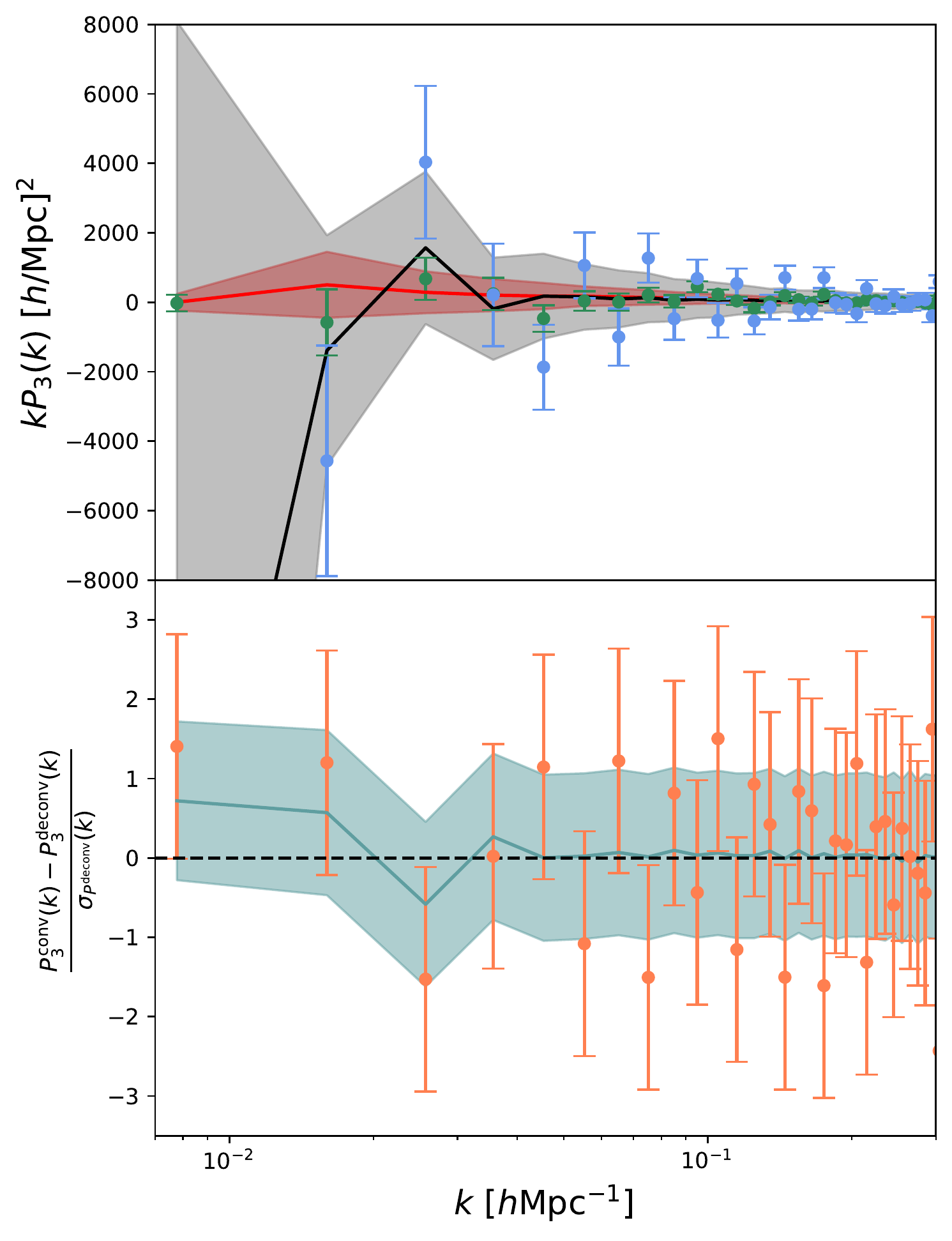}
    \includegraphics[width=0.19\textwidth]{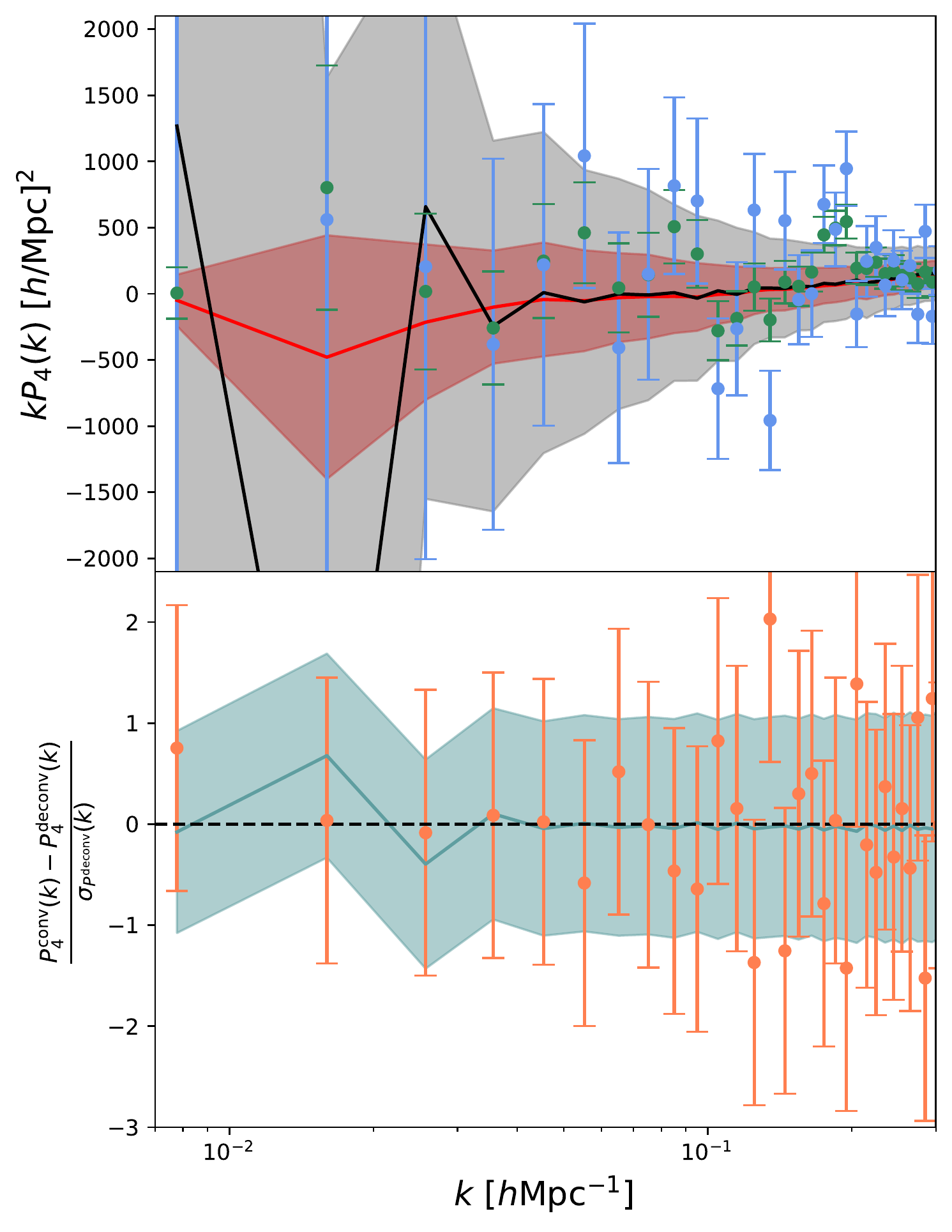}
    \caption{Comparison between the power spectrum multipoles of 6dFGS DR3 measured in the mock catalogs (gray and red shaded area) and in the data (data points). The results before deconvolution are shown as the red shaded area and solid red line (mocks) and the green data points. The deconvolved results are shown as the gray shaded area and solid black line (mocks) and the blue data points. The residuals in the lower panel show that wide-angle and window function effects are sub-dominant in 6dFGS.}
    \label{fig:decon_6dfgs}
\end{figure}

\begin{figure}[h]
    \centering
    \includegraphics[width=0.19\textwidth]{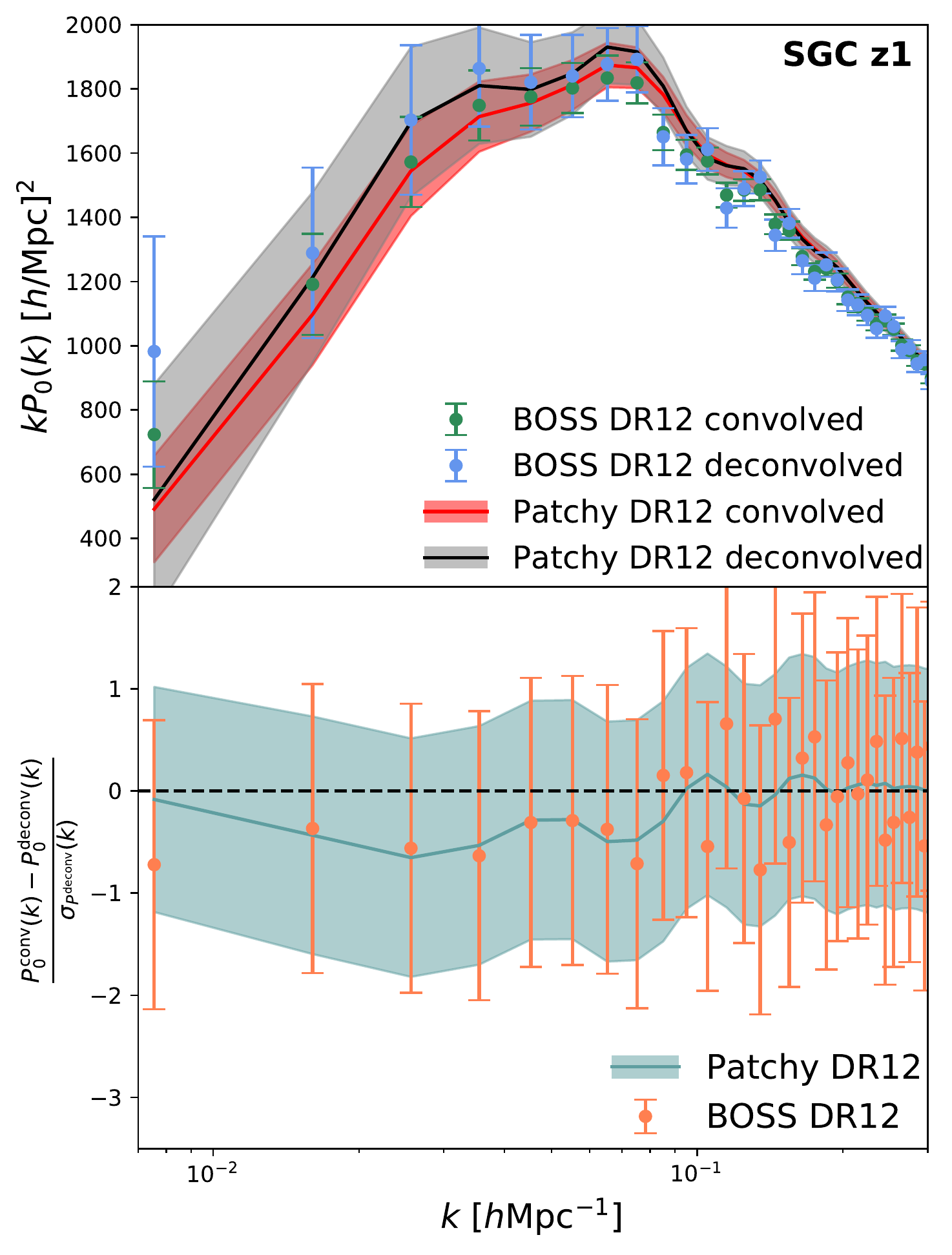}
    \includegraphics[width=0.19\textwidth]{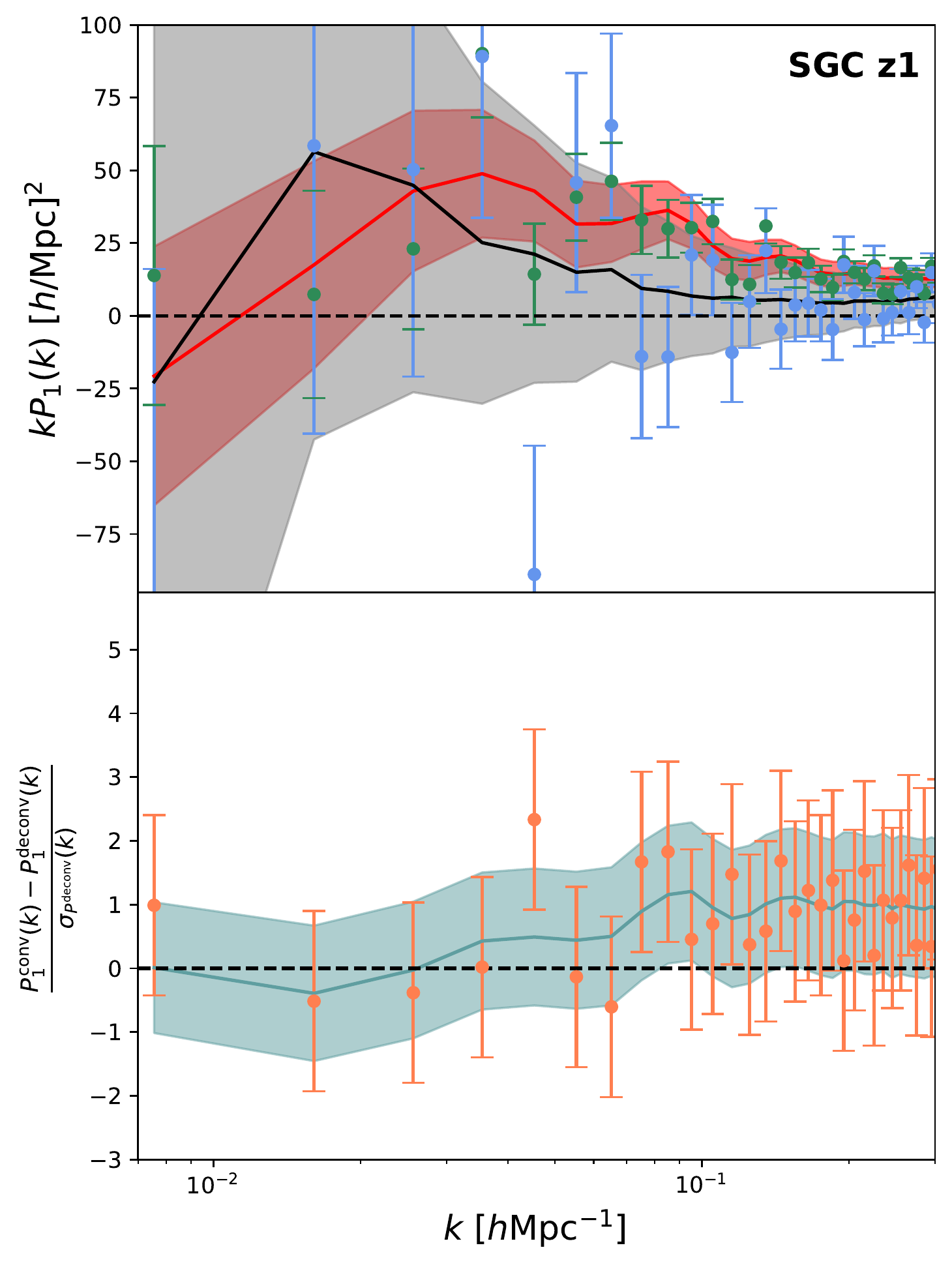}
    \includegraphics[width=0.19\textwidth]{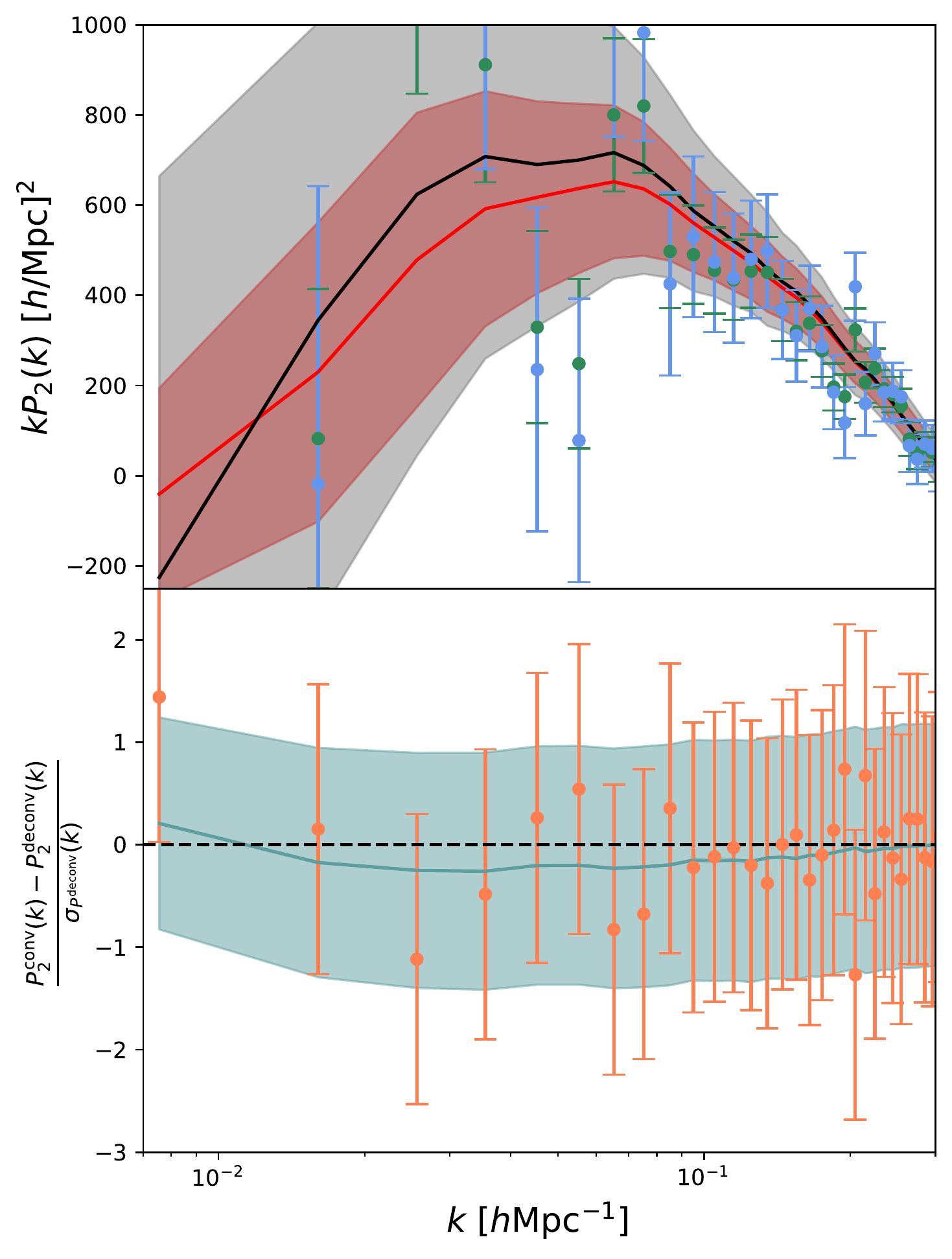}
    \includegraphics[width=0.19\textwidth]{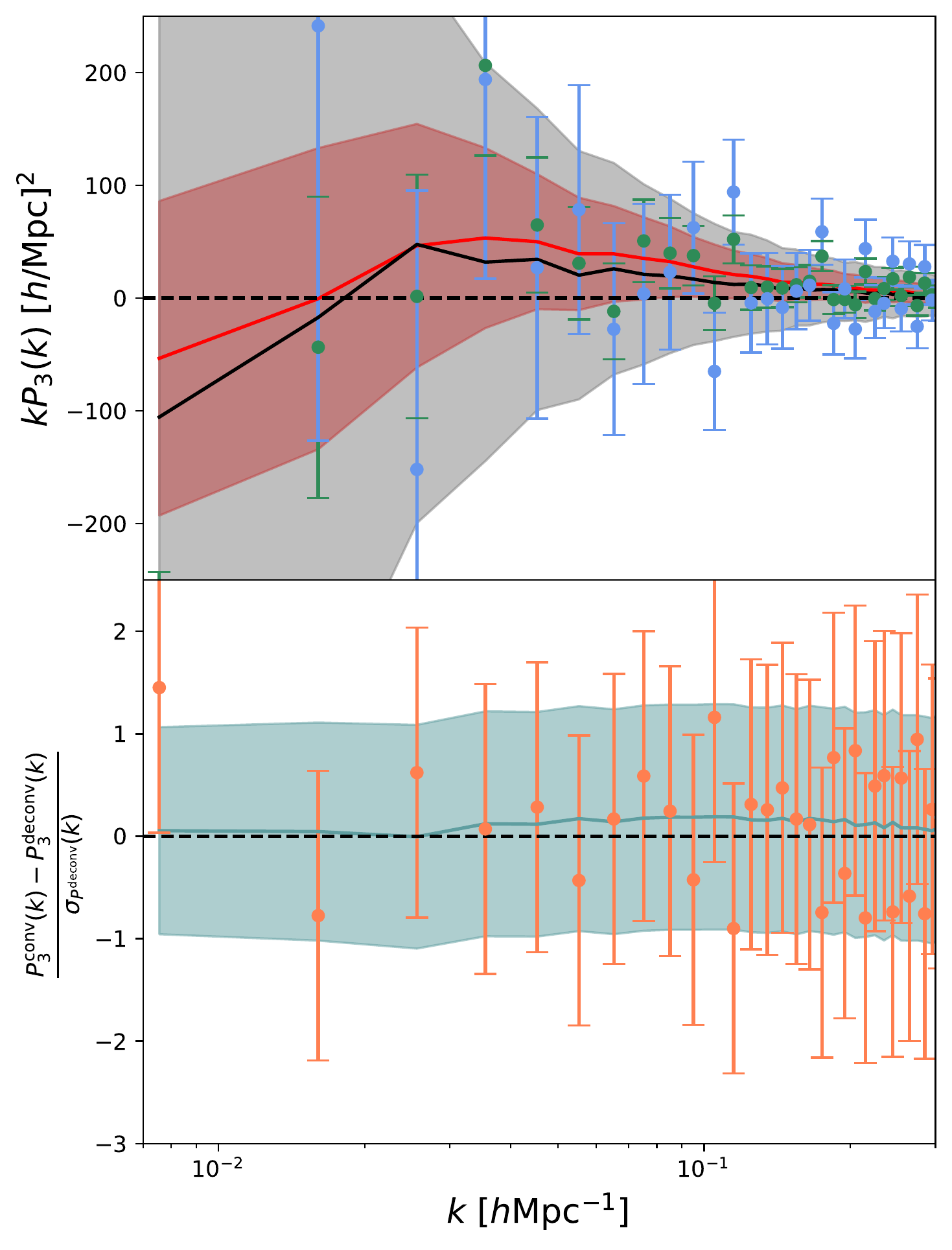}
    \includegraphics[width=0.19\textwidth]{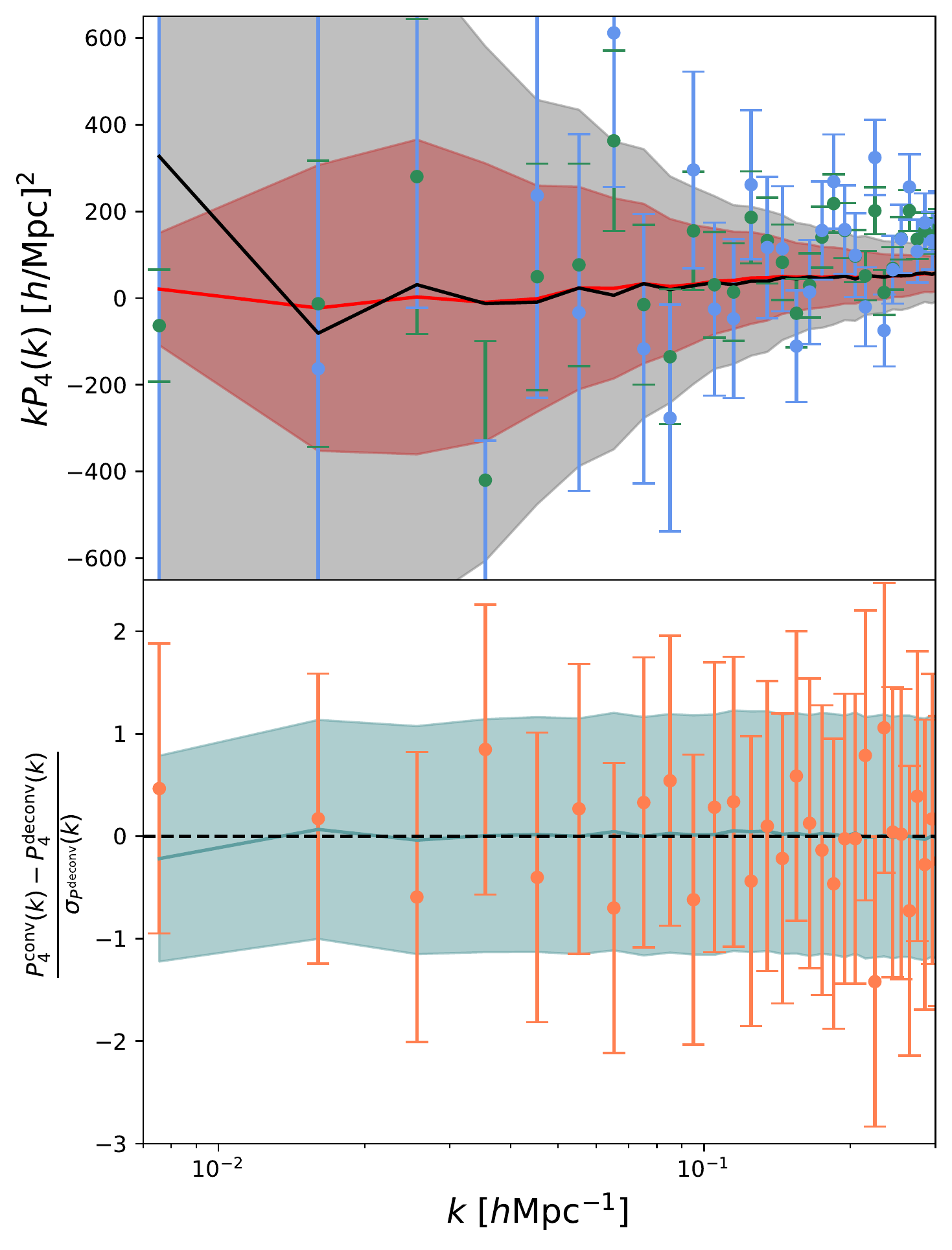}\\
    \includegraphics[width=0.19\textwidth]{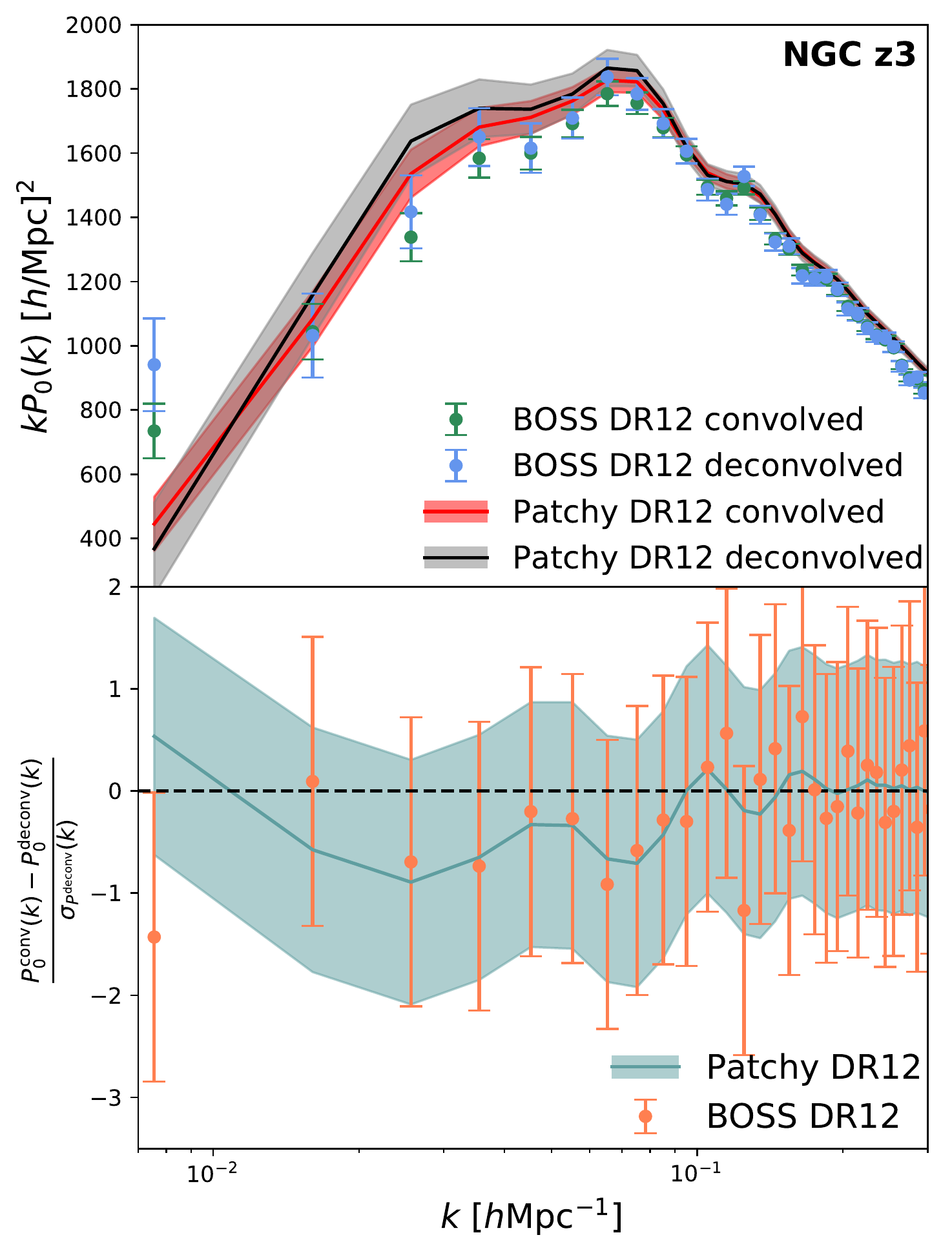}
    \includegraphics[width=0.19\textwidth]{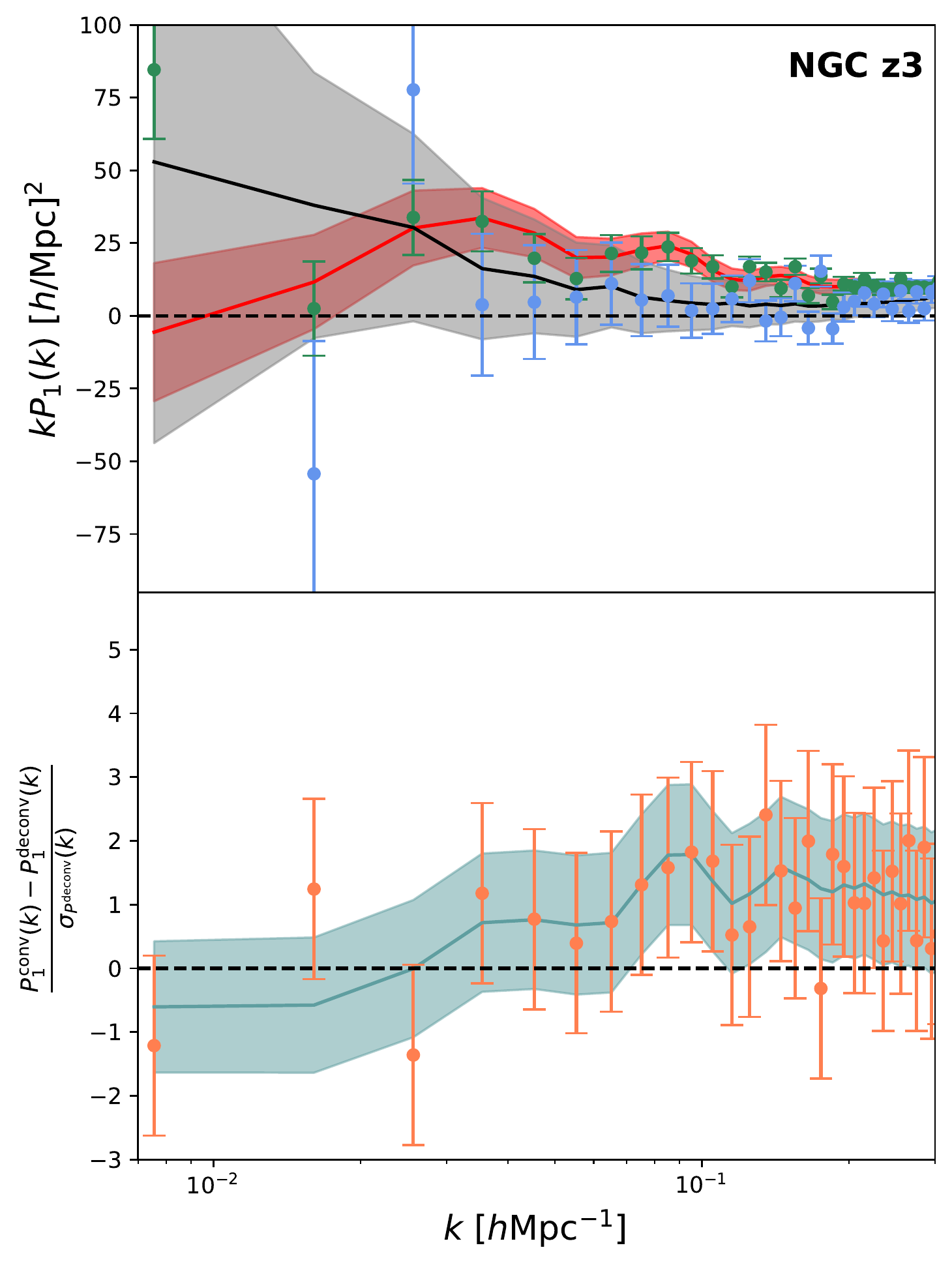}
    \includegraphics[width=0.19\textwidth]{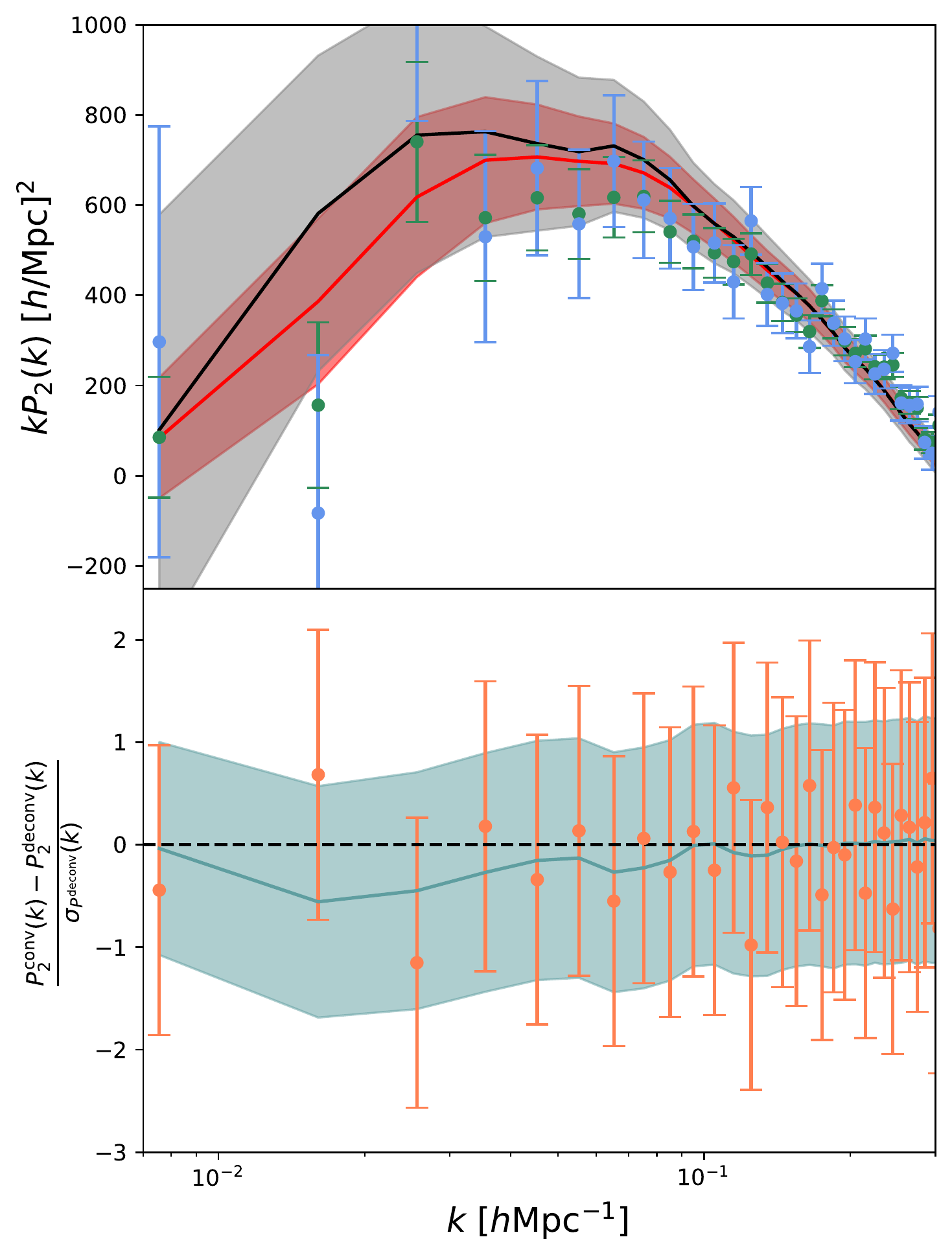}
    \includegraphics[width=0.19\textwidth]{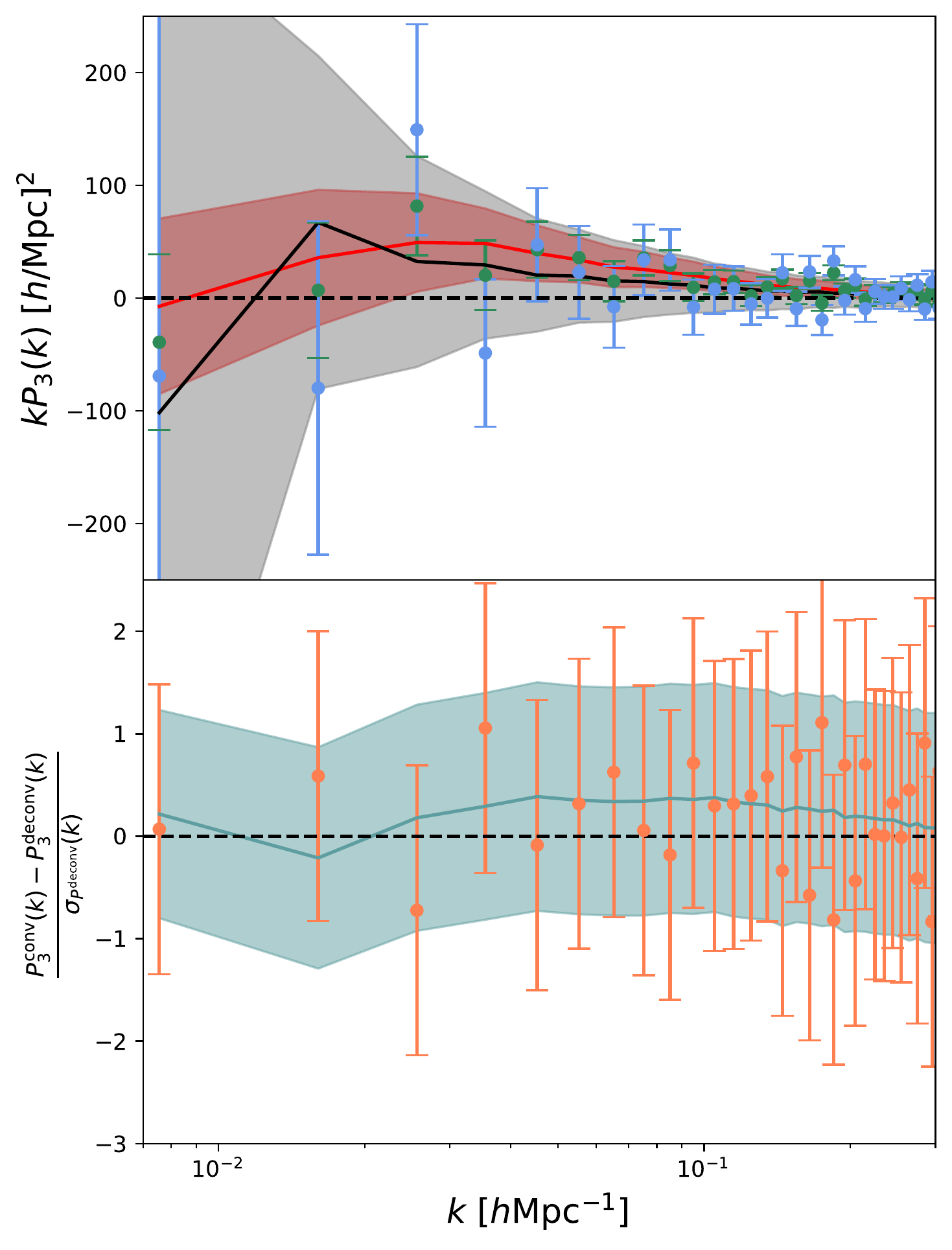}
    \includegraphics[width=0.19\textwidth]{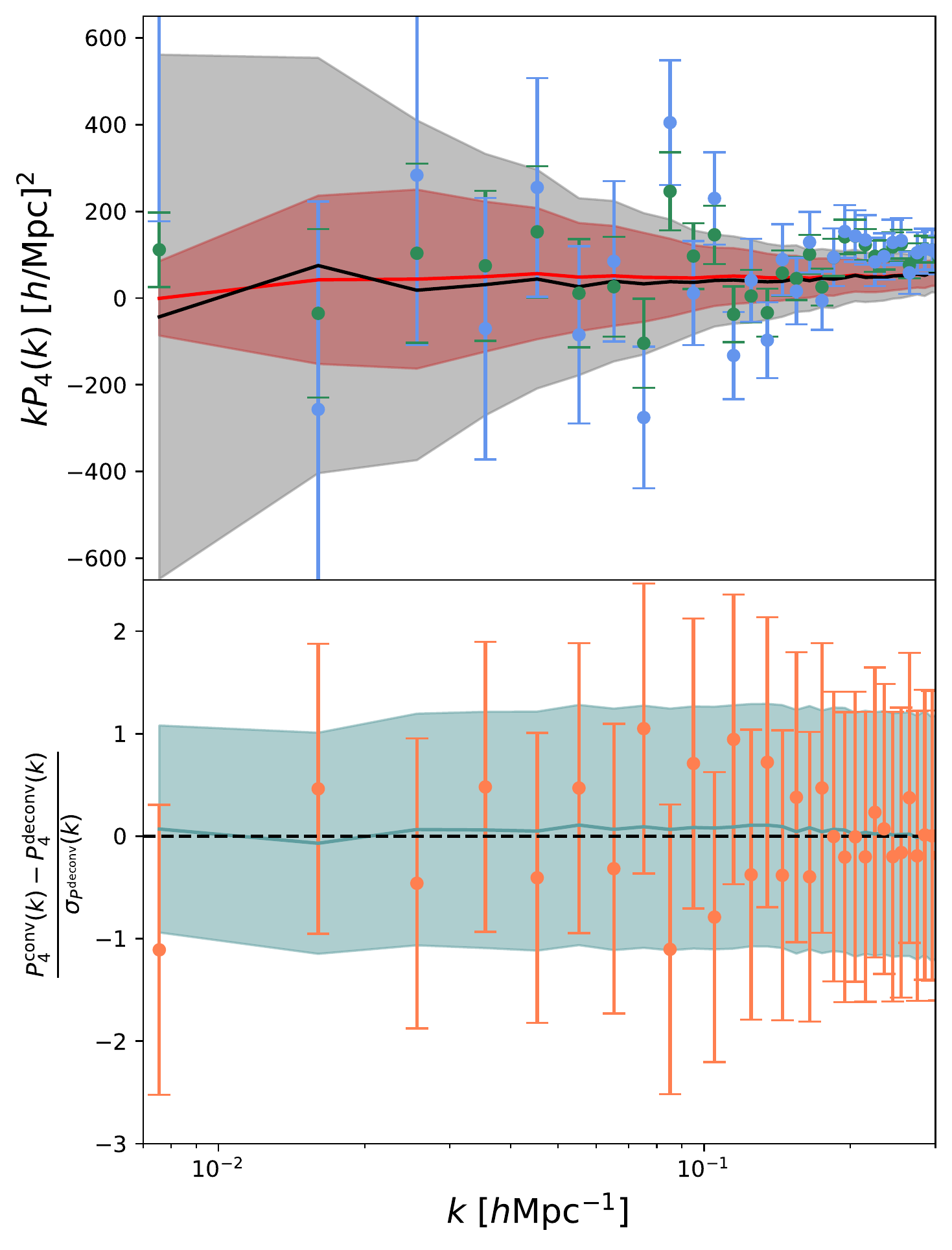}\\
    \includegraphics[width=0.19\textwidth]{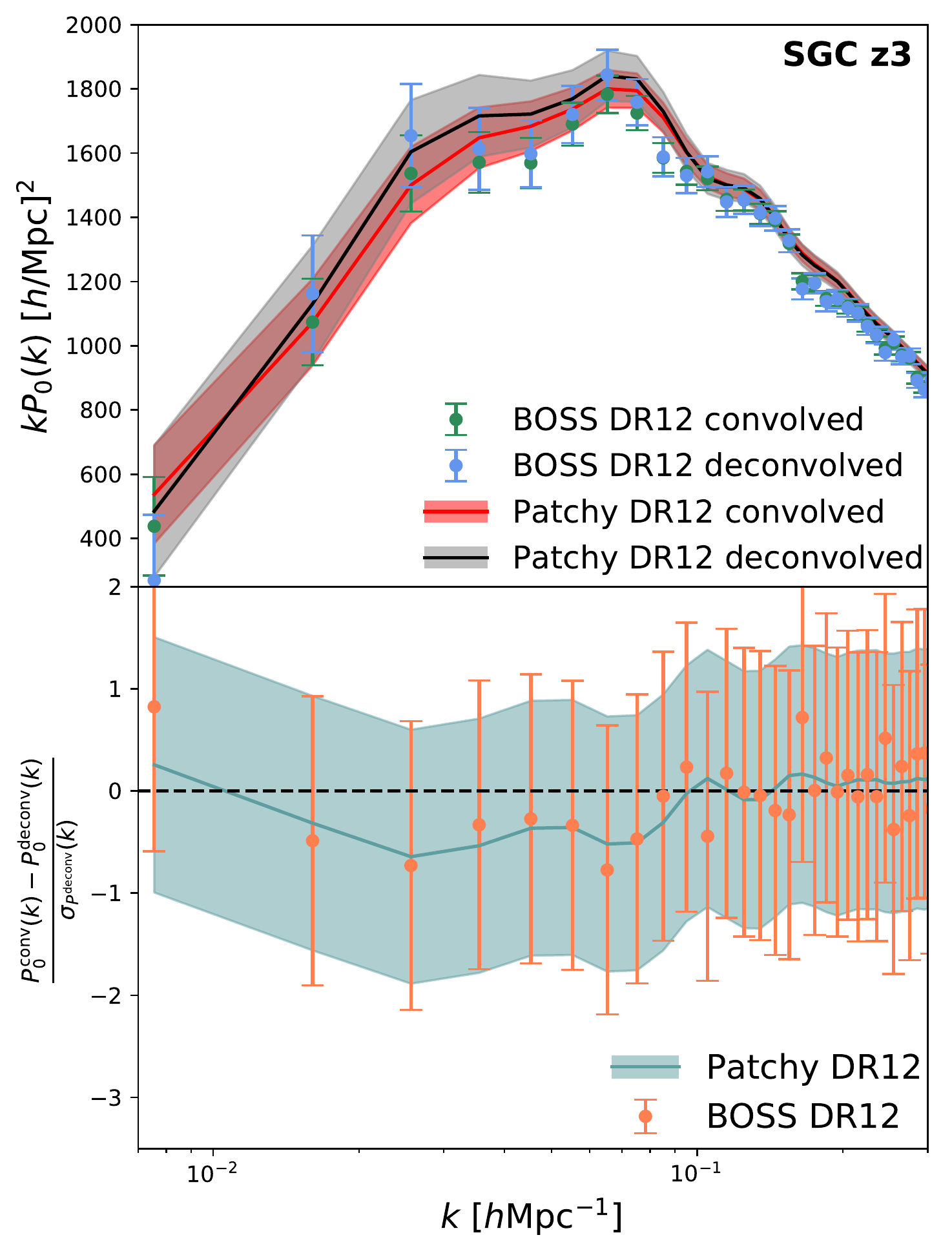}
    \includegraphics[width=0.19\textwidth]{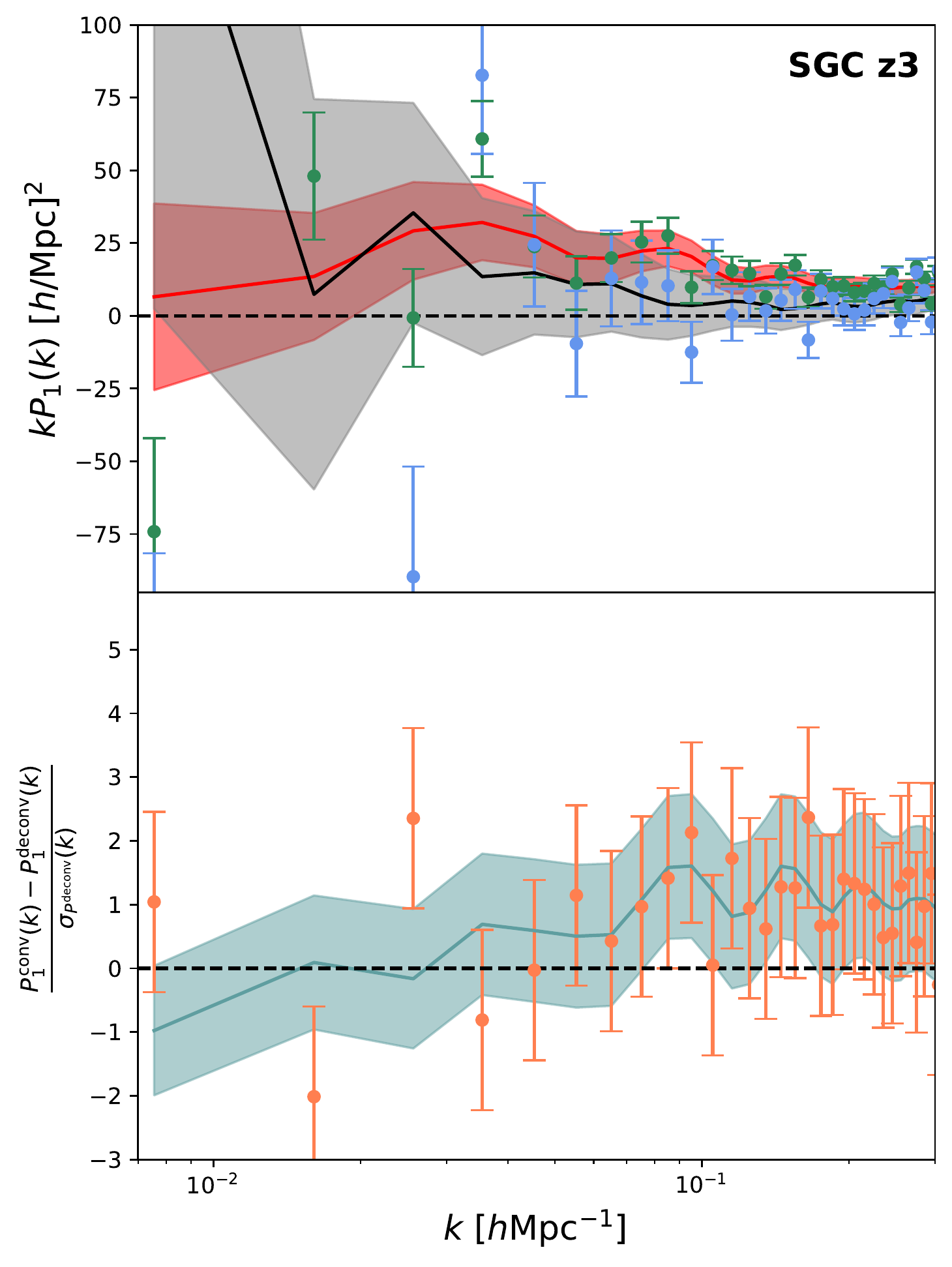}
    \includegraphics[width=0.19\textwidth]{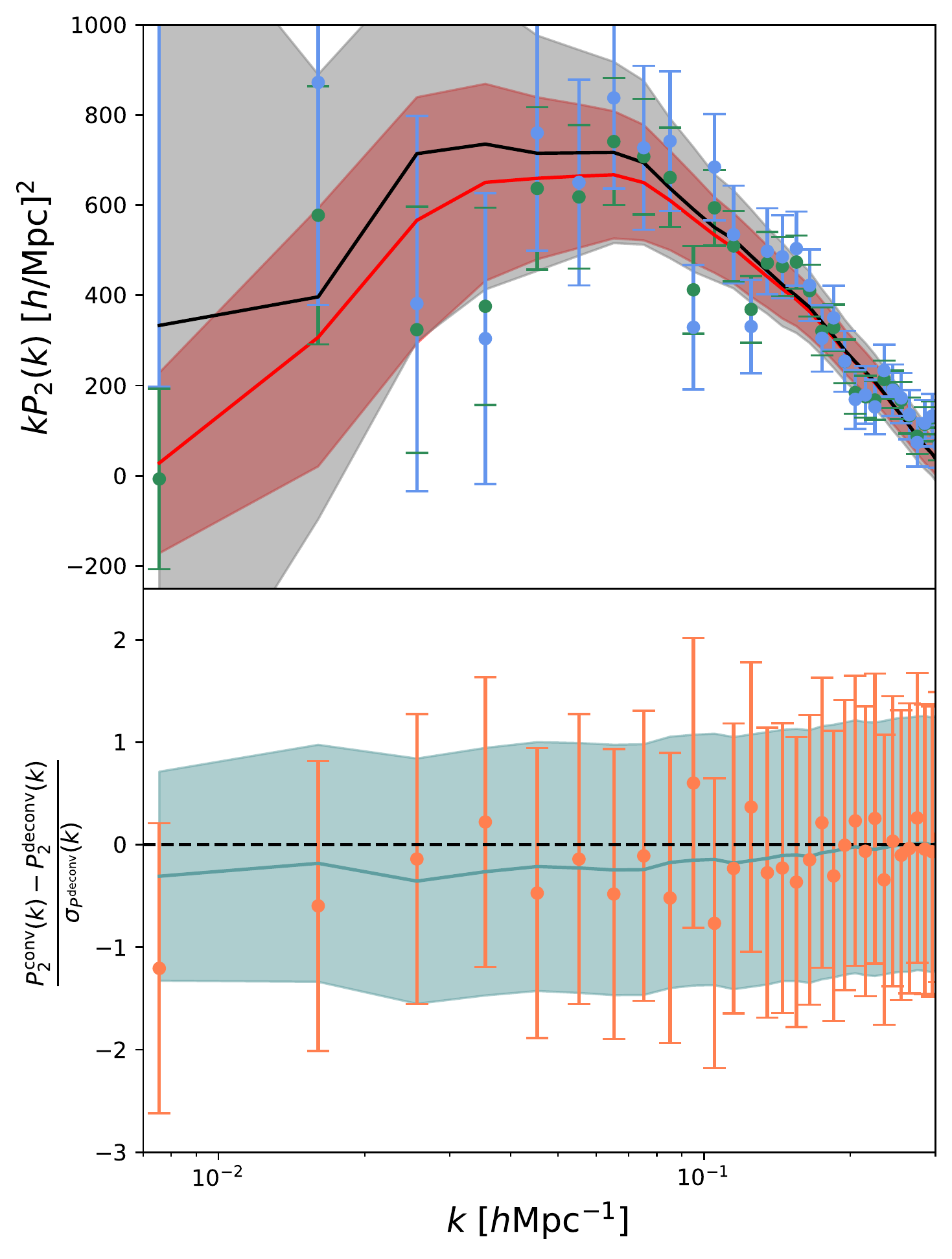}
    \includegraphics[width=0.19\textwidth]{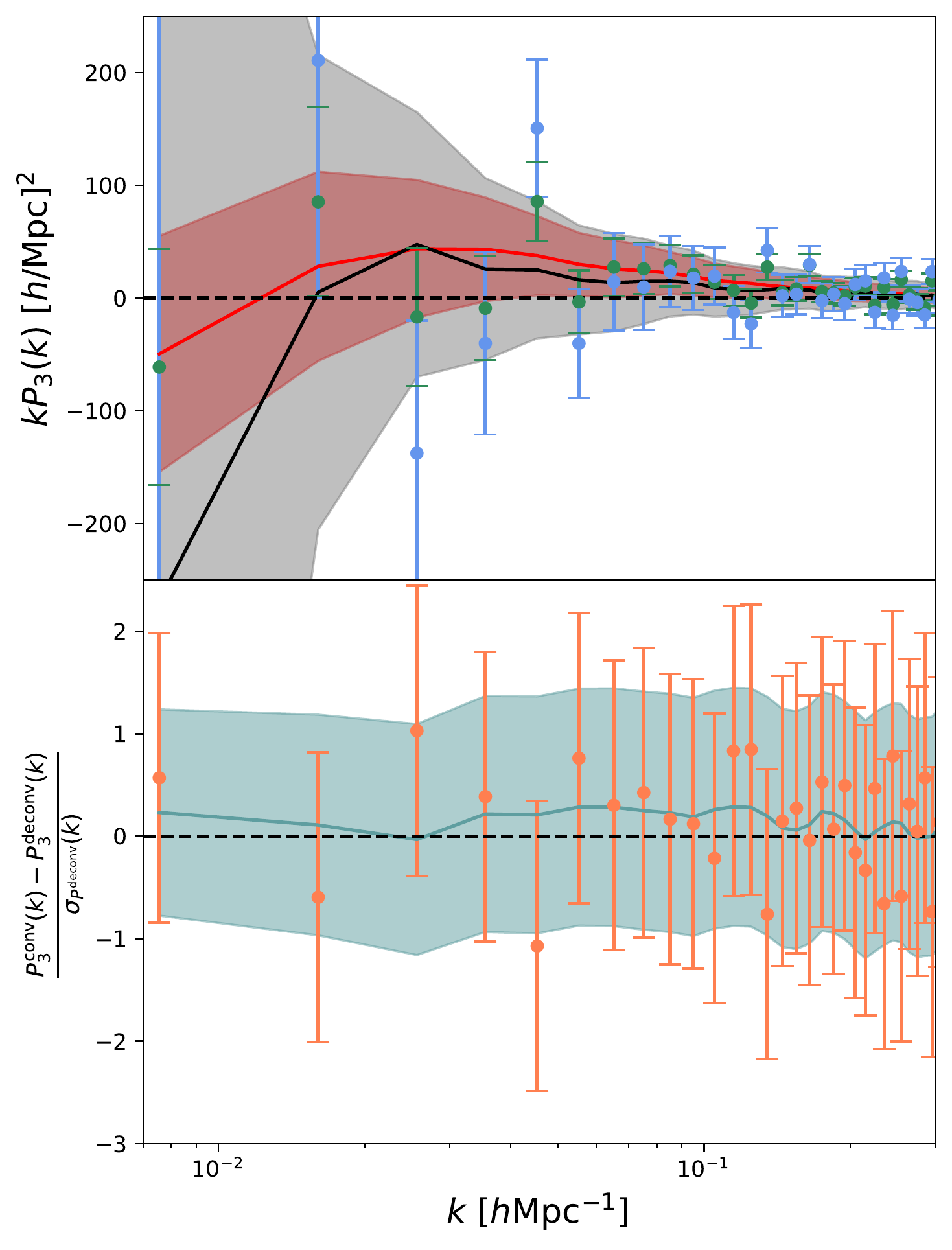}
    \includegraphics[width=0.19\textwidth]{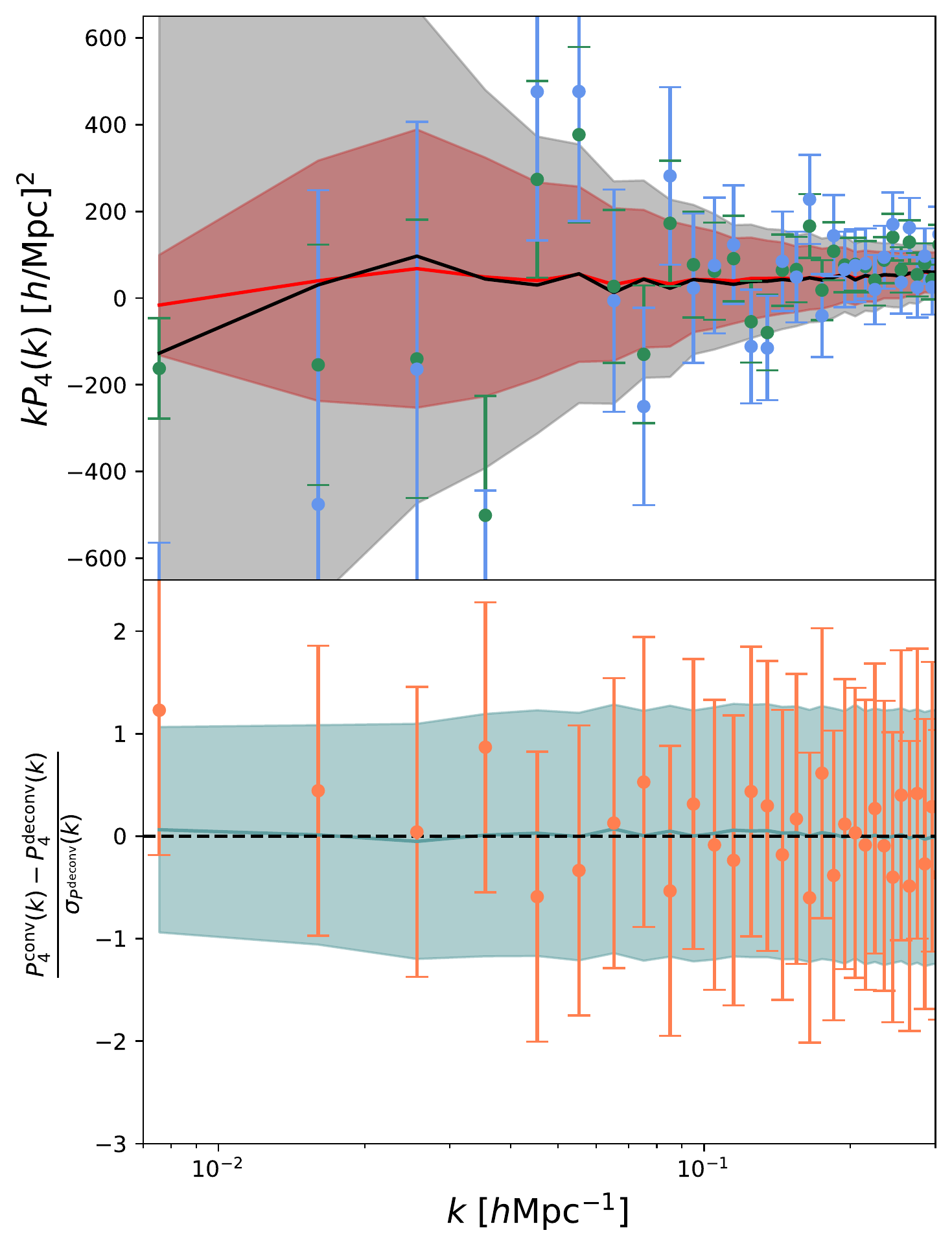}
    \caption{Comparison between the power spectrum multipoles of BOSS DR12 measured in the mock catalogs (gray and red shaded area) and in the data (data points). The results before deconvolution are shown as the red shaded area and solid red line (mocks) and the green data points. The deconvolved results are shown as the gray shaded area and solid black line (mocks) and the blue data points. The equivalent BOSS DR12 NGC results for the low redshift bin are included in the main text (see \fig{fig:even_decon_patchy_NGC_z1} and \ref{fig:odd_decon_patchy_NGC_z1}).}
    \label{fig:decon_boss}
\end{figure}

\begin{figure}[h]
    \centering
    \includegraphics[width=0.19\textwidth]{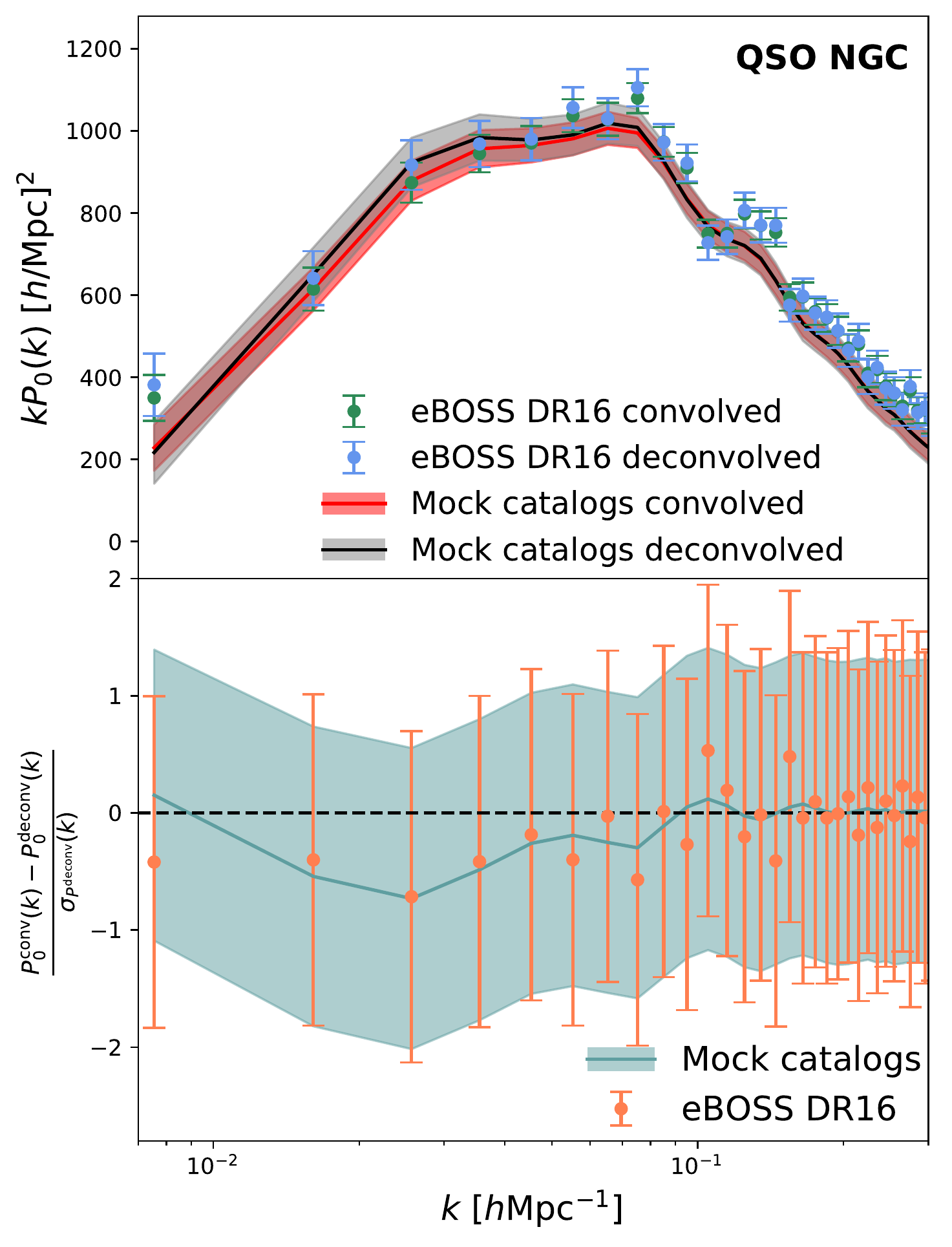}
    \includegraphics[width=0.19\textwidth]{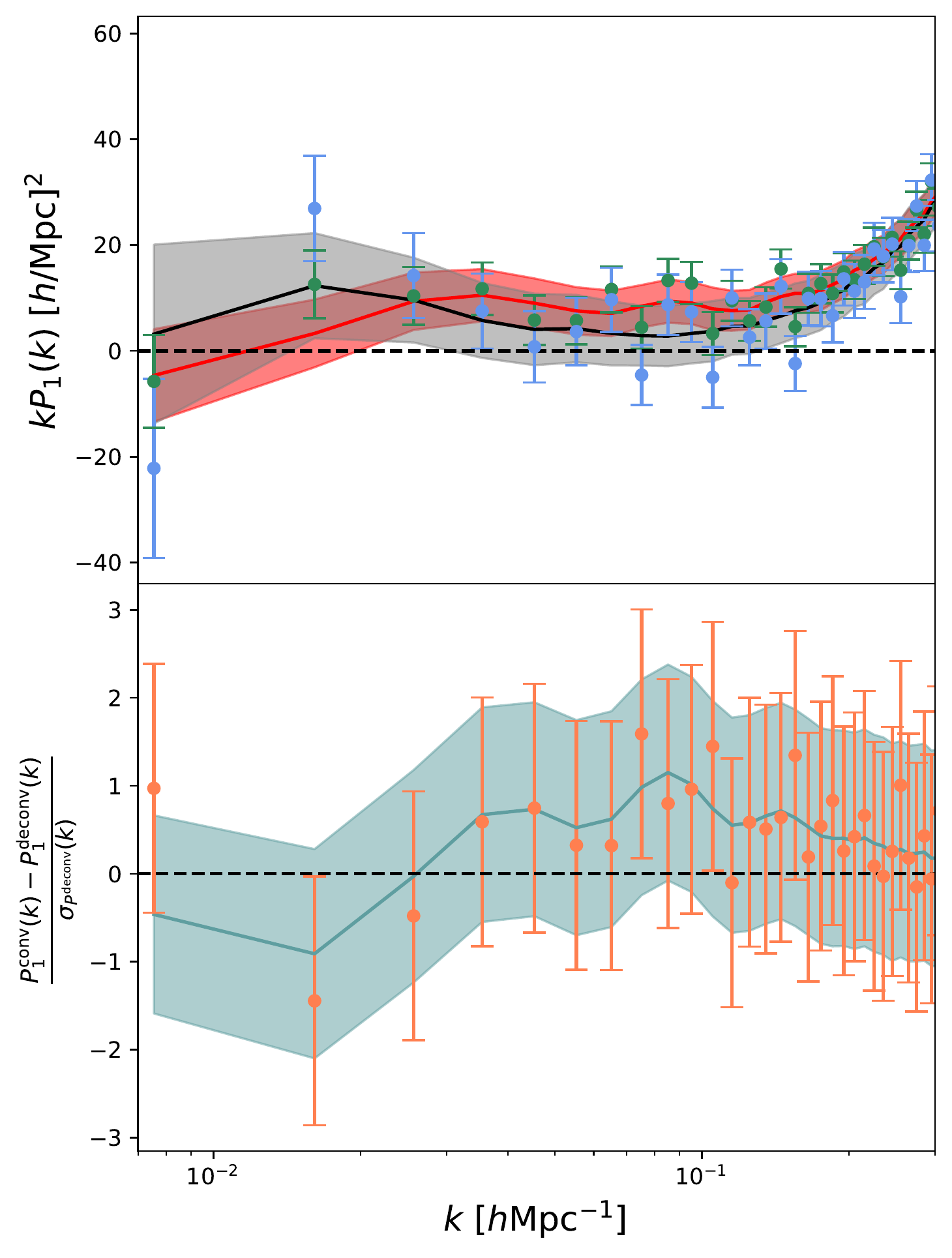}
    \includegraphics[width=0.19\textwidth]{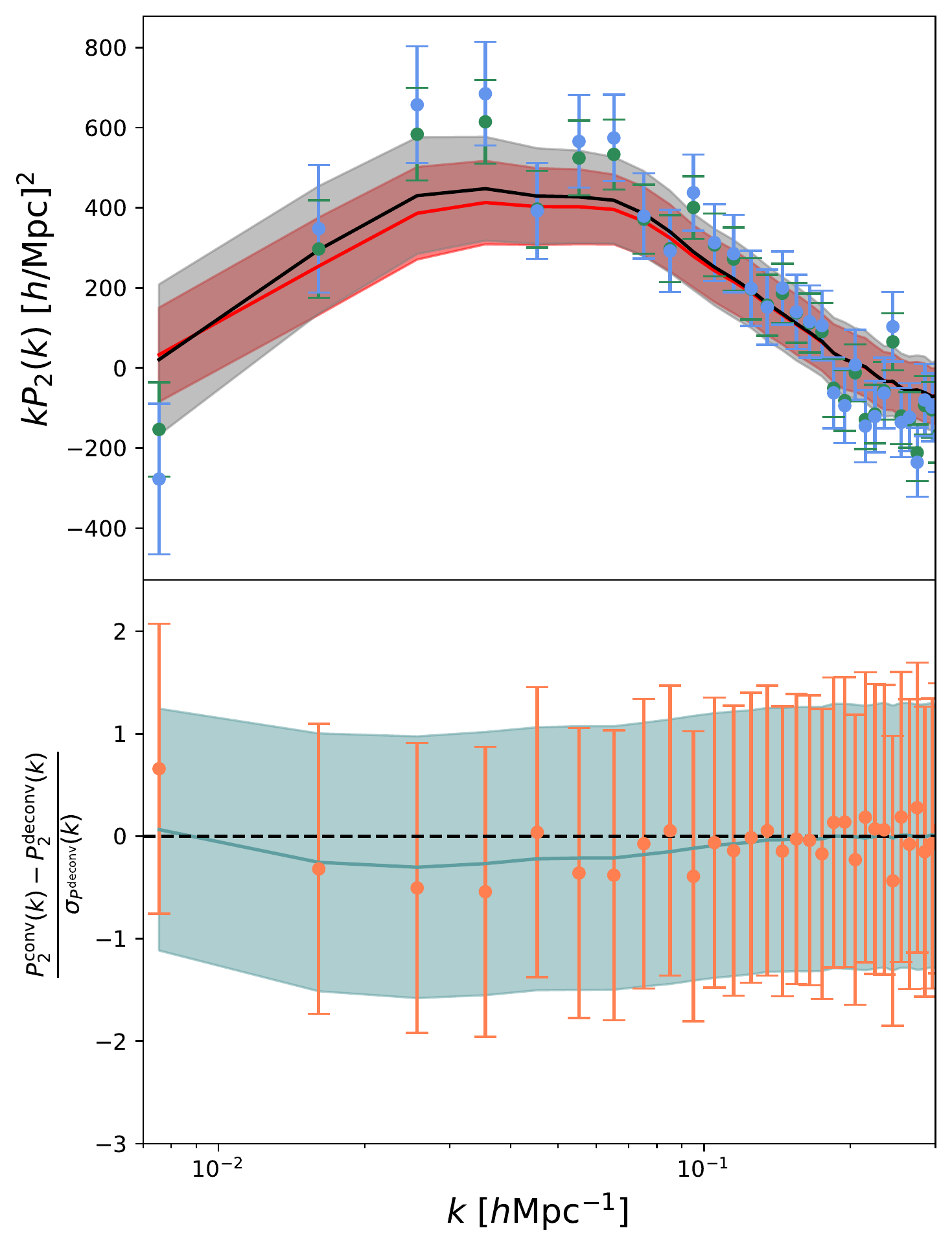}
    \includegraphics[width=0.19\textwidth]{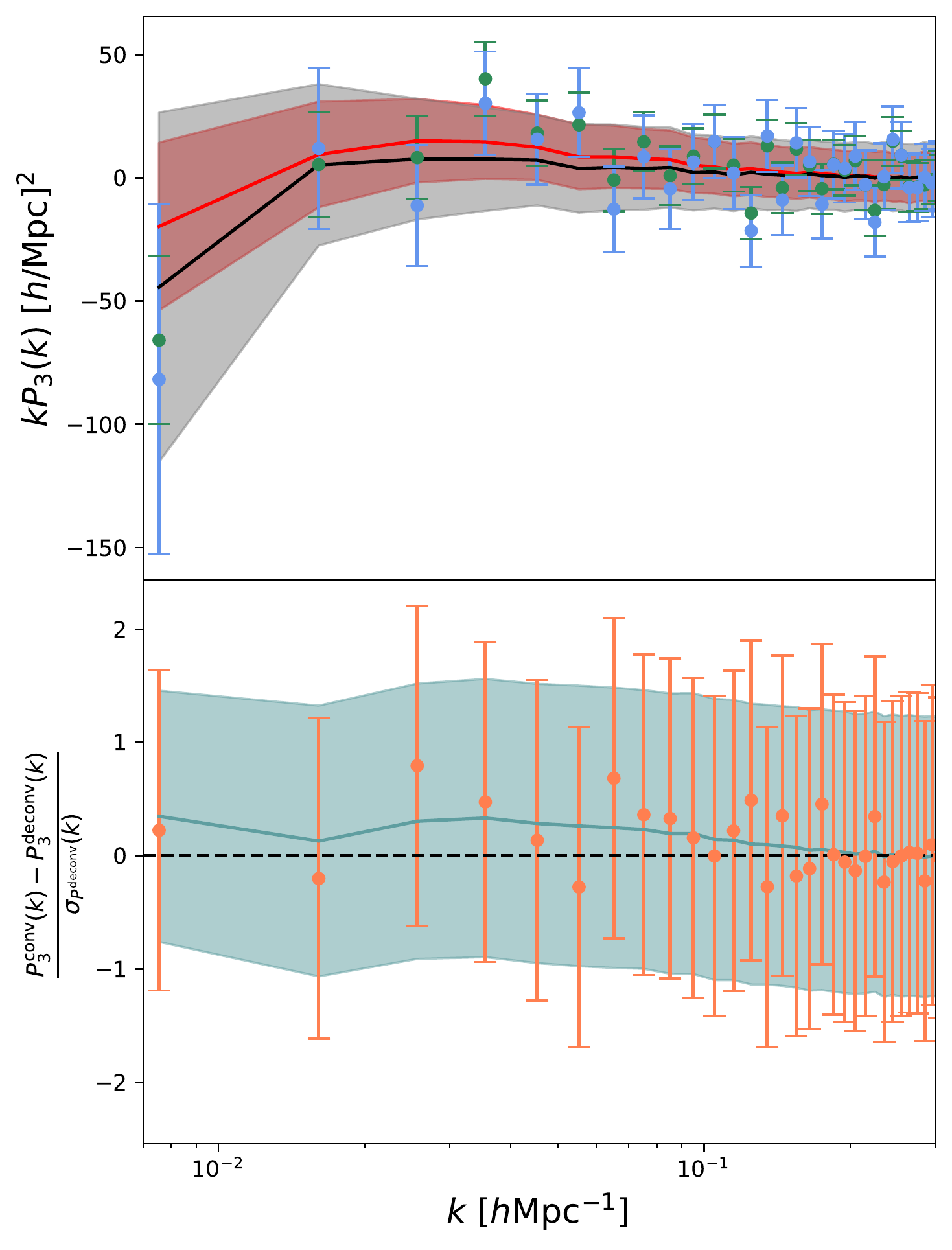}
    \includegraphics[width=0.19\textwidth]{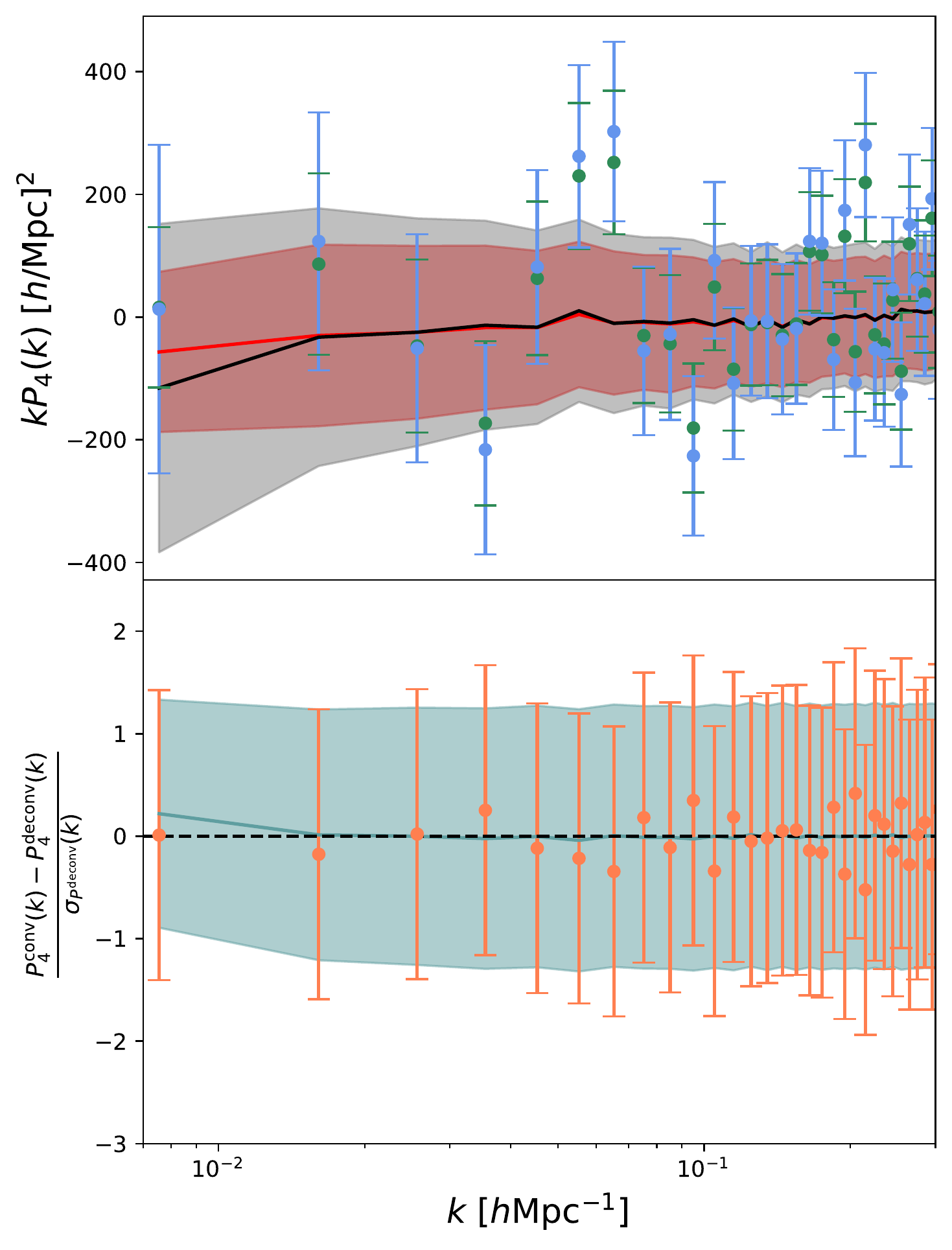}\\
    \includegraphics[width=0.19\textwidth]{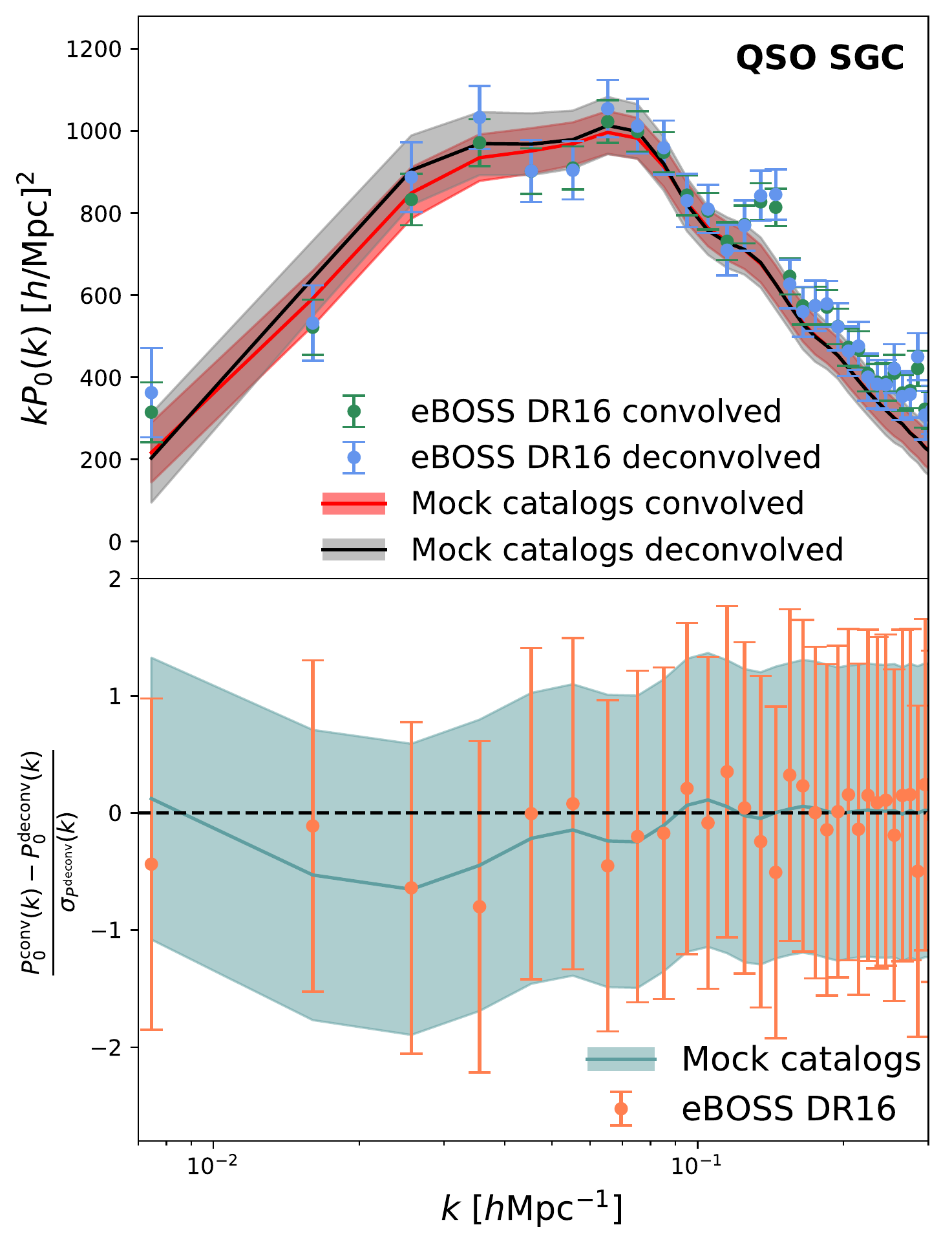}
    \includegraphics[width=0.19\textwidth]{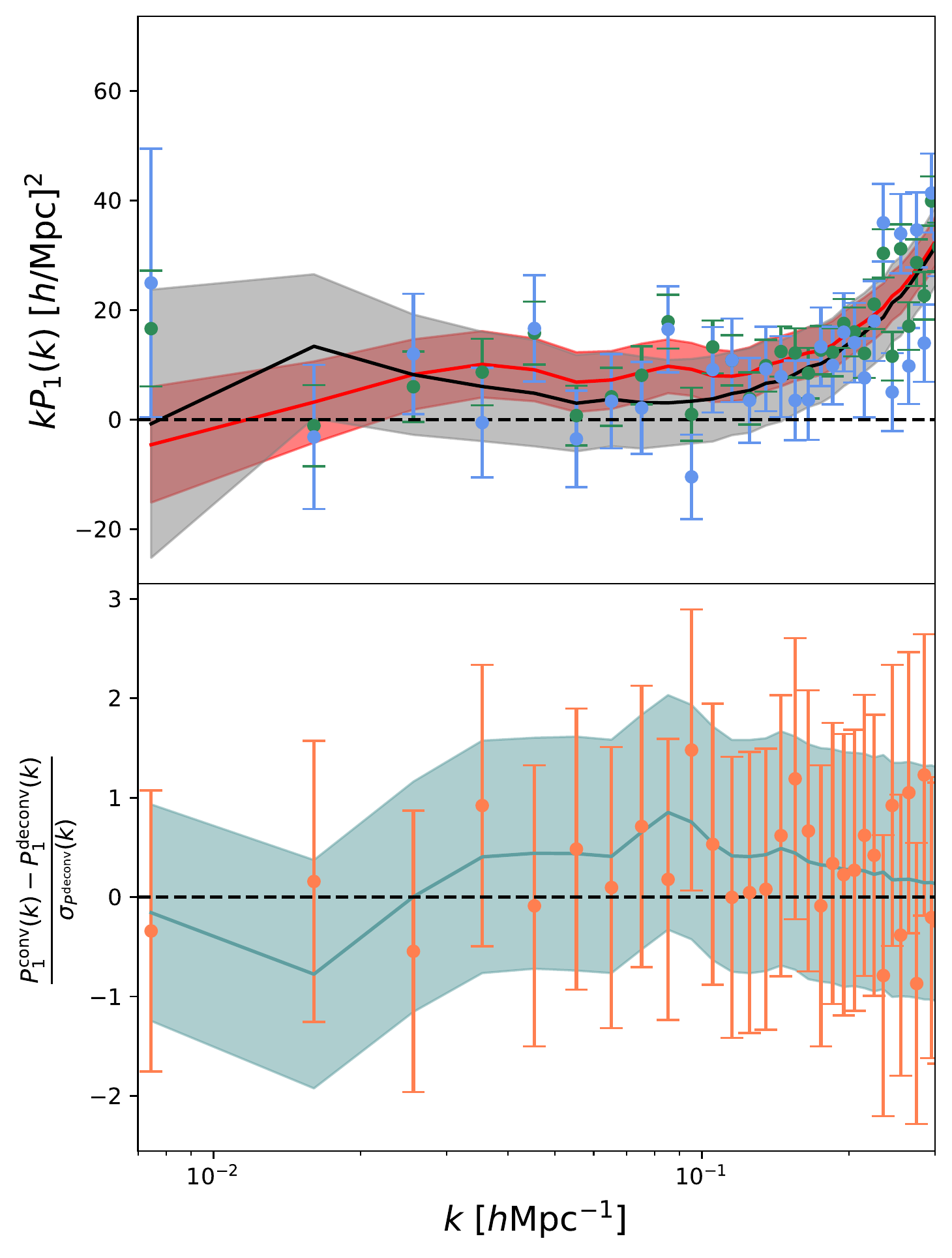}
    \includegraphics[width=0.19\textwidth]{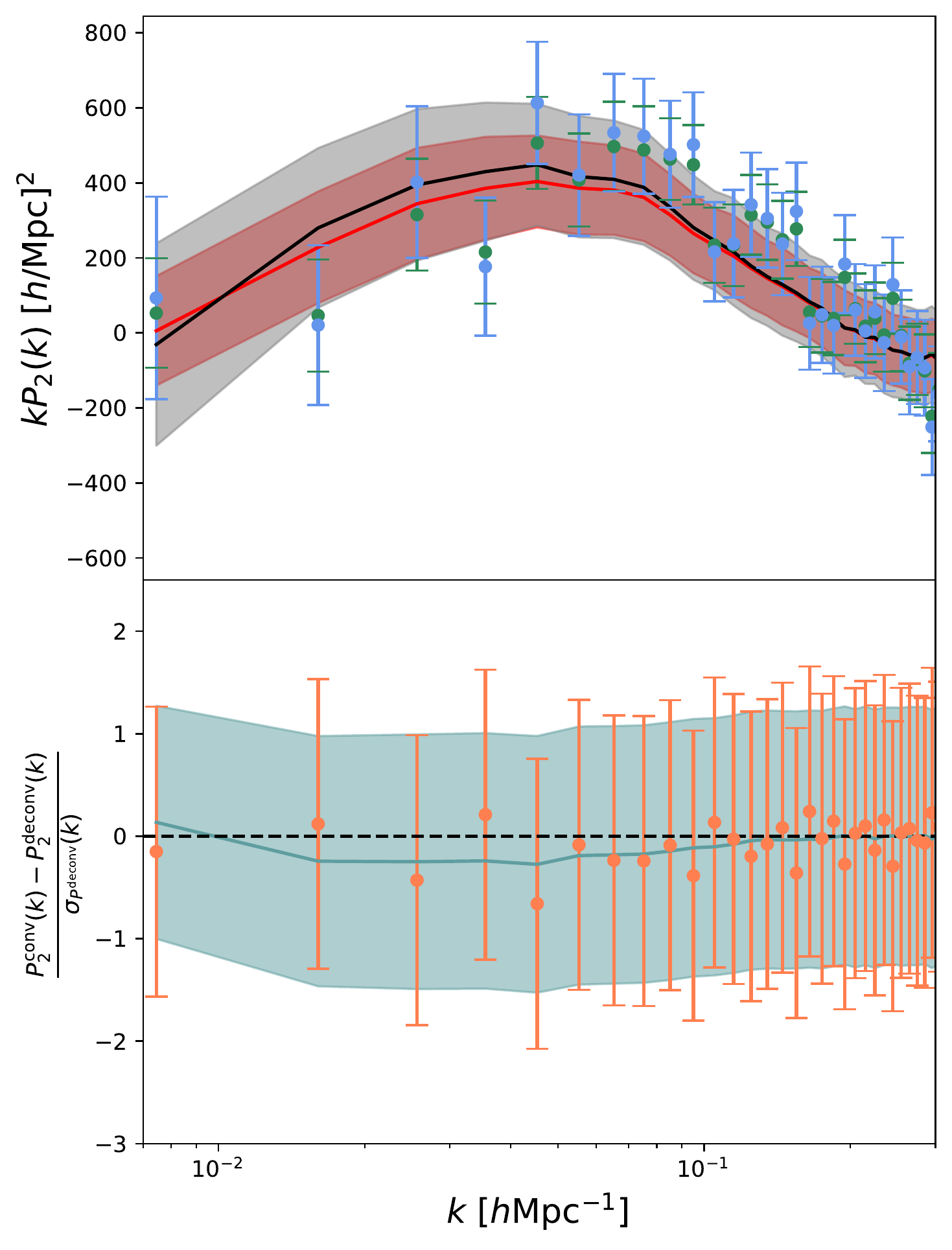}
    \includegraphics[width=0.19\textwidth]{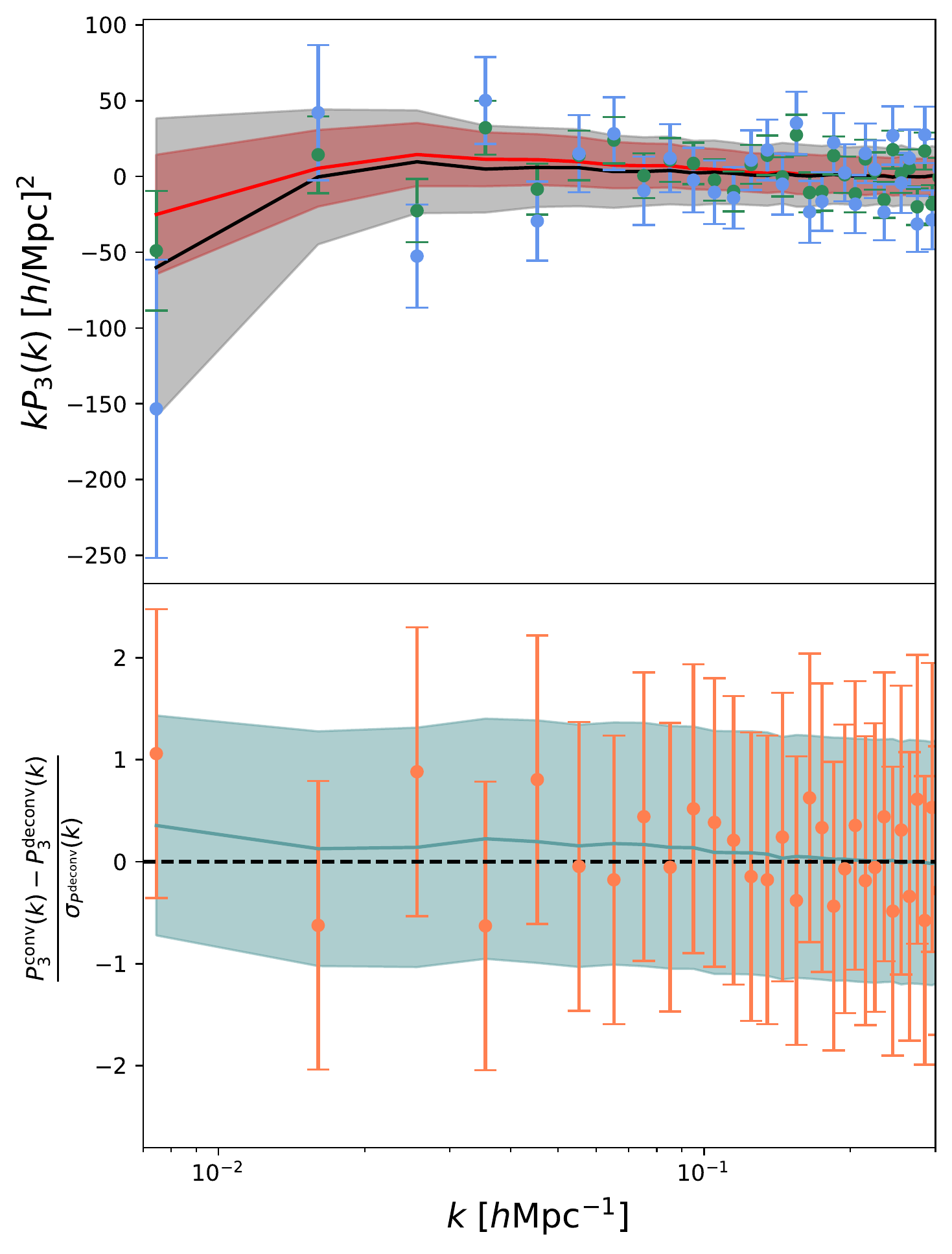}
    \includegraphics[width=0.19\textwidth]{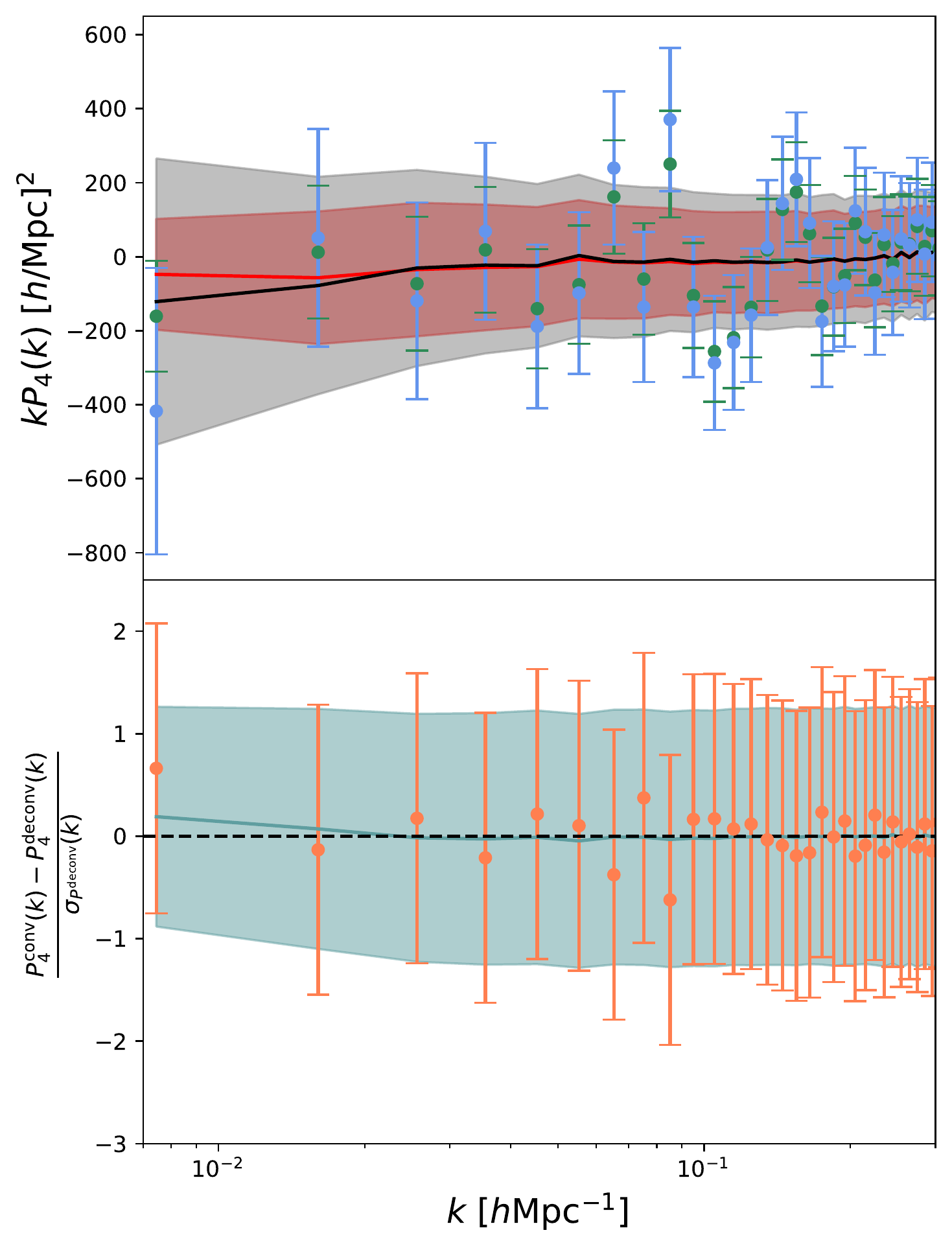}
    \caption{Comparison between the power spectrum multipoles of the eBOSS DR16 QSO sample measured in the mock catalogs (gray and red shaded area) and in the data (data points). The results before deconvolution are shown as the red shaded area and solid red line (mocks) and the green data points. The deconvolved results are shown as the gray shaded area and solid black line (mocks) and the blue data points. The increasing in the dipole power spectrum at high $k$ seems to be an aliasing effect, even though the Nyquist frequency is twice as high as the scale range plotted here ($k_{\rm Ny} > 0.6\kMpc$).}
    \label{fig:decon_eBOSS}
\end{figure}

Figure~\ref{fig:decon_6dfgs}, \ref{fig:decon_boss} and \ref{fig:decon_eBOSS} show the comparisons for the convolved and deconvolved power spectra for 6dFGS DR3, BOSS DR12 and the eBOSS DR16 QSO samples. Similar plots for the low redshift bin of BOSS DR12 NGC are included in the main text (see \fig{fig:even_decon_patchy_NGC_z1} and \ref{fig:odd_decon_patchy_NGC_z1}). The remaining plots are included here rather than the main text, to not interrupt the flow of the paper. 

\end{document}